\documentclass[preprint]{revtex4-1}

\usepackage[colorlinks=true,urlcolor=blue,citecolor=red,linkcolor=red,bookmarks=true]{hyperref}
\usepackage{natbib}

\usepackage[utf8]{inputenc}
\usepackage[T1]{fontenc}
\usepackage[normalem]{ulem} % Carlo readded this package
\usepackage{epstopdf, epsfig}

\usepackage{amsmath,amssymb,amsfonts}
\usepackage{graphicx}
\usepackage{xspace}
\usepackage{textcomp}
\usepackage{color}
\usepackage{tikz}
\usetikzlibrary{patterns}
\usetikzlibrary{calc}

\tikzset{
    ncbar angle/.initial=90,
    ncbar/.style={
        to path=(\tikztostart)
        -- ($(\tikztostart)!#1!\pgfkeysvalueof{/tikz/ncbar angle}:(\tikztotarget)$)
        -- ($(\tikztotarget)!($(\tikztostart)!#1!\pgfkeysvalueof{/tikz/ncbar angle}:(\tikztotarget)$)!\pgfkeysvalueof{/tikz/ncbar angle}:(\tikztostart)$)
        -- (\tikztotarget)
    },
    ncbar/.default=0.5cm,
}

\tikzset{square left brace/.style={ncbar=0.5cm}}
\tikzset{square right brace/.style={ncbar=-0.5cm}}

\tikzset{round left paren/.style={ncbar=0.5cm,out=120,in=-120}}
\tikzset{round right paren/.style={ncbar=0.5cm,out=60,in=-60}}

\newif\ifhyper
\hypertrue
\ifhyper
\hypersetup{
  citecolor = {green},
  urlcolor = {blue} 
} 

\newcommand\df[2]{\frac{\delta #1}{\delta #2}}
\newcommand\ddf[3]{\frac{\delta^2 #1}{\delta #2 \delta #3}}

\newcommand{\p}{\partial} 
\newcommand{\la}{\langle} 
 
\newcommand{\vx}{\vec{x}}

\newcommand{\vrr}{\vec{r}}
\newcommand{\vv}{\vec{v}}

\newcommand{\vJ}{\vec{J}}

\newcommand{\bq}{{\bf q}}
\newcommand{\bp}{{\bf p}}

\newcommand{\bk}{{\bf k}}
\newcommand{\bx}{{\bf x}}
\newcommand{\by}{{\bf y}}
\newcommand{\bz}{{\bf z}}
\newcommand{\vq}{{\vec{q}}}
\newcommand{\vp}{{\vec{p}}}
\newcommand{\vk}{{\vec{k}}}

\newcommand{\sref}[1]{Sec.~\ref{#1}}

\newcommand{\aref}[1]{Appendix~\ref{#1}}

\newcommand{\Eq}[1]{Eq.~(\ref{#1})}
\newcommand{\eq}[1]{(\ref{#1})}

\newcommand{\ie}{{\it i.e.}\xspace}
\newcommand{\eg}{{\it e.g.}\xspace}

\usepackage{abbrevs}
\usepackage{etoolbox}
\newabbrev\RG{Renormalisation Group (RG)}[RG]
\newabbrev\NPRG{Non-Perturbative (also named functional) Renormalisation Group (NPRG)}[NPRG]
\newabbrev\NS{Navier-Stokes (NS)}[NS]
\newabbrev\SNS{stochastic Navier-Stokes (SNS)}[SNS]
\newabbrev\PI{one particle-irreducible (1-PI)}[1-PI]
\newabbrev\PR{one particle-reducible (1-PR)}[1-PR]
\newabbrev\rhs{right-hand side (r.h.s.)}[r.h.s.]
\newabbrev\lhs{left-hand side (l.h.s.)}[l.h.s.]
\newabbrev\MSRJD{Martin-Siggia-Rose-Janssen-de Dominicis (MSRJD)}[MSRJD]
\newabbrev\UV{ultraviolet (UV)}[UV]
\newabbrev\IR{infrared (IR)}[IR]
\newabbrev\threeD{three-dimensional ($3$D)}[$3$D]
\newabbrev\twoD{two-dimensional ($2$D)}[$2$D]
\newabbrev\EAA{Effective Average Action (EAA)}[EAA]
\newabbrev\NLO{next-to-leading order (NLO)}[NLO]

\def\mean#1{\left< #1 \right>}

\robustify{\RG}
\robustify{\NPRG}
\robustify{\NS}
\robustify{\SNS}
\robustify{\PI}
\robustify{\PR}
\robustify{\rhs}
\robustify{\lhs}
\robustify{\MSRJD}
\robustify{\UV}
\robustify{\IR}
\robustify{\twoD}
\robustify{\threeD}
\robustify{\EAA}
\robustify{\NLO}

\makeatletter
\renewcommand\maybe@space@{%
  \maybe@ictrue % <= this is new
  \expandafter   \@tfor
    \expandafter \reserved@a
    \expandafter :%
    \expandafter =%
                 \nospacelist
                 \do \t@st@ic
  \ifmaybe@ic % <= this is new
    \space
  \fi
}
\makeatother

\begin{document}

\title{Stationary, isotropic and homogeneous two-dimensional turbulence: a first non-perturbative renormalization group approach}

\ResetAbbrevs{All}

\author{Malo Tarpin}
\affiliation{Universit\'e Grenoble Alpes and CNRS, LPMMC,  38000 Grenoble, France }

\author{L\'eonie Canet}
\email{leonie.canet@grenoble.cnrs.fr}
\affiliation{Universit\'e Grenoble Alpes and CNRS, LPMMC,  38000 Grenoble, France }

\author{Carlo Pagani}
\affiliation{Institute für Physik (WA THEP), Johannes-Gutenberg-Universität,Staudingerweg 7, 55099 Mainz, Germany}

\author{Nicol\'as Wschebor}
\affiliation{Instituto de F\'isica, Facultad de Ingenier\'ia, Universidad de la Rep\'ublica, J.H.y Reissig 565, 11000 Montevideo, Uruguay}

\begin{abstract}

We study the statistical properties of stationary, isotropic and homogeneous turbulence in \twoD flows, focusing
 on the direct cascade, that is on wave-numbers large compared to the integral scale, where both energy and enstrophy are provided to the fluid.
 Our starting point is the \twoD Navier-Stokes equation in the presence of a stochastic forcing, or more precisely 
  the associated field theory. We unveil two  extended symmetries of the \NS action which were not identified yet, one related 
 to time-dependent (or time-gauged) shifts of the response fields and existing in both \twoD and 3D, 
 and the other to  time-gauged rotations and specific to \twoD flows. We derive the corresponding Ward identities, and exploit them  within the non-perturbative renormalization group formalism, and the large wave-number expansion scheme developed in [Phys. Fluids {\bf 30}, 055102 (2018)]. We consider the flow equation for a generalized $n$-point correlation function, and  calculate 
 its leading order term in the large wave-number expansion. At this order, the resulting  flow equation can be closed exactly.
  We  solve  the fixed point equation for the 2-point function, which yields
   its explicit  time dependence,  for both small 
 and large time delays in the stationary turbulent state. On the other hand, at equal times, the leading order term vanishes, so we compute the  next-to-leading order term.
 We find that the flow equations for simultaneous $n$-point correlation functions are not fully constrained by the set of extended symmetries, and discuss the consequences.

\end{abstract}

\maketitle

\section{Introduction}

The \twoD Navier-Stokes equation is a relevant description for large scale atmospheric and oceanic flows, rotating fluids, or magnetically forced stratified turbulence \cite{Tabeling02,Boffetta12}.  In  \twoD, not only the energy, but also  the enstrophy (squared vorticity) is conserved, which yields the existence of a double cascade, spreading on two distinct inertial ranges, as early predicted by Kraichnan \cite{Kraichnan67}, Leith \cite{Leith68} and Batchelor \cite{Batchelor69}. The energy flows from the integral scale $L$, where both energy and enstrophy are injected, towards the large scales
 where it is dissipated by  some mechanism, such as an Eckman friction at scale $L_0$, while 
  the enstrophy flows towards the small scales, where it is dissipated by viscosity at the Kolmogorov scale $\eta$. In the forced-dissipative stationary regime, the inverse cascade of energy is characterized by Kolmogorov scalings with  a $k^{-5/3}$ decay of the energy spectrum. In the direct cascade of enstrophy,
 the scalings can also be deduced from Kolmogorov types of arguments, which yield a  $k^{-3}$ decay of the energy spectrum \cite{Kraichnan67,Leith68}. This scaling was later corrected by Kraichnan himself,  who suggested the presence of logarithmic corrections to the energy spectrum of the form $k^{-3}\ln(L k)^{-1/3}$ to ensure the  constancy of the enstrophy flux  \cite{Kraichnan71}. 
 Moreover, exact relations for the equal-time three-point correlations, analogous to the -4/5 law in \threeD turbulence, can also be derived in \twoD, exploiting  the conservation of both energy and enstrophy  in their respective inertial ranges \cite{Bernard99,Eyink96}. 

 %The double cascade scenario was confirmed by early numerical simulations \cite{Borue94,Babiano95}, although a high resolution is required to converge to the $k^{-3}$ spectrum in the absence of hyperviscosity \cite{Gotoh98}.
 In many respects, the understanding of \twoD turbulence appears more advanced than its \threeD counterpart. For instance, 
 exact bounds on the exponents of the structure functions are known \cite{Eyink96}.    
 The small-scales statistics of the vorticity in the direct cascade 
    was  investigated by Falkovich and Lebedev \cite{Falkovich94,Falkovich94b}. They  found that the Kolmogorov-like exponents
  of the  $2n$-point structure functions of the vorticity
  are not modified by intermittency effects, but that the power-laws are corrected by logarithms, following $\langle \omega^n(\vrr_1)\omega^n(\vrr_2)\rangle \propto \ln(L/|\vrr_1-\vrr_2|)^{2n/3}$. 
  This result   received support from experimental measurements in  electromagnetically forced conducting fluid layers turbulence \cite{Paret99} and in flowing soap films \cite{Bruneau05}. These works all indicate that    
 there is   no substancial intermittency in the small-scale statistics of \twoD turbulence, at least in the absence of an Ekman friction. In the presence of such a term, the exponents of the structure functions are changed by intermittency corrections, which depend on the friction coefficient  \cite{Nam00,Bernard00,Boffetta02}.  Let us also mention that perturbative \RG techniques have been applied to study \twoD turbulence in the presence of a power-law forcing  \cite{Honkonen98,Mayo05,Mazzino2009}. We refer the reader to reviews on \twoD turbulence for a more exhaustive account
  \cite{Tabeling02,Lesieur08,Eyink96,Boffetta12}.

  However, the complete characterization of the statistical properties of \twoD turbulence remains a fundamental quest.
  In particular, the previous results concentrate  on the structure functions, which are equal-time quantities, but the time dependence of velocity or vorticity correlations is also of fundamental interest.  In this respect, recent theoretical works have shown that the 
 time dependence  of  correlation (and response) functions could be calculated within the \NPRG formalism
 \cite{Canet17,Tarpin18}. This framework allows one to compute statistical properties of turbulence from ``first principles'', in the sense that it is based on the
  forced \NS equation and does not require phenomenological inputs \footnote{Let us point out that from the mathematical point of view,
  it is implicitly assumed in the present manuscript that, for reasonable boundary conditions, there exists a unique weak solution of the forced Navier-Stokes equation, except possibly for some configurations of the forcing with probability zero.}.
 It was exploited in \threeD to obtain the exact  time dependence of $n$-point generalized velocity correlation functions at leading order at
 large wave-numbers  at non-equal times \cite{Tarpin18}. The case of equal times is much more involved and its complete analysis in \threeD is still lacking.

  The purpose of this paper is to use the \NPRG formalism to investigate  \twoD turbulence. The outcome is three-fold.  
  We identify two new extended symmetries of the \twoD \NS action and derive the associated Ward identities.
  At leading order in wave-numbers, we show that the flow equation for a generic $n$-point correlation function in the stream formulation can be closed exactly as in the velocity formulation in \threeD, and we derive the fixed-point solution for the $2$-point correlation. The corresponding predictions can be tested in experiments or numerical simulations.  Moreover, we present a first step in the analysis of  equal-time correlations in \twoD. We compute the \NLO term in the large wave-number expansion,  in order to probe the presence of intermittency corrections at equal times.  Interestingly, almost all the terms are controlled by
the (extended) symmetries. These  controlled terms  turn out to vanish at equal times, which means that,
  as  the leading terms,
they cannot generate intermittency effects for simultaneous correlations. However, the symmetries do not seem to be sufficient to completly constrain the flow equation at \NLO, and the only remaining term could be non-zero and be responsible for intermittency.  Our analysis suggests that the corresponding effects are anyhow weaker than in \threeD, since almost all the terms are controlled and vanish in \twoD because of the specific time-gauged rotation symmetry, which does not occur in \threeD. This is in accordance with standard observations, which support weak or no intermittency in the direct cascade in \twoD turbulence.

  The remainder of the paper is organized as follows.  The field theory associated with the \twoD \NS equation in the presence of a stochastic forcing is revisited in \sref{sec:FieldTheory}, where its extended symmetries are analyzed in details.
  The \NPRG formalism to study this field theory, developed in \cite{Canet15,Canet16,Tarpin18}, is briefly presented in \sref{sec:NPRG}, and
  the Ward identities related to the extended symmetries are derived in this framework. We then consider the flow equation for a generalized $n$-point correlation function in the stream formulation, and explain the principles of the  large wave-number expansion.
  In \sref{sec:timedependence}, the leading order term in this expansion is calculated exactly, and the corresponding fixed-point solution
 is obtained for the two-point function,  yielding the general form of its time dependence. 
   The leading order term vanishes at equal time, as in \threeD. In \sref{sec:equaltime}, we calculate the \NLO term in the large wave-number expansion, focusing on coinciding times, and discuss the results.

\section{Field theory for the stream function and its symmetries}
\label{sec:FieldTheory}

We consider the \NS equation in the presence of an external stirring force $f^{\rm inj}_\alpha$ and of an energy damping force $f^{\rm damp}_\alpha$, in order to sustain a stationary turbulent regime 
\begin{equation}
 \p_t v_\alpha  + v_\beta \p_\beta v_\alpha = \nu \p^2 v_\alpha -  \p_\alpha p + f^{\rm damp}_\alpha+ f^{\rm inj}_\alpha\,,
\end{equation}
where $\nu$ is the kinematic viscosity and $p$ is the pressure divided by the density of the fluid, and where the velocity, pressure and  force fields depend on the space-time coordinate $(t,\vx)$. This equation is supplemented by the incompressibility constraint
\begin{equation}
 \p_\alpha v_\alpha = 0\,.
\end{equation}
Since in \twoD the energy is transferred towards the large scales, the energy damping force is necessary to provide a dissipation mechanism
 at the largest scales. This damping can be achieved by a linear Ekman friction, $\vec f^{\rm damp} = -\alpha \vv$,
 which models the friction exerted on the bulk by the surrounding layers in which the \twoD flow is embedded. In this work, we consider
 a non-local generalization of this term with a characteristic length scale $L_0$
  \begin{equation}
 \vec  f^{\rm damp}(t,\vx) = -\int_{\vx\,'} R_{L_0^{\text{-}1}}(|\vx-\vx\,'|) \vv(t,\vx\,')\,.
\label{eq:fdamp}
\end{equation}
 The function $R_{L_0^{\text{-}1}}$ is chosen such that its Fourier transform vanishes exponentially at large wave-numbers $k \gg L_0^{-1}$. This term can be interpreted as an effective friction acting only at the boundaries of the fluid, while not affecting the small scales, contrary to the Eckman friction.

For the derivation of the associated field theory  and since universality is expected with respect to the precise form of the stirring force
 (as long as its characteristic distance scale $L$ is $L \gg \eta$),
  one usually chooses a stochastic forcing with Gaussian distribution of variance
\begin{equation}
 \mean{f^{\rm inj}_\alpha(t,\vx) f^{\rm inj}_\beta(t',\vx\,')}= D_{\alpha\beta}(t-t',\vx-\vx\,') = 2\, \delta_{\alpha\beta}\, \delta(t-t')N_{L^{\text{-}1}}(|\vx-\vx\,'|)\,.
\label{eq:finj}
\end{equation}
 The mapping to a field theory is then 
 achieved using the \MSRJD response-field formalism, developed in \cite{Martin73, Janssen76, Dominicis76}. Whereas in most derivations in the context of turbulence, the profile $N$  has to be a power-law \cite{Forster77,Dominicis79,Honkonen98,Adzhemyan99,Zhou10}, 
 it can within the \NPRG be shaped as a realistic large distance-scale forcing. The function $N_{L^{\text{-}1}}$ is hence chosen such that its Fourier transform is smooth, is peaked at the scale $L^{\text{-}1}$, is zero at  vanishing wave-number and decays exponentially at large wave-number \footnote{Note that one does not need to impose the forcing to be solenoidal, as  the incompressibility is explicitly enforced along each realization of the flow. As a consequence, it can be chosen diagonal in component space without loss of generality}.
 The \MSRJD formalism with the non-local terms \eq{eq:fdamp} and \eq{eq:finj} is presented in \cite{Canet15,Canet16,Tarpin18}. 
 It yields the partition function for the velocity, pressure, response  fields, under the form 
\begin{equation}
\mathcal{Z}[\vec J, \vec{ \bar J}, K, \bar K] =  \int \mathcal{D}[\vv,  \,\vec{\bar v},\,p, \,\bar p]\,
 \, e^{-\mathcal{S}_v[\vv,\vec{\bar v},p,\bar p] -\Delta \mathcal{S}_{v}[\vv,\vec{\bar v}] } e^{\int_{\bx}\{ \vJ\cdot \vv+\vec{\bar J}\cdot \vec{\bar v}+K p+\bar K\bar p\}} \, ,
\label{eq:Z}
\end{equation}
with the notation $\bx \equiv (t,\vx)$ and $\int_{\bx} = \int d^d \vx dt$, and similarly in the following $\bp \equiv(\omega,\vp)$ and 
 $\int_{\bp}\equiv \int \frac{d^d \vp}{(2\pi)^d}\frac{d\omega}{2\pi}$. In the \MSRJD formalism,
 the response velocity $\vec{\bar v}$ and response pressure $\bar p$ are introduced as the Lagrange multipliers of the equation of motion and of the incompressibility constraint respectively, and  $\vec J$, $\vec{\bar J}$, $K$, and $\bar K$ are the sources for the four fields \footnote{Note that the standard derivation of the \MSRJD action for \NS implicitly assumes existence and unicity of weak solutions of the Navier-Stokes equation, which is
   is a delicate issue from a mathematical point of view \cite{Buckmaster17}. The assumption is actually a little weaker than strict uniqueness, since for a typical set of initial conditions, there may exist a set of velocity configurations spoiling unicity, as long as they are of zero measure with respect to the realizations of the forcing.}.  The \NS action for the velocity  is obtained as
\begin{align}
 \mathcal{ S}_v[\vv,\vec{\bar v},p,\bar p] &= \int_{\bx}\Big\{\bar v_\alpha(\bx)\Big[ \p_t 
v_\alpha(\bx) 
 -\nu \nabla^2 v_\alpha(\bx) +v_\beta (\bx)\p_\beta v_\alpha(\bx)+\frac 1\rho \p_\alpha p (\bx) \Big] + \bar p (\bx)\,\p_\alpha v_\alpha(\bx)\Big\}\nonumber\\
\Delta \mathcal{ S}_v[\vv,\vec{\bar v}] &= \int_{t,\vx,\vx\,'} \Big\{ \bar v_\alpha(t,\vec x) R_{L_0^{\text{-}1}}(|\vx-\vx\,'|) v_\alpha(t,\vx\,')- \bar v_\alpha(t,\vec x) N_{L^{\text{-}1}}(|\vx-\vx\,'|)\bar v_\alpha(t,\vx\,')\Big\}\, ,
\label{eq:ActionSNS}
\end{align}
where the quadratic non-local part has been separated for latter purposes. These expressions hold in a generic dimension $d$. We now specialize to two dimensions, and introduce the stream function formulation.

\subsection{Action for the stream function}

In \twoD, the incompressibility constraint allows one to express the velocity as
\begin{equation}
v_\alpha = \epsilon_{\alpha\beta} \p_\beta \psi
\label{eq:defStream}
\end{equation}
where $\psi$ is a pseudo-scalar field, called the stream function, and
 $\epsilon_{\alpha\beta}$ are the components of the antisymmetric tensor
 with two indices and with $\epsilon_{12}=1$, which satisfies the \twoD identity
\begin{equation}
 \epsilon_{\alpha\gamma}\epsilon_{\beta\gamma}=\delta_{\alpha\beta}\,.
\label{eq:ideps}
\end{equation}
 The stream function is related to the vorticity field through a Laplacian:
\begin{equation}
 \omega = \epsilon_{\alpha\beta} \p_\alpha v_\beta = \epsilon_{\alpha\beta} \epsilon_{\beta\gamma} \p_\alpha \p_\gamma \psi = - \p^2 \psi \, .
\end{equation}
 From the field theory \eq{eq:Z} and \eq{eq:ActionSNS}, setting the sources $K$ and $\bar K$ to zero, integrating
over the pressure fields $p$ and $\bar p$, and using the resulting incompressibility contraints for $\vv$ and $\bar \vv$, the \NS action for the velocity  can be expressed as an action for the stream function, as
\begin{align}
 \mathcal{S}_{\psi}[\psi, \bar \psi] &= \int_{\bx} \p_\alpha \bar \psi(\bx) \Big[ \p_t \p_\alpha \psi(\bx) - \nu \nabla^2 \p_\alpha \psi(\bx) +\epsilon_{\beta\gamma} \,\p_\gamma \psi(\bx) \,\p_\beta \p_\alpha \psi(\bx) \Big ] \nonumber\\
 \Delta \mathcal{S}_{\psi}[\psi, \bar \psi] &= \int_{t,\vx,\vx'} \Big\{\p_\alpha  \bar\psi(t, \vx) R_{L_0^{\text{-}1}}(|\vec x-\vec x'|) \p_\alpha' \psi (t, \vx') - \p_\alpha \bar\psi(t, \vx) N_{L^{\text{-}1}}(|\vec x-\vec x'|) \p_\alpha' \bar \psi (t, \vx')  \Big\}\,,
\label{eq:ActionNSstream}
\end{align}
where the response stream is related to the response velocity through
\begin{equation}
 \bar v_\alpha  = \epsilon_{\alpha\beta} \p_\beta \bar \psi\,.
\label{eq:defbarStream}
\end{equation}
 One can then introduce a source for the stream function (resp. for the response stream) ${J = -\epsilon_{\alpha\beta}\p_\beta J_\alpha}$ (resp.  ${\bar J = -\epsilon_{\alpha\beta}\p_\beta \bar J_\alpha}$)  in the partition function, to obtain the moments of $\psi$ (resp. $\bar \psi$)
 as functional derivatives with respect to $J$ (resp. $\bar J$).
 Let us point out that this action is often obtained by taking the curl of the \NS equation before
 casting it into a functional integral~\cite{Honkonen98,Mayo05}. Here, this operation comes as the consequence of the incompressibility constraint for $\bar \vv$. This shows that in \twoD, the velocity field action \eq{eq:ActionSNS} and the stream function
one \eq{eq:ActionNSstream} are equivalent.

\subsection{Symmetries and extended symmetries}
\label{sec:Sym}

The action \eq{eq:ActionNSstream} possesses several symmetries. In this work, we consider not only the exact symmetries of this action, but also its extended symmetries.
 We define an extended symmetry as a change of variables in the partition function which does not leave the action strictly invariant, but
  which induces a variation  of the action linear in the fields. The key point is that one can derive from these extended symmetries  Ward identities which
  are more  general than their non-extended versions. The Ward identities are exact relations between
 different correlation (or vertex) functions in specific configurations. Typically, the extended symmetries considered below are time-dependent generalizations of the original exact symmetries. As a consequence, the corresponding Ward identities are valid for arbitrary frequency
 instead of holding only at zero frequency. These identities are very useful in general within 
  a field-theoretical framework, and in particular within the \NPRG. 
 Let us list all the symmetries and extended symmetries of the action  \eq{eq:ActionNSstream}, denoting generically $\eta$ or $\bar{\eta}$ their (scalar or vectorial) parameter: 
\begin{equation}
\begin{array}{l l l l}
a)\quad \delta\psi &= \eta(t)\,, &\bar{a}) \quad \delta\bar\psi = \bar\eta(t)\\
b)\quad \delta\psi &= 0 \,, &\hphantom{\bar{a}) \quad}\delta\bar\psi = x_\alpha \bar\eta_\alpha(t)\\
c)\quad \delta\psi &= 0 \,, &\hphantom{\bar{a}) \quad}\delta\bar\psi =\frac{x^2}{2}\bar\eta(t)\\
d)\quad  \delta\psi &= \epsilon_{\alpha\beta} x_\alpha \dot\eta_\beta(t) + \eta_\alpha(t)\p_\alpha \psi\,, 
&\hphantom{\bar{a}) \quad}\delta\bar\psi =\eta_\alpha(t)\p_\alpha \bar\psi\\
e)\quad \delta\psi &=-\dot\eta(t)\frac{x^2}{2}+\eta(t)\epsilon_{\alpha\beta}x_\beta\p_\alpha\psi \,, 
&\hphantom{\bar{a}) \quad}\delta\bar\psi =\eta(t)\epsilon_{\alpha\beta}x_\beta\p_\alpha \bar\psi
\end{array}
\label{eq:liststreamsym}
\end{equation}
The symmetries $a)$ and $\bar a)$ are  exact symmetries which just follow from the definitions \eq{eq:defStream} and \eq{eq:defbarStream}. Indeed, the stream function and response stream are defined up to a constant function of time, and the functional integral does not fix this gauge invariance.
 The symmetries $b)$ and $d)$ correspond to known extended symmetries of the velocity action: $d)$ is the time-gauged (or time-dependent) Galilean symmetry \citep{Adzhemyan94,Adzhemyan99,Antonov96,Berera07}, and $b)$ is a time-gauged shift of the response fields, first unveiled in \cite{Canet15}. 
 On the other hand, the symmetries $c)$ and $e)$ are extended symmetries that were not, to the best of our knowledge,  identified yet. 
 The symmetry $d)$ corresponds to a different time-gauged shift of the response field, which is also an extended symmetry of the \threeD \NS action (see below), while the symmetry $e)$ can be interpreted as a time-gauged rotation, which is only realized in \twoD.

Let us expound in more details these extended symmetries, starting with the well-known time-gauged Galilean symmetry $d)$.
 Since the action  \eq{eq:ActionNSstream} is invariant under global Galilean transformation, the only non-zero variations stem from
 the time derivative.  One can check that under this transformation, 
\begin{equation}
 \delta ( \mathcal{S}_\psi + \Delta \mathcal{S}_\psi)= \int_{\bx} \p_\alpha \bar \psi \Big[ \epsilon_{\alpha\beta}\ddot{\eta}_\beta(t) + \dot{\eta}_\beta(t) \p_\beta\p_\alpha\psi + 
\epsilon_{\alpha\beta}\epsilon_{\beta\gamma} \dot{\eta}_\gamma(t) \p_\beta\p_\alpha \psi  \Big] = 0\,,
\end{equation}
using the identity \eq{eq:ideps}. Interestingly, the time-gauged Galilean symmetry is an exact symmetry in the stream formulation,
 whereas it is an extended one for the velocity action, because the gauge degree of freedom is fixed in the latter. 

Let us now consider  the symmetries $b)$ and $c)$. A general space-time shift of the response field
$ \bar \psi(t,\vx) \to \bar \psi(t,\vx)  + \bar \eta(t,\vx) $
 yields the first order variation of the action 
\begin{align}
\delta ( \mathcal{S}_\psi + \Delta \mathcal{S}_\psi) =& -\int_{\bx} \bar \eta(\bx) \Bigg\{\p^2 ( \p_t - \nu \p^2) \psi + \p_\alpha \p_\beta (\epsilon_{\beta\gamma} \p_\gamma \psi \p_\alpha \psi)\nonumber\\
& + \int_{\vx'} \Big[ R_{L_0^{\text{-}1}}(|\vec x-\vec x'|) {\p'}^2 \psi (t, \vx') - 2 N_{L^{\text{-}1}}(|\vec x-\vec x'|) {\p'}^2 \bar \psi (t, \vx') \Big] \Bigg\} \, .
\end{align}
As in the velocity formulation, the non-linear term of this variation, stemming from the interaction, may vanish for some particular space
 dependence of $\bar \eta$.
The choice $\bar \eta(\bx) = \bar \eta(t)$ is simply the gauge-invariance $\bar{a})$ for which  $\delta (\mathcal{S}_\psi + \Delta \mathcal{S}_\psi)= 0$. The choice $\bar \eta(\bx) = x_\alpha \bar \eta_\alpha(t)$ corresponds to the known time-gauged shift of the response fields (velocity and pressure) in the velocity formulation \cite{Canet15}. For this choice 
 $ \delta ( \mathcal{S}_\psi + \Delta \mathcal{S}_\psi)$ also vanishes, which means that the time-gauged shift is an exact symmetry in the stream formulation, while it is an extended one in the velocity formulation, as for the time-gauged Galilean symmetry.

 We here uncover another  transformation which leads to the new extended symmetry $c)$: $\bar \eta(t,\vx) = \frac{x^2}{2} \bar \eta(t)$. For this choice, the variation stemming from the interaction cancels by antisymmetry of $\epsilon_{\alpha\beta}$. Let us emphasize that this symmetry is not specific to \twoD, and can be expressed also in \threeD in the
velocity formulation, where it corresponds to a shift linear in space: 
\begin{equation}
\delta \bar v_\alpha =  \epsilon_{\alpha\beta\gamma} x_\beta \eta_\gamma(t)\, ,\quad \delta \bar p = v_\alpha  \epsilon_{\alpha\beta\gamma} x_\beta \eta_\gamma(t)\,,
\label{eq:xshiftv}
\end{equation}
where the $\epsilon_{\alpha\beta\gamma}$ are the components of the fully antisymmetric tensor with three indices. One can indeed check,  by 
 performing this change of variables in the velocity action \eq{eq:ActionSNS}, and after some integration by parts and use of the anti-symmetry of $\epsilon_{\alpha\beta\gamma}$, that the corresponding variation reads
\begin{align}
 &\delta (\mathcal{ S}_v + \Delta \mathcal{ S}_v) 
= \int_{\bx} \epsilon_{\alpha\beta\gamma} x_\beta \eta_\gamma(t) \Big\{ \p_t  v_\alpha(\bx) 
 + \int_{\vec x'}   R_{L_0^{\text{-}1}}(|\vec x-\vec x'|) v_\alpha(t,\vec  x') - 2  N_{L^{\text{-}1}}(|\vec x-\vec x'|)\bar v_\alpha(t,\vec  x')\Big\}\, , \nonumber
\end{align}
 and is linear in the fields. The transformation \eq{eq:xshiftv} is thus an extended symmetry of the \NS velocity action in both  \twoD and \threeD.
  This new symmetry yields Ward identities, given below in the \twoD stream formulation, that could be useful in the study of the \threeD \NS turbulence as well.
We did not find higher order space dependence of $\bar \eta(t,\vx)$   which induces a linear variation of the action in the fields. 

Finally, we now consider the extended symmetry $e)$, which is specific to \twoD. This new
 symmetry can be interpreted as a time-gauged rotation in the same way as extended Galilean symmetry is a time-gauged translation in space.
 Using the anti-symmetry of $\epsilon_{\alpha\beta}$, one can check that the variation under the transformation $e)$ is indeed linear is the field
 \begin{equation}
\delta ( \mathcal{S}_\psi + \Delta \mathcal{S}_\psi) = 2 \int_{\bx} \eta(t) \Big[ \p_t^2 \bar \psi - \p_t \bar \psi \int_{\vx'} R_{L_0^{\text{-}1}}(|\vec x-\vec x'|)\Big]\, .
\end{equation}
This symmetry can be expressed in the velocity formulation as well, at the cost of introducing a non-local
shift of the pressure. %{\red It $\to$ This} is probably related to the conservation of  vorticity in \twoD.

These extended symmetries can be translated into Ward identities. As they are exploited in the present work within the \NPRG framework, we 
 derive them in terms of the \EAA $\Gamma_\kappa$ defined in the next section. They  essentially coincide
 with the Ward identities for the standard effective action 
 $\Gamma$ usually defined in field theory, but for the terms associated with the variation of the non-local quadratic parts of the action. These terms are subtracted in the \NPRG formalism (see \aref{app:WardGen}), but they can be straightforwardly included to deduce the Ward identities in terms of $\Gamma$ \cite{Canet15}.  Let us now briefly introduce the \EAA, and then establish these Ward identities.

\section{\NPRG formalism and Ward identities}
\label{sec:NPRG}

\subsection{Formulation of the \NPRG}

The \NPRG is a modern implementation of
 Wilson's original idea of the \RG \cite{Wilson74}, conceived to efficiently average over fluctuations, even when they
 develop at all scales, as in standard critical phenomena \cite{Berges02,Kopietz10,Delamotte12}. 
 It is a powerful method to compute the properties of strongly
correlated systems,  which can reach high precision levels \cite{Canet03b,Benitez12} 
and   yield fully non-perturbative results, 
at equilibrium \cite{Grater95,Tissier06,Essafi11} and also out of
equilibrium  \cite{Canet04a,Canet05,Canet10,Canet11a,Berges12}, restricting to a few classical statistical
physics applications. 
The \NS field theory was first studied
 using \NPRG methods in \cite{Tomassini97,Monasterio12,Canet16}, and we here follow the formalism developed in \cite{Canet16,Tarpin18}.

The core idea of the \NPRG is to organize the integration of the fluctuations by adding to the action a non-local quadratic term, called the regulator and noted   $\Delta \mathcal{ S}_\kappa$. The role of the regulator is to freeze the degrees of freedom of the field with wave-number below the renormalization scale $\kappa$ to their mean-field value, while not affecting  the degrees of freedom with wave-number above $\kappa$, with the additional requirement that the transition between these two regimes is sufficiently smooth.  By varying $\kappa$ from the \UV cutoff of the theory, where the regulator ensure that mean-field is exact, to its \IR cutoff, one  smoothly integrates over the fluctuations of the fields. If the starting action is at a critical point, the resulting flow reaches a fixed point, from which universal properties of the field theory can be calculated.

It turns out that in the Navier-Stokes action~\eq{eq:ActionNSstream}, terms which can play the role of regulators are already present for physical reasons. Indeed, the functions  $N_{L^{\text{-}1}}$ and $R_{L_0^{\text{-}1}}$ associated with forcing and large-scale dissipation satisfy all the requirements to act as regulators of the theory. Their Fourier transform are smooth functions, which vanish exponentially for  wave-numbers large compared to $L_0^{\text{-}1}$ or $L^{\text{-}1}$, and which regularize the fluctuating fields for small wave-numbers (see \cite{Berges02,Canet16}).
Thus, identifying their typical  scale, $L_0^{-1}$ and $L^{-1}$, with the \RG scale $\kappa$ yields a regulator $\Delta \mathcal{ S}_\kappa$ 
 for the \NS field theory. 
Since we are interested in the direct cascade where wave-numbers are larger than both $L^{-1}$ and  $L_0^{-1}$, we simply set
  $L^{-1} = L_0^{-1} = \kappa$.  To study the inverse cascade, which corresponds to wave-numbers between  $L_0^{-1}$ and  $L^{-1}$, the scale $L$ should be kept fixed while setting $L_0^{-1} = \kappa$ (this is left for future work). 

In the presence of the regulator $\Delta \mathcal{ S}_\kappa$, the generating functional ${\cal Z}$  of the correlation functions 
becomes scale dependent
\begin{equation}
 \mathcal{ Z}_\kappa[J, \bar J] = \int \mathcal{D}\psi  \mathcal{D}\bar\psi \, e^{-\mathcal{S}_\psi -\Delta \mathcal{S}_{\psi}} e^{\int_{\bx}\{J\psi +\bar J\bar \psi\}} \, .
\end{equation}
 The average of the stream function (and response stream) 
 can be obtained through functional derivatives of $\mathcal{ W}_\kappa = \ln \mathcal{ Z}_\kappa$ with respect to the sources as
\begin{equation}
 \Psi ({\bx}) = \mean{\psi({\bf x})} = \frac{\delta \mathcal{ 
W}_{\kappa}}{\delta J({\bf x})}  \, \, , \, \,
  \bar \Psi({\bf x}) = \mean{\bar \psi({\bf x})} = \frac{\delta 
\mathcal{ W}_{\kappa}}{\delta \bar J({\bf x})}\, .
\end{equation} 
When the renormalization scale $\kappa$ varies, $\mathcal{ W}_\kappa$ evolves according to the following exact flow equation (which is similar to the Polchinski equation \cite{Polchinski84}):
\begin{equation}
 \p_\kappa \mathcal{ W}_\kappa = - \frac{1}{2}\, \int_{\bx,\by}\! \p_\kappa [\mathcal{ R}_\kappa]_{ij}({\bx-\by}) \,\Big\{\frac{\delta^2 \mathcal{ W}_\kappa}{\delta \mathrm{j}_{i}({\bx}) \delta \mathrm{j}_{j}({\by})} + \frac{\delta \mathcal{ W}_\kappa}{\delta  \mathrm{j}_{i}({\bx})} \frac{\delta \mathcal{ W}_\kappa}{\delta \mathrm{j}_{j}({\by})}\Big\}\label{eq:dkw} \, ,
\end{equation}
where $i,j \in\{1,2\}$ with   $\mathrm{j}_{1}=J$ and  $\mathrm{j}_{2}=\bar J$.
The \EAA $\Gamma_\kappa$ 
(which is the generating functional of \PI correlation functions~\cite{Amit84,zinnjustin89})
 is defined as the Legendre transform
 of $\mathcal{ W}_\kappa$, up to the regulator term:
\begin{equation}
\Gamma_\kappa[\Psi,\bar\Psi] + \mathcal{ W}_\kappa[J,\bar J] = \int_{\bx} \! \Big\{ J\,\Psi+\bar J\,\bar \Psi\Big\}  - \Delta \mathcal{S}_\kappa[\Psi, \bar \Psi] \, .
\label{eq:legendre}
\end{equation}
The flow of  $\Gamma_\kappa$ with the \RG scale $\kappa$ is given by the Wetterich equation \cite{Wetterich93}
\begin{equation}
\p_\kappa \Gamma_\kappa = \frac{1}{2}\, \int_{\bx,\by}\! 
 \p_\kappa [\mathcal{ R}_\kappa]_{ij}(|\bx-\by|)   \Big[\Gamma_\kappa^{(2)} + \mathcal{ R}_\kappa\Big]_{ji}^{-1}({\by,\bx})\, ,
\label{eq:dkgam}
\end{equation}
 where  $\Gamma_\kappa^{(2)}$ is the Hessian of $\Gamma_\kappa$ and the regulator matrix  $[\mathcal{ R}_\kappa]$ is defined as
\begin{equation}
 [\mathcal{ R}_\kappa]_{ij}({\bx-\by}) = \frac{\delta^2 \Delta \mathcal{ S}_\kappa}{\delta \varphi_i({\bx}) \delta \varphi_j({\by})} \, ,
\label{eq:calR}
\end{equation}
with $i,j \in \{1, 2\}$ and $\varphi_1=\Psi$, $\varphi_2=\bar \Psi$.
 The \RG flow equation \eq{eq:dkgam} is also exact. Its initial condition corresponds to the `microscopic' model, 
 which is  $\mathcal{S}_\psi$ in \eq{eq:ActionNSstream}. The flow is hence initiated at a very large wave-number $\Lambda$
  at which the continuous description of the fluid dynamics in terms of \NS equation starts to be valid. At this scale, one can show that $\Gamma_\Lambda$ identifies with the bare action $\Gamma_\Lambda= \mathcal{S}_\psi$, since no fluctuation is yet incorporated. When $\kappa\to L^{\text{-}1}$, the regulator reaches its original value and one obtains the actual properties of the model, when all fluctuations up to the physical \IR cutoff have been integrated over.  \Eq{eq:dkgam} provides the exact interpolation between these two scales.

\subsection{Definition of generalized correlation functions}

The functional $\mathcal{ W}_\kappa$  is the generating functional of connected 
correlation functions, which  correspond to the cumulants 
for  a field theory~\cite{Amit84,zinnjustin89}.
  The $n$-point generalized connected correlation functions can be obtained as functional derivatives of $\mathcal{ W}_\kappa$ with respect to the sources
\begin{equation}
 G_{i_1\dots i_{n}}^{(n)}[\{\bx_\ell\}_{1 \leq \ell \leq n};\mathrm{j}] =  \frac{\delta^{n} \mathcal{W}_\kappa}{\delta \mathrm{j}_{i_1}({\bx_1}) \dots \delta  \mathrm{j}_{i_{n}}({\bx_{n}})} \, ,
\end{equation}
where  $i_k=1,2$ with $ \mathrm{j}_1=J$ and $\mathrm{j}_2=\bar J$ as before. They are called generalized because they include derivatives with respect to response fields, which are related to correlations with the forcing \cite{Canet16}.
 Note that in this definition, $G_{i_1\dots i_{n}}^{(n)}$ is still a functional of the sources, which is materialized by the square brackets and the explicit $\mathrm{j}$ dependency.
 Let us  also introduce the notation   $G^{(m,\bar{m})}[{\bx_1}, \dots, {\bx_{m+\bar{m}}};\mathrm{j}]$ where the $m$ first derivatives are with respect
 to $J$ and the $\bar{m}$ last with respect to  $\bar J$.
 We indicate that a correlation function is evaluated at zero sources using the notation
\begin{equation}
 G_{i_1\dots i_{n}}^{(n)}(\{\bx_\ell\}_{1 \leq \ell \leq n})\equiv G_{i_1\dots i_{n}}^{(n)}[\{\bx_\ell\};\mathrm{j}=0]
\end{equation}
(and accordingly for $G^{(m,\bar m)}$). The Fourier transforms of these functions are defined as:
\begin{equation}
\tilde G^{(n)}(\{\bp_\ell\}_{1 \leq \ell \leq n}) = \int_{\{\bx_\ell\}} G^{(n)}(\{\bx_\ell\}_{1 \leq \ell \leq n}) e^{- i \sum_{k=1}^n (\vp_k \cdot \vx_k - \omega_k t_k)}\, ,
\end{equation} 
 or similarly extracting the delta function of conservation of the total wave-vector and frequency
\begin{equation}
\tilde G^{(n)}(\{\bp_\ell\}_{1 \leq \ell \leq n}) = (2\pi)^{d+1}\delta^{d}(\sum_{k=1}^n \vp_k) \delta(\sum_{k=1}^n \omega_k) \bar G^{(n)}(\{\bp_\ell\}_{1 \leq \ell \leq n-1})\,. \label{eq:tilde-bar}
\end{equation}

The \EAA is the generating functionals  of \PI correlation functions, also called vertex functions. 
 This means that  any vertex functions  can be obtained by taking functional derivatives of $\Gamma_\kappa$ with respect to the average fields
 $\Psi$ and $\bar\Psi$. 
The $n$-point vertex (\PI) functions are defined using the same conventions as for the connected correlation functions:
\begin{equation}
 \Gamma_{i_1\dots i_{n}}^{(n)}[{\bx_1}, \dots, {\bx_n};\varphi]=  \frac{\delta^{n} \Gamma_\kappa}{\delta \varphi_{i_1}({\bx_1}) \dots \delta\varphi_{i_{n}}({\bx_{n}})} \, ,
\end{equation}
where $i_k=1,2$ with $\varphi_{1}=\Psi$ and $\varphi_{1}=\bar \Psi$, or alternatively $\Gamma^{(m,\bar m)}[{\bx_1}, \dots, {\bx_{m+\bar m}};\varphi]$. 
Accordingly, we define
$\Gamma_{i_1\dots i_{n}}^{(n)}({\bx_1}, \dots, {\bx_n})$ and $ \Gamma^{(m,\bar m)}({\bx_1}, \dots, {\bx_{m+\bar m}})$ as the previous vertex functions evaluated at zero fields.
Finally we define  the Fourier transforms before and after extracting the delta function of conservation of wave-vector and frequency, 
 $\tilde \Gamma^{(n)}(\{\bp_\ell\}_{1 \leq \ell \leq n})$ and $\bar \Gamma^{(n)}(\{\bp_\ell\}_{1 \leq \ell \leq n-1})$,
 respectively.
 The knowledge of the set of connected correlation functions or of the set of vertex functions is equivalent.
 Indeed, they are inter-related: any $n$-point connected correlation function $G^{(n)}$ can be constructed as a sum of tree diagrams whose vertices are the \PI functions $\Gamma^{(k)}$, $2<k\leq n$ 
 and whose edges are the propagators $G^{(2)}$ \cite{Amit84,zinnjustin89}.

\subsection{Ward identities for the vertex functions in the stream formulation}

In  \sref{sec:timedependence} and \sref{sec:equaltime}, we present calculations within the large wave-number expansion.
 As explained in \aref{app:flowD}, a key ingredient in these calculations in the existence of Ward identities
  for the vertex functions. We hence give below
 the Ward identities associated with the extended symmetries of the \NS action in terms of the vertex functions, 
 and within the \NPRG framework.  Ward identities for the connected correlation functions can be derived in the same way. 

 The list \eq{eq:liststreamsym} of the extended symmetries in the stream function formulation only contains
  continuous changes of variables which are at most linear in the field, and so are the corresponding variations of 
  the action. In this case, they can be translated readily into Ward identities that the \EAA must verify
   along the \RG flow. These identities simply express that the \EAA possesses the same
    symmetry as $\mathcal{S}_\psi$, except for the non-invariant terms which are not renormalized.
     The general derivation was presented in \cite{Canet15} and is summarized in \aref{app:WardGen}. It is shown in particular that in the \NPRG framework,
       the regulator terms do not enter these identities when the change of variable is a shift of the fields or when it leaves the regulators invariant, which is the case for all the symmetries presented in \sref{sec:Sym}.

The functional Ward identities for the \EAA associated with the extended symmetries
 \eq{eq:liststreamsym} are:
\begin{align}
 & a) \quad \int_{\vx} \df{\Gamma_\kappa}{\Psi(\bx)} = 0, \quad\quad\quad \quad\quad\quad \bar{a})  \int_{\vx} \df{\Gamma_\kappa}{\bar \Psi(\bx)} = 0\nonumber\\
 & b) \quad \int_{\vx} x_\alpha \df{\Gamma_\kappa}{\bar \Psi(\bx)} = 0\nonumber\\
 & c) \quad \int_{\vx} \frac{x^2}{2} \df{\Gamma_\kappa}{\bar \Psi(\bx)} = - 2 \int_{\vx} \p_t \Psi\nonumber\\
 & d) \quad \int_{\vx} \Big\{ \big(- \epsilon_{\beta\alpha} x_\beta \p_t + \p_\alpha \Psi\big) \df{\Gamma_\kappa}{\Psi(\bx)} + \p_\alpha \bar\Psi \df{\Gamma_\kappa}{\bar \Psi(\bx)} \Big\} = 0 \nonumber\\
 & e) \quad \int_{\vx} \Big\{ \big(\frac{x^2}{2}\p_t + \epsilon_{\alpha\beta}x_\beta\p_\alpha\Psi\big) \df{\Gamma_\kappa}{\Psi(\bx)} + \epsilon_{\alpha\beta} x_\beta\p_\alpha \bar \Psi \df{\Gamma_\kappa}{\bar \Psi(\bx)} \Big\} = 2 \int_{\vx} \p_t^2 \bar \Psi\,.
\label{eq:listwardstream}
\end{align}
From these functional identities, one can derive  a hierarchy of identities for the vertex functions $\Gamma^{(m,\bar{m})}$, by taking 
 the corresponding functional derivatives  and evaluating them at zero fields.
  In the Fourier space, they read:
\begin{align}
 & a),\bar{a}) \quad \bar \Gamma_\kappa^{(m,\bar{m})} (\dots, \varpi, \vq, \dots)\Big|_{\vq=0} = 0 \nonumber\\
 & b) \quad \frac{\p}{\p q^i} \bar \Gamma_\kappa^{(m,\bar{m}+1)}( \{\bp_\ell\}_{1 \leq \ell \leq m},  \varpi, \vq, \{\bp_\ell\}_{1 \leq \ell \leq \bar{m}-1} ) \Big|_{\vq=0} = 0\nonumber\\
 & c) \quad \frac{\p^2}{\p q^2} \bar \Gamma_\kappa^{(m,\bar{m}+1)}(\{\bp_\ell\}_{1 \leq \ell \leq m},  \varpi, \vq, \{\bp_\ell\}_{1 \leq \ell \leq \bar{m}-1} ) \Big|_{\vq=0} = 0\nonumber\\
 &\quad \quad \quad \quad \text{except } \frac{\p^2}{\p q^2} \bar \Gamma_\kappa^{(1,1)}(\varpi, \vq) \Big|_{\vq=0} = - 4 i \varpi \nonumber\\
 & d) \quad \frac{\p}{\p q^i} \bar \Gamma_\kappa^{(m+1,\bar{m})}(\varpi,\vq,\{\bp_\ell\}_{1 \leq \ell \leq m+\bar{m}-1})\Big|_{\vq=0}= i \epsilon_{\alpha\beta}\tilde{\mathcal{ D}}_\beta(\varpi)
   \bar \Gamma_\kappa^{(m,\bar{m})}(\{\bp_\ell\})\nonumber\\
 & e) \quad \frac{\p^2}{\p q^2} \bar \Gamma_\kappa^{(m+1,\bar{m})}({\varpi,\vq},\{\bp_\ell\}_{1 \leq \ell \leq m+\bar{m}-1}) \Big|_{\vq=0}= \tilde{\mathcal{R}}(\varpi) \bar \Gamma_\kappa^{(m,\bar{m})}(\{\bp_\ell\}) \, ,
\label{eq:listwardstreambar}
\end{align}
where we have introduced the two  operators $\tilde{\mathcal{D}}_\alpha(\varpi)$ and  $\tilde{\mathcal{R}}(\varpi)$ defined as:
\begin{align}
 &\tilde{\mathcal{D}}_\alpha(\varpi) F( \{\bp_\ell\}_{1 \leq \ell \leq n}) \equiv - \sum_{k=1}^{n} p_k^\alpha
    \Bigg[  \frac{F(\{\bp_\ell\}_{1 \leq \ell \leq k-1}, \omega_k + \varpi,\vp_k, \{\bp_\ell\}_{k+1 \leq \ell \leq n})- F(\{\bp_\ell\})}{\varpi}\Bigg]\nonumber\\
&\tilde{\mathcal{R}}(\varpi) F( \{\bp_\ell\}_{1 \leq \ell \leq n}) \equiv 2 i \epsilon_{\alpha\beta} \sum_{k=1}^{n} p_k^\alpha \frac{\p}{\p p_k^\beta} 
    \Bigg[  \frac{F(\{\bp_\ell\}_{1 \leq \ell \leq k-1}, \omega_k + \varpi,\vp_k, \{\bp_\ell\}_{k+1 \leq \ell \leq n})- F(\{\bp_\ell\})}{\varpi}\Bigg]\, .
\end{align}
The derivation of the identities $d)$ and $e)$ is reported in \aref{app:WardG} and \aref{app:WardS} respectively.

\subsection{Large wave-number expansion of the flow equations}

Let us explain the principles of the large wave-number expansion.
 The flow equation \eq{eq:dkgam} is exact, but it is not closed. Indeed, the flow equation
 for a generic $n$-point function $\Gamma_\kappa^{(n)}$ can be deduced by taking the corresponding functional derivatives of  \eq{eq:dkgam},  and 
  it involves $(n+1)$ and $(n+2)$ vertex functions.  As such, one has to consider  an infinite hierarchy of flow equations.
For example, the flow equation for the two-point function is given in the Fourier space by
\begin{align}
 \p_s \bar \Gamma_{mn}^{(2)}(\bp) &= \int_\bq \p_s{\mathcal{R}}_{ij}(\bq) \bar G^{(2)}_{jk}(\bq) \Big[ -\frac{1}{2} \bar \Gamma_{klmn}^{(4)}(\bq,-\bq,\bp) \nonumber\\
  &+ \bar \Gamma_{kms}^{(3)}(\bq, \bp)  \bar G^{(2)}_{st}(\bq + \bp) \bar \Gamma_{tnl}^{(3)}(\bq + \bp, - \bp) \Big]  \bar G^{(2)}_{li}(\bq)\, ,
\end{align}
 which depends on the 3- and 4-point vertices. 
The \rhs is represented diagrammatically in Fig. 1, where the dashed circles are the vertex functions, the thick lines are propagators and the  cross is the derivative of the regulator. 
  \begin{figure}[h]
  \centering
\begin{tikzpicture}
  \node[left, scale=1] at (0,0) {$\partial_{\kappa} \Gamma_{\kappa}^{(2)}(\bp) = \;-\displaystyle\frac{1}{2}$};
  \draw[line width=0.3mm] (0.7,0.8) circle (0.8);
  \draw (0.7,1.6) node {$\boldsymbol{\times}$};
  \draw (0.0,1.6) node {$\bq$};
  \draw (1.3,1.6) node {$-\bq$};
  \draw (0.7,0.0) -- (0.1,-0.6);
  \draw (0.7,0.0) -- (1.3,-0.6);
  \draw (1.3,-0.9) node {$-\bp$};
  \draw (0.0,-0.9) node {$\bp$};
  \filldraw[white,opacity=1] (0.7,0.0) circle (0.6);
  \filldraw[pattern=north east lines] (0.7,0.0) circle (0.6);
  \node[scale=1] at (1.6,0) {$+$};
  \draw (2.3,0) -- (3.3,0);
  \draw (2.2,0.2) node {$\bp$};
  \draw[line width=0.3mm] (4.5,0) circle (1);
  \draw (4.5,-1.3) node {$\bp+\bq$};
  \draw (4.5,1) node {$\boldsymbol{\times}$};
  \draw (3.8,1) node {$\bq$};
  \draw (5.1,1) node {$-\bq$};
  \draw (5.7,0) -- (6.7,0);
  \draw (6.7,0.2) node {$-\bp$};
  \filldraw[white,opacity=1] (3.3,0) circle (0.6);
  \filldraw[pattern=north east lines] (3.3,0) circle (0.6);
  \filldraw[white,opacity=1] (5.7,0) circle (0.6);
  \filldraw[pattern=north east lines] (5.7,0) circle (0.6);
\end{tikzpicture}
  \label{diag:FlotGamma2}
  \caption{Diagrammatic representation of the flow of $\Gamma_\kappa^{(2)}$.}
 \end{figure}
The \rhs involves the integrated, or internal,  wave-vector and frequency  $\bq$ circulating in the loops besides the external wave-vector and frequency $\bp$ at which the vertex function on the \lhs is evaluated.

In most applications, this hierarchy is closed by simply truncating higher-order vertices, or proposing an ansatz for $\Gamma_\kappa$ \cite{Berges02}. An alternative strategy,  pioneered in~\cite{Blaizot06, Benitez12} and called the BMW approximation scheme, consists
 in expanding these vertices in the internal wave-vector $\vq$. This approximation relies on the two following properties of the regulator: on the one hand, its insertion in the integration loop on the \rhs of \eq{eq:dkgam} cuts off the internal
 wave-number $|\vq|$ to values   $|\vq| \lesssim \kappa$.
As a  consequence, if the system is probed at a wave-number scale $|\vp|$ much larger than the renormalization scale, $p \gg \kappa$, there is a clear separation of scales in the flow equations: $q/p \ll 1$. On the other hand, the presence of the regulator ensures that the
vertex functions are  smooth at any finite scale $\kappa>0$, which allows one to perform a Taylor expansion in powers of $\vq$. The underlying idea is that, close to a fixed point, the vertex functions are expected to depend on the internal wave-number only through ratio of the type $q/p$, which means that the expansion at $q\simeq 0$  is expected to be equivalent to an expansion at $p\to \infty$. 
 This expansion becomes exact in the limit of infinite wave-numbers, and the error at finite but large $p$ is  small.
 In fact, this expansion was found to be a reliable approximation for arbitrary  momenta \cite{Blaizot06, Benitez12}.

 The BMW strategy has turned out to be very successful in the context of turbulence, since the expanded flow equations can be closed  at zero fields     thanks to the Ward identities, whereas it generically requires to keep a whole dependence in background fields. 
 This was first noticed in~\cite{Canet16} for the two-point function, and generalized in~\cite{Tarpin18}
 where the exact leading order term in the large wave-number expansion of the flow equation of an arbitrary $n$-point correlation function
  was obtained in \threeD.  The striking feature of these flow equations is that they do not exhibit the decoupling property usually expected 
 for flow equations, \eg in  standard critical phenomena. The decoupling property ensures that the existence of a fixed point 
 entails standard scale invariance  for $\kappa$ much smaller than any non-exceptional wave-number (see \cite{Tarpin18}). In \threeD turbulence, the violation of the decoupling property yields
 a breaking of standard scale invariance, which is manifest in the time dependence of generic correlation functions.
  This breaking was related in \cite{Canet17,Tarpin18}, at least for small time delays, to the  sweeping effect. The latter is the random advection of small-scale velocities by large-scale structures in the turbulent flow, and is well-known phenomenologically \cite{Kraichnan59,Tennekes75}.
  
In the following, we calculate the flow equation of a generalized $n$-point correlation function in the stream formulation
 within the large wave-number expansion.
In \sref{sec:timedependence}, we derive the leading order, and derive the general form of the
 time dependence of the two-point function. We show that a similar breaking of standard scale invariance  occurs in \twoD turbulence, but
  its explicit expression is different. 
 However, the (leading order) non-decoupling term of the flow equation turns out to vanish at equal time, as in \threeD, such that standard scale invariance, 
 which means Kolmogorov-Kraichnan  scaling, is recovered at this  order for equal-time quantities. 
 Hence, in  \sref{sec:equaltime}, we calculate the \NLO
 term in the large wave-number expansion of the flow equation, focusing on equal times, in order to seek for possible intermittency corrections to simultaneous quantities, such as structure functions. This calculation is specific to \twoD since it exploits the new extended symmetry related to time-gauged rotations.

\section{Time dependence of generic correlation functions}
\label{sec:timedependence}

In this section, we express the flow equation for a generic generalized $n$-point correlation function in the stream formulation, and we compute its exact leading term
 in the large wave-number expansion for non-equal time delays. We then derive the corresponding fixed-point solutions for the two-point function, and show that its time dependence explicitly breaks scale invariance.

 \subsection{Flow equation for generic correlation functions at leading order}
\label{sec:FlowD}

The flow equation for a generic connected correlation function of the stream and response stream functions $G_\psi^{(n)}$ is obtained by taking $n$ functional derivatives of \eq{eq:dkw} with respect to the sources $\mathrm{j}_{i_k}$, $1 \leq k \leq n$, which yields
\begin{align}
 \p_\kappa G^{(n)}_{\psi,i_1\dots i_{n}}[\{\bx_\ell\}_{1 \leq \ell \leq n};\mathrm{j}] &= - \frac{1}{2}\,  \int_{\by_1,\by_2}\! \p_\kappa [\mathcal{R}_\kappa]_{ij}({\by_1-\by_2}) \,\Bigg\{ G^{(n+2)}_{\psi,ij i_1\dots i_{n}}[\by_1,\by_2,\{\bx_\ell\};\mathrm{j}]
\nonumber\\
 & + \sum_{(\{i_\ell\}_1, \{i_\ell\}_2) \atop \#1 + \#2=n} G^{(\#1+1)}_{\psi,i \{i_\ell\}_1}[{\by_1,\{\bx_{\ell}\}_1};\mathrm{j}] G^{(\#2+1)}_{\psi,j \{i_\ell\}_2}  [{\by_2,\{\bx_{\ell}\}_2};\mathrm{j}]\Bigg\} \, .
\label{eq:dsGn}
\end{align}
where  the indices $i_\ell \in \{1,2\}$ stand for the sources  $J$ or $\bar J$,
 and $(\{i_\ell\}_1, \{i_\ell\}_2)$ indicates all the possible bipartitions of the $n$ indices $\{i_\ell\}_{1 \leq \ell \leq n}$, and $(\{\bx_{\ell}\}_1,\{\bx_{\ell}\}_2)$ the corresponding bipartition in coordinates. 
Finally, $\#1$ and $\#2$ are the cardinals of $\{i_\ell\}_1$ (resp. $\{i_\ell\}_2$).

We now consider the large wave-number expansion of \eq{eq:dsGn}. The calculation of the leading order term is 
 formally the same as in \threeD in the velocity formulation \cite{Tarpin18}. Nevertheless, the derivation in the stream formulation is reported in \aref{app:flowD}
 for completeness. We first show that in the limit of large wave-numbers, that is when all the  $\vp_\ell$, $1 \leq \ell \leq n$ are large compared to $\kappa$, as well
as all their partial sums, the flow equation \eq{eq:dsGn} reduces to
\begin{equation}
 \p_\kappa \tilde G^{(n)}_{\psi,i_1\dots i_{n}}(\{\bp_\ell\}_{1 \leq \ell \leq n}) =  \frac{1}{2}\,  \int_{\bq_1,\bq_2}\!
 \tilde \p_\kappa \tilde  G^{(2)}_{\psi,ij}(-\bq_1,-\bq_2)\left[ \frac{\delta^2}{\delta \varphi_i(\bq_1)\delta\varphi_j(\bq_2)} 
 \tilde G^{(n)}_{\psi,i_1\dots i_{n}}[\{\bp_\ell\};\mathrm{j}] \right]_{\varphi=0}\, ,\label{eq:flowGnstream}
\end{equation}
up to terms tending to zero faster than any power of the $p_\ell$. In the velocity formulation, 
 the leading order term is obtained  by setting $q_1, q_2$ to zero in the equivalent terms in bracket.
 In the stream formulation, because of 
the gauge  symmetry $a)$ in \eq{eq:liststreamsym}, there is no information at this order. Indeed, the symmetry $a)$ implies
 that the vertex functions are zero if one of the  wave-numbers is set to zero, according to \eq{eq:listwardstreambar}, and thus
\begin{equation}
 \left[ \frac{\delta^2}{\delta \varphi_i(\bq_1)\delta\varphi_j(\bq_2)} 
 \tilde G^{(n)}_{\psi,i_1\dots i_{n}}[\{\bp_\ell\}_{1 \leq \ell \leq n};\mathrm{j}] \right]_{\varphi=0}\Bigg|_{\vq_1 = \vq_2 = 0} = 0 \, .
\end{equation}
As a consequence, the first non-zero contribution comes from the second order term in  the  $q_1, q_2$ expansion (since the odd terms
of this expansion vanishes  by parity of $\tilde \p_\kappa \tilde  G^{(2)}_{\psi,ij}(-\bq_1,-\bq_2)$). It reads
\begin{align}
 \p_\kappa \tilde G^{(n)}_{\psi,i_1\dots i_{n}}(\{\bp_\ell\}_{1 \leq \ell \leq n}) &=  \frac{1}{2}\,  \int_{\bq_1,\bq_2}\!
 \tilde \p_\kappa \tilde  G^{(2)}_{\psi,ij}(-\bq_1,-\bq_2) \nonumber\\
 &\times \frac{q_a^\alpha q_b^\beta}{2} \frac{\p^2}{\p q_a^\alpha \p q_b^\beta} \left[ \frac{\delta^2}{\delta \varphi_i(\bq_1)\delta\varphi_j(\bq_2)} 
 \tilde G^{(n)}_{\psi,i_1\dots i_{n}}[\{\bp_\ell\};\mathrm{j}] \right]_{\varphi=0}\Bigg|_{\vq_1=\vq_2 = 0} \,,
\label{eq:flowleading}
\end{align}
where $a,\,b$ take value in $\{1,2\}$. As in the velocity formulation, one can then show that this flow equation can be closed using the Ward identities (see \aref{app:flowD}), which yields
\begin{equation}
 \p_\kappa \tilde G^{(n)}_{\psi,i_1\dots i_{n}}(\{\bp_\ell\}_{1 \leq \ell \leq n}) 
 = \frac{1}{2}\,  \int_{\bq_1,\bq_2}\!
 \tilde \p_\kappa \tilde  G^{(2)}_{v_\mu v_\nu}(-\bq_1,-\bq_2)\tilde{\mathcal{D}}_\mu(\varpi_1) \tilde{\mathcal{D}}_\nu(\varpi_2)
 \tilde G^{(n)}_{\psi,i_1\dots i_{n}}(\{\bp_\ell\})\, .
\label{eq:dsgleading}
\end{equation}
 One can check, using the correspondence
\begin{equation}
 \tilde G^{(n)}_{v, {k_1}\cdots {k_n}}(\{\omega_k,\vp_k\}) = (i)^n\,\epsilon_{k_1 \ell_1} p_1^{\ell_1}\cdots \epsilon_{k_n \ell_n} p_n^{\ell_n}\tilde G^{(n)}_{\psi,i_1\dots i_{n}}(\{\omega_k,\vp_k\})
\end{equation}
that the same result as in the velocity formulation, obtained in \cite{Tarpin18}, is recovered.

\subsection{Fixed-point and conservation of enstrophy in the direct cascade}
\label{sec:DIM}

Our goal is to describe the universal properties of the direct enstrophy cascade in the forced-dissipative stationary regime.
 This regime, characterized by some form of scale invariance, corresponds to a fixed point in the \RG framework 
(see \cite{Tarpin18} for details). In order to study the fixed point, it is convenient to work with dimensionless variables,
 denoted with a hat symbol, using the \RG scale $\kappa$ as unit of wave-numbers. We introduce the  dimensionless forcing profile  through
\begin{equation}
 N_\kappa(\vq) = D_\kappa\; ({|\vq|}/{\kappa})^2 \hat n({|\vq|}/{\kappa}) \, ,
\end{equation}
where $D_\kappa$ is a scale dependent forcing amplitude, and $\hat n$ is the specific forcing profile, which fulfills the requirements stated in \sref{sec:NPRG} (smoothness and fast decay at large wave-numbers). 
Similarly, the  dimensionless effective friction can be defined through
    \begin{equation}
 R_\kappa(\vq) = \nu_\kappa\; \vq\,^2 \hat r({|\vq|}/{\kappa})\, ,
\end{equation}
with $\nu_\kappa$ a scale dependent coefficient and $\hat r$ is also a smooth and fastly decaying function. 
As the action \eq{eq:ActionNSstream} is dimensionless, one deduces that $\psi\bar{\psi}$ is also dimensionless, and 
that dimensionless frequencies can be defined according to
 $\omega = \kappa^2 \nu_\kappa \hat \omega$.
One then obtains that the dimensionless response stream is given by $\bar\psi= \kappa (D_\kappa^{-1}\nu_\kappa)^{1/2}\hat{\bar{\psi}}$,  and 
 the dimensionless stream function by $\psi=  \kappa^{-1} (D_\kappa \nu_\kappa^{-1})^{1/2} \hat \psi$. 
  At the fixed point, the coefficients $D_\kappa$ and $\nu_\kappa$ are generically expected to behave as power-laws
   $D_\kappa\sim \kappa^{-\eta_D^*}$ and $\nu_\kappa \sim \kappa^{-\eta_\nu^*}$, where the exponents are the fixed-point values
    of the functions
  \begin{equation}
  \eta_D(\kappa)= -\kappa\p_\kappa \ln D_\kappa,
  \, , \quad \quad \quad \eta_\nu(\kappa)= -\kappa\p_\kappa \ln \nu_\kappa\,.
  \end{equation}  
     One deduces that the dynamical critical exponent $z$, which characterizes  the scaling between space and time as
      $\omega\sim|\vp|^z$, is given by  $z=2-\eta_\nu^*$. 
 The two exponents $\eta_D^*$ and $\eta_\nu^*$ are not independent. Their relation follows from the Galilean invariance.
 Let us temporarily introduce a coupling $\lambda$  in front 
 of the non-linear advection term in the action \eq{eq:ActionNSstream}. The Galilean invariance entails that this coupling is not
  renormalized, that is $\p_\kappa \lambda=0$, or equivalently, the flow equation for
 the dimensionless coupling, defined through $\lambda = k D_\kappa^{-1/2} \nu_\kappa^{3/2} \hat \lambda_\kappa$, is exactly 
\begin{equation}
 \kappa \p_\kappa \hat \lambda_\kappa= - \frac 1 2 \hat \lambda_\kappa \, (2 +\eta_D(\kappa) -3 \eta_\nu(\kappa)) \, .
\label{eq:dslam}
\end{equation}
 This  implies that at any non-Gaussian fixed point ($\hat \lambda_*\neq 0$), the exponent $\eta_\nu^*$ 
  is given by
  \begin{equation}
\eta_\nu^* = -(2+\eta_D^*)/3.
\label{eq:etan} 
\end{equation}
In \threeD, the value of the exponent $\eta_D^*$ is fixed by requiring that the mean rate of energy injection (and dissipation)
 $\varepsilon$ is constant \cite{Canet16}. In \twoD, this constraint only gives some bounds on the exponent, and one 
   is led to also analyze the mean rate of enstrophy injection $\varepsilon_{\omega}$ (and dissipation) \cite{Canet16}.
The latter can be expressed  as \cite{Lesieur08}
\begin{align}
   \varepsilon_{\omega} &= \big\langle (\vec\nabla \times 
\vec f\;)(t,\vx)\cdot \vec \omega(t,\vx)  \big\rangle =  \lim_{\delta t\to 0^+}\int_{\omega,\vq} q^4\, N(\vq)\, G_\psi^{(1,1)}(\omega,\vq) e^{-i\omega \delta t}\nonumber\\
  &=  D_\kappa \kappa^{4}   \lim_{\delta t\to 0^+} \int_{\hat\omega, \hat\vq} 
\hat q^6\, \hat n(\hat q)\, \hat G_\psi^{(1,1)}(\hat \omega,\hat \vq) 
e^{-i\hat\omega \hat{\delta t}}.
\label{eq:def-eps}
\end{align}
 The properties of $\hat n$ ensures that the integral is both \UV and \IR finite, 
 and we denote $\Omega^{-1}$ the value of this integral, which is non-universal. 
  To obtain a constant ($\kappa$ independent) mean enstrophy injection rate thus imposes $\eta_D^*=4$.
 The identity \eq{eq:etan} then yields $\eta_\nu^*=2$. 
 As a consequence,  the leading scaling between space and time vanishes,
 that is $z=0$, which is very peculiar and occurs only in $d=2$.   
This scaling is thus governed by  sub-leading  logarithms, as $\omega \sim (\ln (|\vp|/\Lambda))^\delta$, where $\Lambda$ is for instance the \UV scale, and $\delta$ an exponent to be determined. In order  to account for this sub-leading behavior, one is led 
 to include logarithmic corrections in the scale-dependent coefficients, such that they behave at a fixed point as
\begin{equation}
 \nu_\kappa \sim \kappa ^{-\eta^*_\nu} \,(\ln (\kappa/\Lambda))^{\gamma^*_\nu}\, ,\quad\quad D_\kappa \sim \kappa ^{-\eta^*_D} \,(\ln (\kappa/\Lambda))^{\gamma^*_D}\, .
\label{eq:nulog}
\end{equation}
This sub-leading behavior modifies the flow equation \eq{eq:dslam} as follows
\begin{equation}
 \kappa \p_\kappa \hat \lambda_\kappa= - \frac 1 2 \hat \lambda_\kappa \, (2 +\eta_D -3 \eta_\nu) + \frac{1}{2}(\gamma_D-3\gamma_\nu)(\ln(\kappa/\Lambda))^{-1} \,\hat \lambda_\kappa\, .
\end{equation}
Since the corrections to the fixed point are expected to decay faster than a logarithmic, we  require that $\gamma_D-3\gamma_\nu=0$ 
 at a non-Gaussian fixed point, which yields $\gamma_D=3\gamma_\nu\equiv 3\gamma$.
 The value of $\gamma$ is a priori not fixed, and should be computed by integrating the flow equations. 
 In fact, if  one assumes that there is no intermittency correction, that is that dimensional scalings are not modified,
  then $\gamma$ can be determined by consistency, following Kraichnan's argument. This is
  done in \sref{sec:Kraichnan}. In the following, we make no assumptions  on the presence or not of intermittency,
  and keep an undetermined exponent $\gamma$.

 In order to define dimensionless correlation functions, let us first express the forcing amplitude at the integral scale $D_{L^{-1}}$
 as a function of the enstrophy rate using \eq{eq:def-eps} evaluated at $\kappa=L^{-1}$: $\varepsilon_\omega = D_{L^{-1}} L^{-4}\Omega^{-1}$. Assuming the log-corrected power-law behavior \eq{eq:nulog} on the whole inertial range, one can relate 
 the coefficient $D_\kappa$  to its value at the integral scale as
\begin{equation}
 D_\kappa =  D_{L^{-1}} (\kappa L)^{-4} \Big(\frac{\ln(\kappa/\Lambda)}{\ln(L^{-1}/\Lambda)}\Big)^{3 \gamma} \simeq   \varepsilon_\omega \, \Omega\, \kappa^{-4} \Big|\frac{\ln(\kappa/\Lambda)}{\ln(L\Lambda)}\Big|^{3\gamma} \, \equiv D_0 \, \varepsilon_\omega \,\kappa^{-4}\, s^{3\gamma} ,
\end{equation}
where we have introduced the ``\RG time'' $s=\ln(\kappa/\Lambda)$ and the non-universal constant $D_0 =\Omega |\ln(L\Lambda)|^{-3\gamma}$.
The coefficient $\nu_\kappa$ can be related in the same way to its value at the dissipative scale $\eta =(\nu^3/ \varepsilon_{\omega})^{1/6}$ as
\begin{equation}
 \nu_\kappa = \nu_{\eta^{-1}}  (\kappa \eta)^{-2} \Big(\frac{\ln(\kappa/\Lambda)}{\ln(\eta^{-1}/\Lambda)}\Big)^{\gamma} \simeq  \varepsilon_{\omega}^{1/3} \kappa^{-2} \Big|\frac{\ln(\kappa/\Lambda)}{\ln(\eta \Lambda)}\Big|^{\gamma}\,\equiv \nu_0 \, \varepsilon_{\omega}^{1/3} \kappa^{-2} s^\gamma  ,
\end{equation}
where we have identified $\nu\simeq \nu_{\eta^{-1}}$, \ie neglected its evolution  between the microscopic scale $\Lambda$ and the dissipative one $\eta^{-1}$, and defined  $\nu_0 = |\ln(\eta \Lambda)|^{-\gamma}$.

\subsection{Solution at the fixed point for the two-point function}

In this section, we derive the expression of the two-point correlation function in the stream formulation $C_\psi\equiv G^{(2,0)}_\psi$,  obtained as the solution of the leading order flow equation at the fixed point. The flow equation at leading order for the mixed time-wavevector two-point correlation function can be deduced from the general flow \eq{eq:dsgleading} by performing the inverse Fourier transform on the frequency. It reads
\begin{equation}
\kappa \partial_\kappa  C_\psi(t, \vp)=-\frac{1}{2}  p^2 C_\psi(t,\vp) \int_{\varpi} 
 \frac{\cos(\varpi t) -1}{\varpi^2}\, J_\kappa(\varpi)\, ,
\label{eq:flowCt}
\end{equation}
where $J_\kappa(\varpi)$ can be expressed as
 \begin{align}
J_\kappa(\varpi) &= -\int_\vq \tilde \p_\kappa \bar C(\varpi,\vq) =- 2  \,\int_{\vq} \Bigg\{ \kappa \p_\kappa N_\kappa(\vq) \,|\bar G(\varpi,\vq)|^2 -  \kappa \p_\kappa R_\kappa(\vq)\,\bar C(\varpi,\vq) \Re \big[\bar G(\varpi,\vq)\big]\Bigg\}\, ,
 \label{eq:Ikomega}
\end{align}
with  $\bar C$ and $\bar G$ the transverse parts of the two-point correlation and response function of the velocity, $\bar G^{(2,0)}_{v,\mu\nu}({\bp}) =P_{\mu\nu}^\perp(\vp) \bar C({\bp})$ and $\bar G^{(1,1)}_{v,\mu\nu}({\bp}) =P_{\mu\nu}^\perp(\vp) \bar G({\bp})$. 
The remarkable feature of this equation is that $\kappa \partial_\kappa  C_\psi/C_\psi$ does not vanish at large wave-numbers, which means that  there is no decoupling of the large wave-numbers. As shown in  \cite{Canet16,Tarpin18}, this leads to a violation of standard scale invariance,  which is manifest in the solutions derived below.

This flow equation can be simplified in both the regime of large  and small time delays as \cite{Tarpin18}
\begin{equation}
 \kappa \partial_\kappa  C_\psi(t, \vp) = C_\psi(t, \vp) \times\displaystyle\left\{  \begin{array}{l l l }
 \frac{I^0_\kappa}{4}\,t^2\, p^2\, +{\cal O}(p_{\rm max})  \,  \quad\quad\quad &I^0_\kappa = \int_{\varpi} J_\kappa(\varpi)  \quad\quad\quad & t \ll 1\nonumber\\
  \frac{I^\infty_\kappa}{4}|t|\, p^2\,  +{\cal O}(p_{\rm max})  \,    \quad\quad\quad  &I^\infty_\kappa = J_\kappa(0)  \quad\quad\quad  & t \gg 1
\end{array} \right. \, ,
\end{equation} 
where the  ${\cal O}(p_{\rm max})$ explicitly indicates the  contributions beyond the leading order neglected in this flow equation. 
In order to study the fixed point, and using the dimensional analysis of \sref{sec:DIM}, we define the dimensionless correlation function
\begin{equation}
 C_\psi(t,\vp) =  \frac{D_0}{\nu_0}\frac{\varepsilon_\omega^{2/3}}{\kappa^{6}}\, s^{2\gamma} \,\hat C_s\left(\hat t = 
 \nu_0\, \varepsilon_{\omega}^{1/3}\,s^\gamma \,t, \hat p ={p}/{\kappa}\right)\, ,
\end{equation}
  such that one obtains the dimensionless flow equation
\begin{equation}
 \Big[\p_s  - 6 +\frac{2\gamma}{s}-\hat p \p_{\hat p} +\frac{\gamma}{s}\hat t \p_{\hat t}\Big] \hat C_s(\hat t,\hat \vp)= \,\hat C_s(\hat t,\hat \vp) \times \left\{ 
\begin{array}{l l l }
 &   \hat \alpha^{0}_s\,\hat p^{2} \, \hat t^2 +{\cal O}(\hat p_{\rm max}) \quad\quad &  t\ll 1\\
 &    \hat \alpha^{\infty}_s\,\hat p^{2} \, |\hat t| +{\cal O}(\hat p_{\rm max}) &     t\gg 1  \end{array} \right.  \, ,
\label{eq:Xfix}
\end{equation}
 where $\hat \alpha^{0,\infty}_s=D_0 \nu_0^{-3}  \hat I^{0,\infty}_s/4$, with the dimensionless integrals 
$\hat I^0_s \equiv D_\kappa^{-1} \nu_\kappa I^0_\kappa$ and $\hat I^\infty_s \equiv \kappa^2 D_\kappa^{-1} \nu_\kappa^2  I^\infty_\kappa$.
In order to simplify the flow equation, we search for a solution of the form 
\begin{equation}
 \hat C_s(\hat t,\hat \vp) = \frac{1}{\hat p^6}\ln(\hat p)^{-2\gamma} \tilde C_s(\tilde t = \ln(\hat p)^{-\gamma}\hat t ,\hat \vp)\, ,
\end{equation}
which satisfies the flow equation
\begin{align}
 \Big[\p_s   +2\gamma\Big(\frac{1}{s}+\frac{1}{\ln (p/\Lambda) - s}\Big) &-\hat p \p_{\hat p} +\gamma\Big(\frac{1}{s}+\frac{1}{\ln (p/\Lambda) - s}\Big)\tilde t \p_{\tilde t}\Big] \tilde C_s(\tilde t,\hat \vp)= \nonumber\\
 &\tilde C_s(\tilde t,\hat \vp) \times \left\{ 
\begin{array}{l  l }
 &   \hat \alpha^{0}_s\,\hat p^{2} \, \tilde t^2 \, \ln(\hat p)^{2\gamma} +{\cal O}(\hat p_{\rm max})\\
 &   \hat \alpha^{\infty}_s\,\hat p^{2} \, |\tilde t| \, \ln(\hat p)^{\gamma}+{\cal O}(\hat p_{\rm max})
\end{array} \right.  \, .
\end{align}
When approaching the fixed point, in the limit $s\to -\infty$, the term $\ln(p/\Lambda)$ is negligible compared to $s$  for any given external wave-number $p$, such that the terms proportional to $\gamma$ vanish.
  Moreover,  the fixed point corresponds by definition to $\p_s \hat C_s =0$, and dimensionless quantities reaching a constant value
 $\hat \alpha^0_s \to \hat \alpha^0_*$ and $\tilde \alpha^\infty_s \to \hat \alpha^\infty_*$. The fixed-point solution can be 
 obtained by integrating the resulting fixed-point equation as
\begin{equation}
 \ln(\tilde C_*(\tilde t,\hat \vp)) = \left\{  \begin{array}{l l l }
 &   -\tilde t^2 \,\hat \alpha^{0}_*\, \int_0^{\hat p}  x \ln(x)^{2\gamma} dx\,  + \hat F_0(\tilde t) + {\cal O}(\hat p_{\rm max})\quad\quad    \quad\quad & t\ll 1\\
 &  - |\tilde t| \,\hat \alpha^{\infty}_*\,\int_0^{\hat p}  x\ln(x)^{\gamma} dx  \,   + \hat F_\infty(\tilde t)  + {\cal O}(\hat p_{\rm max}) &   t\gg 1  \end{array} \right.  \, ,
\end{equation}
where $\hat F_{0,\infty}$ are universal scaling functions, that can not be computed from the large wave-number regime alone, but which can be determined
 by (numerical) integration of the complete flow equation.
One then deduces the dimensionful physical two-point correlation function
\begin{align}
   C_\psi(t,\vp) &= C_0 \frac{ \varepsilon_\omega^{2/3}}{p^{6}}\ln(p L)^{-2\gamma} \hat {\cal F}_{0,\infty}\Big(\bar{\nu}_0\varepsilon_{\omega}^{1/3} \,t \ln(p L)^{-\gamma}\Big)
 \nonumber\\ 
 &\times\left\{  \begin{array}{l l l }
 &   \exp(- \beta^0_L \, t^2  \int_0^{pL}  x \ln(x)^{2\gamma} dx\,  + {\cal O}(p_{\rm max}L))    \quad\quad & t\ll 1\\
 &     \exp(- \beta_L^\infty\, |t|  \int_0^{pL}  x \ln(x)^{\gamma} dx\,+ {\cal O}(p_{\rm max}L))  &   t\gg 1  \end{array} \right.  \, ,
\label{eq:solC}
\end{align}
with $C_0 =D_0\nu_0^{-1}|\ln(L\Lambda )|^{2\gamma}$, $\bar{\nu}_0 = \nu_0 |\ln(L\Lambda)|^{\gamma}$, $\beta_L^0 = \varepsilon_\omega^{2/3} L^2   \bar{\nu}_0^2 \hat \alpha^0_*$, $\beta_L^\infty = \varepsilon_\omega^{1/3}\bar{\nu}_0 L^2 \hat \alpha^\infty_*$, and $\hat {\cal F}_{0,\infty}=\exp(\hat F_{0,\infty}$).
 The leading order term in the exponential, typically of order $p^2$, is exact. It provides the  decorrelation time of the two-point function
 in the two regimes of small and large time-delays.
This term involves an explicit dependence in the integral scale $L$, and thus breaks standard scale invariance. 
 In \threeD, one obtains a Gaussian dependence in $tp$ for large $p$ and small $t$ (no logarithms), which  is usually interpreted
  as a consequence of the sweeping effect \cite{Tarpin18}. It was early
  predicted  by Kraichnan within the DIA approximation \citep{Kraichnan59}, and later confirmed by \RG approaches under some assumptions on  the effective viscosity \cite{Antonov94}. It has been  observed in many numerical simulations of the \NS equation \citep{Orszag72,Sanada92,He04,Favier10,Canet17} as well as in experiments \citep{Poulain06} in \threeD. 
 The solution \eq{eq:solC} shows that the effect of sweeping takes a modified form in \twoD, where it is corrected by a logarithm. Moreover,   it indicates that a crossover  to a $|t|$ dependence  occurs at long time delays, as predicted in \threeD \cite{Tarpin18}.
 
 The term $\hat {\cal F}_{0,\infty}/p^6$ in  \eq{eq:solC}  corresponds to  the solution that would be obtained assuming standard scale invariance (corrected by the sub-leading logarithm in \twoD). As clear in \Eq{eq:solC}, it is not exact in this calculation, since  the contribution ${\cal O}(p_{\rm max}L)$ of the neglected sub-leading terms in the flow equation could modify this scaling solution, and possibly 
 generate  intermittency corrections. This has to be assessed by computing the \NLO term in the flow equation, which is the purpose of \sref{sec:equaltime}. We first briefly discuss the generalization to $n$-point correlations.

\subsection{Time-dependence of $n$-point function at the fixed point}

 The flow equation \eq{eq:flowleading} for a $n$-point function at leading order in the large wave-number expansion 
 can be expressed in a time-wavevector representation as
\begin{align}
 \p_\kappa  G^{(n)}_{\psi,i_1\dots i_{n}} ({t_1,\vp_{1}},\cdots,{t_{n-1},\vp_{n-1}}) &= \frac{1}{2}\,   G^{(n)}_{\psi,i_1\dots i_{n}}  ({t_1,\vp_{1}},\cdots,{t_{n-1},\vp_{n-1}}) \nonumber\\
 & \times\sum_{k,\ell} \vp_k \cdot \vp_\ell \int_{\varpi}\! J_\kappa(\varpi)\,
\frac{e^{i\varpi(t_k - t_\ell )} - e^{i\varpi t_k} - e^{-i\varpi t_\ell} +1}{\varpi^2} \, .
\label{eq:dup}
\end{align}
This equation can be simplified in both the limits of small and large time delays, as \cite{Tarpin18}
\begin{align}
 \p_\kappa  G^{(n)}_{\psi,i_1\dots i_{n}} (\{t_i,\vp_i\}) &= G^{(n)}_{\psi,i_1\dots i_{n}} (\{t_i,\vp_i\}) \nonumber\\
&\times\left\{\begin{array}{l l l}
&\frac{I^0_\kappa}{4}\,\Big|\sum_{\ell}t_\ell\, p_\ell\Big|^2 + {\cal O}(p_{\rm max}) & t_k \ll 1\\
& \frac{I^\infty_\kappa}{4} \sum_{k,\ell} \vp_k \cdot\vp_\ell \big(|t_k|+|t_\ell| -|t_k-t_\ell|\big) + {\cal O}(p_{\rm max})\quad\quad &t_k \gg 1
\end{array}\right. \, , 
\label{eq:Gnsol}
\end{align}
In order to find the fixed point solution, one introduces the dimensionless $n$-point function, specifying the number $m$ of $\psi$ and $\bar{m}$ of $\bar{\psi}$ fields, as
 \begin{equation}
  G^{(m,\bar{m})}_{\psi,i_1\dots i_{m+\bar{m}}} (\{t_i,\vp_i\}) =\Big(\frac{D_0}{\nu_0}\Big)^{\frac{m-\bar{m}}{2}} \varepsilon_\omega^{\frac{m-\bar{m}}{3}} \frac{s^{\gamma(m-\bar{m})}}{\kappa^{4m-2}}  \hat G^{(m,\bar{m})}_{\psi,i_1\dots i_{m+\bar{m}}} \Big(\big\{\hat{t}_i = \nu_0 \varepsilon_\omega^{1/3} s^\gamma t_i,\hat{\vp}_i = {\vp_i}/\kappa\big\}\Big)\, .
 \end{equation}
The dimensionless function hence satisfies the flow equation
\begin{align}
 \Bigg[ \p_s &-4m +2+(m-\bar{m})\frac{\gamma}{s} - \sum_{k=1}^{m+\bar{m}-1} \hat p_k \p_{\hat p_k}+  \frac{\gamma}{s} \sum_{k=1}^{m+\bar{m}-1} \hat t_k \p_{\hat t_k}\Bigg] \hat G^{(m,\bar{m})}_{\psi,i_1\dots i_{n}} = \nonumber\\
   &\hat G^{(m,\bar{m})}_{\psi,i_1\dots i_{m+\bar{m}}}\times \left\{\begin{array}{l l l}
&{\hat \alpha^0_\kappa}\,\Big|\sum_{\ell}\hat t_\ell\, \hat\vp_\ell\Big|^2+ {\cal O}(\hat p_{\rm max}) \quad\quad\quad & t_k \ll 1\\
& {\hat \alpha^\infty_\kappa}\,  \sum_{k,\ell} \hat\vp_k \cdot\hat\vp_\ell  \big(|\hat t_k|+|\hat t_\ell| -|\hat t_k-\hat t_\ell|\big) + {\cal O}(\hat p_{\rm max}) \quad\quad &t_k \gg 1
\end{array}\right. \, ,
\label{eq:fixedpointGn}
\end{align}
 As for the two-point function, the explicit $s$
 dependence can be removed by searching for a solution of a particular form. The 
 corresponding fixed-point solutions, at un-equal times, will have a similar behavior as
  the solutions obtained in \threeD in \cite{Tarpin18}, modified by the logarithmic corrections.
We leave for future work their explicit derivation, and rather focus on the simultaneous functions in the following.

\section{$N$-point correlation function at equal times}
\label{sec:equaltime}
 
The \NLO term in the large wave-number expansion of the flow equation \eq{eq:flowleading} is calculated in  \sref{eq:nexttoleading}.
 Before  studying this term, let us assume that it decouples at equal-time, which means that the ${\cal O}(p_{\rm max}L)$ term
 in \eq{eq:Gnsol} is also zero at equal time, and expound the consequences for the two-point function.

\subsection{Logarithmic corrections assuming no intermittency}
\label{sec:Kraichnan}

In this section, we hence focus on the two-point function $C_\psi$, and  assume decoupling at equal time, which means that  there is no intermittency, and the exponent of the power-law in \eq{eq:solC} (corresponding to Kolmogorov-Kraichnan scaling) is  exact. 
 In this case, as suggested by Kraichnan \cite{Kraichnan71}, logarithmic corrections are needed in order to ensure consistency with the 
 hypothesis of a constant enstrophy flux in the inertial range. Let us unfold Kraichnan's argument within the present formalism, which will fix the 
 value of the exponent $\gamma$ of the logarithm. We  then give explicitly the logarithmic corrections in the time-dependence of the two-point function and in the equal-time $n$-point functions.
 
For this, we first compute the energy spectrum, assuming that the (connected) equal-time two-point function $C_\psi(0,\vx) =\la \psi(t,\vx)\psi(t,0)\rangle_c$ has no intermittency correction.
 Using the previous scaling analysis, one deduces that 
\begin{equation}
 C_\psi(0,\vp) \sim |\vp|^{-6}\, (\ln(|\vp| L))^{-2\gamma}\, .
\end{equation}
 and obtains for the energy spectrum
\begin{equation}
 E(p) = 2 \pi p^3 \, C_\psi(0,\vp) \sim \, p^{-3}\, (\ln(p L))^{-2\gamma}\,.
\end{equation}
We now establish the expression of the flux of enstrophy.
 Let  $T(p)$ be the rate of energy transfer owing to the nonlinear interactions in \NS equation. The nonlinear transfer
 of enstrophy is then given by $p^2 T(p)$. The flux of enstrophy $Z(p)$ is defined as the nonlinear transfer accross a scale $p$
 as $Z(p)=\int_p^{\infty} p'^2 T(p') dp'$. In the direct-cascade range of wave-numbers, the enstrophy flux is estimated to be \cite{Kraichnan71,Lesieur08,Boffetta12}
\begin{equation}
 Z(p) \sim \varpi_p p^3 E(p) \quad\quad \hbox{with} \quad\quad \varpi^2_p \sim \int_{p_{\rm min}}^p dp' \, p'^2 E(p')\,
\end{equation}
where $\varpi_p$ is the characteristic frequency of the distortion of eddies at scale $1/p$ and $p_{\rm min}\sim 1/L$ is the lowest
 turbulent wave-number.  
In our framework, one obtains  $\varpi_p^2 \sim (\ln( p L))^{-2\gamma+1}$, which then yields $Z(p) \sim (\ln(p L))^{-3\gamma +\frac 1 2}$. Requiring a constant enstrophy flux $Z(p)\equiv \varepsilon_{\omega}$ for wave-numbers in the direct cascade thus fixes $\gamma=1/6$.  As expected, this value corresponds to the log-corrected spectrum predicted by Kraichnan 
 \begin{equation}
 E(p) \sim \, p^{-3} (\ln(pL))^{-1/3} \, .
\end{equation}
 Let us emphasize that this reasoning does not prove the existence  of the log-corrections,
 but simply deduce their form under the assumption of absence of  intermittency.
 With this value of the exponent, the integral $\int_0^{pL} x (\ln x)^{\mu_{0,\infty}} dx$, with $\mu_0 \equiv 2\gamma$ and $\mu_\infty \equiv \gamma$, in \Eq{eq:solC} behaves at large $p$ as $(pL)^2 \ln(pL)^{\mu_{0,\infty}}$, with possible superimposed oscillations. Hence one obtains in the exponential 
 \begin{align}
   C_\psi(t,\vp) &\sim \left\{  \begin{array}{l l l }
 &   \exp(- \bar{\beta}^0_L \, t^2 p^2 \ln(pL)^{-1/3} )    \quad\quad & t\ll 1\\
 &     \exp(- \bar{\beta}_L^\infty\, \, |t| p^2 \ln(pL)^{-1/6} )  &   t\gg 1  \end{array} \right.  \, ,
\end{align}
where numerical constants and a factor $L^2$ has been absorbed in the $\bar\beta_L^{0,\infty}$. It would be very interesting to test this prediction in numerical simulations or experiments. If the exponent of the logarithm in the time dependence can be precisely determined (which is certainly difficult), this would constitute another test of the existence of intermittency in the \twoD direct cascade.
 
To make connection with other existing results, let us express the 
  equal-time two-point correlation function of the vorticity
\begin{equation}
  C_\omega(0,\vp) =  p^4 C_{\psi}(0,\vp)  \sim  p^{-2} \ln(p L)^{-1/3}\, .
\end{equation}
In real space, one obtains
\begin{align}
 C_\omega(0,\vrr) = \int_0^\pi d\theta \int dp \frac{1}{p} \ln(p L)^{-1/3} \exp(i |\vp| |\vrr|\cos\theta) = \int dp \frac{1}{p} \ln(p L)^{-1/3} J_0(p r)
\end{align}
where $J_0$ is a Bessel function. The integral on $p$ is cut in the \IR by $1/L$. In the \UV, 
the Bessel function is dominated by values $p \lesssim 1/r$ since it rapidly oscillates around 0 at large $p$,
 which suppresses the integrand.
 One thus obtains   $C^{(2,0)}_\omega(0,\vx) \sim \ln(L/|\vx|)^{2/3}$, which corresponds to the Falkovich and Lebedev prediction \cite{Falkovich94,Falkovich94b}.

Extending this comparison to higher-order simultaneous $n$-point correlations of the vorticity requires further work.
 Let us just give the result in the Fourier space.
 The flow equation \eq{eq:fixedpointGn} reduces at the fixed point, for equal times $t_\ell=0$, and expressed
   for the 
  $m=2n$ ($\bar{m}=0$) vorticity correlation as 
 \begin{equation}
  \Bigg[ -4n +2+ \frac{n}{3s} - \sum_{k=1}^{2n-1} \hat p_k \p_{\hat p_k}\Bigg] \hat G^{(2n,0)}_{\omega,i_1\dots i_{n}} = 0 \, .
 \end{equation}
 One deduces the general solution of this equation as
\begin{equation}
 \hat G^{(2n,0)}_{\omega,i_1\dots i_{2n}}\big(0,\hat\vp_1,\cdots, 0,\hat\vp_{2n-1}\big) =\Bigg( \prod_{k=1}^{2n-1} \hat p_k^{-2} (\ln \hat p_k)^{-1/6}
 \Bigg) \ln|\hat\vp_1+\cdots+\hat \vp_{2n-1} |^{-1/6} \hat  {F}^{(2n)} \, ,
\label{eq:ansatz2}
\end{equation}
where ${\hat F}^{(2n)}$ is a scaling function.
The obtained logarithmic corrections have an overall behavior compatible with Falkovich-Lebedev prediction in real space
$\langle \omega^n(\vrr_1)\omega^n(\vrr_2)\rangle \propto \ln(L/|\vrr_1-\vrr_2|)^{2n/3}$.
 However,  in order to make the statement precise, one needs  to perform the multi-dimensional inverse Fourier 
 transforms of \eq{eq:ansatz2}. This requires to take into account the different integration sectors 
 with great care. We leave the corresponding analysis for further work.

\subsection{$N$-point correlation function at equal times}
\label{eq:nexttoleading}

 At equal time, the exact leading term at large wave-number of the flow equation for a generic $n$-point correlation function, given by the \rhs of \Eq{eq:flowleading}, vanishes. This can be read off directly  from \eq{eq:dsGn}, or equivalently in the frequency space 
  from \eq{eq:dsgleading}. Indeed,  the equal-time correlation function is obtained 
 by integrating over all the external frequencies
\begin{equation}
  \tilde G^{(n)}_{\psi,i_1\dots i_{n}}(\{t=0,\vp_\ell\}_{1 \leq \ell \leq n}) =\int_{\omega_1,\dots\omega_n} \tilde G^{(n)}_{\psi,i_1\dots i_{n}}(\{\omega_\ell,\vp_\ell\}_{1 \leq \ell \leq n})\, .
\end{equation}
 Since the operator $\tilde {\cal D}(\varpi)$ in \eq{eq:dsgleading} acts as a finite difference, the integrated \rhs vanishes upon absorbing the related shifts by a change of variable in the external frequencies. At equal times, the first non-trivial contribution 
 hence comes from the \NLO term in the large-wave number expansion, which is the fourth order term in the $q_1,q_2$ expansion
\begin{align}
 \p_\kappa \int_{\{\omega_\ell\}} &\tilde G^{(n)}_{\psi,i_1\dots i_{n}}(\{\bp_\ell\}_{1 \leq \ell \leq n})  =  \frac{1}{2}\,  \int_{\bq_1,\bq_2}\!
 \tilde \p_\kappa \tilde  G^{(2)}_{\psi,i j}(-\bq_1,-\bq_2) \nonumber\\
 &\times \int_{\{\omega_\ell\}} \frac{q_a^\mu q_b^\nu q_c^\rho q_d^\sigma}{4!} \frac{\p^4}{\p q_a^\mu \p q_b^\nu \p q_c^\rho \p q_d^\sigma} 
 \left[ \frac{\delta^2}{\delta \varphi_i(\bq_1)\delta\varphi_j(\bq_2)} 
 \tilde G^{(n)}_{\psi,i_1\dots i_{n}}[\{\bp_\ell\};\mathrm{j}] \right]_{\varphi=0}\Bigg|_{\vq_1=\vq_2 = 0} \,,
\end{align}
where as before $a,\,b,\,c,\,d$ take value in $\{1,2\}$. 
The detailed calculation of this term is reported in \aref{app:flowR}, we summarize below the main steps. 
 
 First, one can show that among all the different combinations of $\vq_1$ and $\vq_2$ derivatives, only the ones with two $\vq_1$ and two $\vq_2$ survive after the integration
over the external frequencies. The terms with four derivatives with respect to $\vq_1$ vanish when evaluating at $\vq_2=0$ 
 because of the identity $a)$ in \eq{eq:listwardstreambar} related to the gauge symmetry, (and similarly for $\vq_2$). The terms with only one $\vq_1$ derivative (and similarly only one $\vq_2$) vanish as well.
 The reason is that this derivative yields an  overall $\tilde{\mathcal{D}}$ operator as at leading order, 
 and this contribution vanishes when integrating over the external frequencies
 (see \aref{app:flowR}). Only the terms with two derivatives of $\vq_1$ and $\vq_2$ remain, and they can be written, using space translations and rotations invariance of $\tilde \p_\kappa \tilde  G_\psi^{(2)}$, as
\begin{align}
& \p_\kappa \int_{\{\omega_\ell\}}  \tilde G^{(n)}_{\psi,i_1\dots i_n}(\{\bp_\ell\}_{1 \leq \ell \leq n}) =  \frac{1}{2}\,  \int_{\varpi_1,\varpi_2}\!
\tilde K_{ij}(\varpi_1,\varpi_2) \nonumber\\
&\times \int_{\{\omega_\ell\}} \Big(\frac{\p^4}{\p q_1^\mu \p q_1^\mu \p q_2^\nu \p q_2^\nu} + 2 \frac{\p^4}{\p q_1^\mu \p q_2^\mu \p q_1^\nu \p q_2^\nu} \Big)  \Big[ \frac{\delta^2}{\delta \varphi_i(\bq_1)\delta\varphi_j(\bq_2)} 
 \tilde G^{(n)}_{\psi,i_1\dots i_{n}}[\{\bp_\ell\};\mathrm{j}] \Big]_{\varphi=0} \Bigg|_{\vq_1=\vq_2 = 0}\,.
\label{eq:flowfinal}
\end{align} 
with
\begin{equation}
\tilde K_{ij}(\varpi_1, \varpi_2) \equiv \frac{1}{32}\,  \int_{\vq}\!
 \tilde \p_\kappa \tilde  G^{(2)}_{\psi,ij}(-\varpi_1,-\varpi_2, q^2)  (q^2)^2\,.
\end{equation}
The last two parts of \aref{app:flowR} are devoted to show that
\begin{align}
\frac{\p^4}{\p q_1^\mu \p q_1^\mu \p q_2^\nu \p q_2^\nu}  &\Big[ \frac{\delta^2}{\delta \varphi_i(\bq_1)\delta\varphi_j(\bq_2)} 
 \tilde G^{(n)}_{\psi,i_1\dots i_{n}}[\{\bp_\ell\}_{1 \leq \ell \leq n};\mathrm{j}] \Big]_{\varphi=0} \Bigg|_{\vq_1=\vq_2 = 0}\nonumber\\
&=\delta_{i\psi}\delta_{j\psi}\tilde{\mathcal{R}}(\varpi_1)  \tilde{\mathcal{R}}(\varpi_2) \tilde G^{(n)}_{i_1\dots i_{n}}(\{\bp_\ell\})  \,,
\label{eq:uncrossed}
\end{align}
which means that the contribution with the uncrossed derivatives is completly controlled by the extended symmetries and can be closed exactly 
using the corresponding  Ward identities.
It turns out that this term vanishes after integration over frequencies by conservation of angular momentum.
\begin{align}
 &\int_{\{\omega_\ell\}} \tilde{\mathcal{R}}(\varpi_1)   \frac{2 i \epsilon_{\alpha\beta}}{\varpi_2}\sum_{k=1}^n p_k^a \frac{\p}{\p p_k^b} \tilde G^{(n)}_{i_1 \dots i_n}(\dots, \omega_k+\varpi_2, \vp_k,\dots)\nonumber\\
&= \int_{\{\omega_\ell\}} \tilde{\mathcal{R}}(\varpi_1)   \frac{2 i \epsilon_{\alpha\beta}}{\varpi_2} \sum_{k=1}^{n-1} p_k^a \frac{\p}{\p p_k^b} \Big[ \tilde G^{(n)}_{i_1 \dots i_n}(\dots, \omega_k+\varpi_2, \vp_k,\dots) - \tilde G^{(n)}_{i_1 \dots i_n}(\dots,\omega_n+\varpi_2, \vp_n) \Big]\nonumber\\
&=0\,,
\end{align}
and thus, it gives no contribution at equal times.

To summarize, beyond the technical details, one finds that all  the terms which are
controlled by the extended symmetries of the \NS action  vanish after integration over external frequencies. The only remaining
 term, which is not controlled by symmetries is the one with the crossed derivatives
\begin{align}
 \p_\kappa &\int_{\{\omega_\ell\}} \tilde G^{(n)}_{\psi,i_1\dots i_{n}}(\{\bp_\ell\}_{1 \leq \ell \leq n})\nonumber\\
 &=  \int_{\varpi_1,\varpi_2}\!
\tilde K_{ij}(\varpi_1,\varpi_2) \int_{\{\omega_\ell\}} \frac{\p^4}{\p q_1^\mu \p q_2^\mu \p q_1^\nu \p q_2^\nu} \Big[ \frac{\delta^2}{\delta \varphi_i(\bq_1)\delta\varphi_j(\bq_2)} 
 \tilde G^{(n)}_{\psi,i_1\dots i_{n}}[\{\bp_\ell\};\mathrm{j}] \Big]_{\varphi=0} \Bigg|_{\vq_1=\vq_2 = 0}\,. 
\label{eq:crossed}
\end{align}
This term is a priori non zero, and could be a source of intermittency. However, the effect can be expected to be much weaker than in \threeD, since the time-gauged rotation does not hold in \threeD and the corresponding terms do  not vanish a priori.

 It is possible that the crossed contribution 
 \eq{eq:crossed} turns out to be proportional to the uncrossed one \eq{eq:uncrossed}, and thus vanishes, at least in some specific wave-vector configurations,
  but we have not been able to prove it. If  this were case, 
  this would  imply that there is no intermittency in the direct cascade of \twoD turbulence  at equal times.
It is instructive to consider the flow of the two-point function to further comment on this.
The function appearing in square bracket in the \rhs of the flow equation \eq{eq:flowfinal} for $n=2$
is a function of $\vp$, $\vq_1$, and $\vq_2$, and the corresponding frequencies: 
\begin{equation}
\Big[ \frac{\delta^2}{\delta \varphi_i(\bq_1)\delta\varphi_j(\bq_2)} 
 \tilde G^{(2)}_{\psi,i_1, i_{2}}[\bp;\mathrm{j}] \Big]_{\varphi=0} \equiv F(\omega,\vp,\varpi_1,\vq_1,\varpi_2,\vq_2)\,.
\end{equation}
The wave-vector part of $F$ involves only  five independent scalars in \twoD, which, considering the symmetry of exchange $\vq_1\longleftrightarrow \vq_2$ can be chosen as
\begin{equation}
 F(\vp,\vq_1,\vq_2) = {\cal F}(p^2,q_1^2+q_2^2, (q_1^2+q_2^2)^2,\vq_1\cdot\vq_2,\vp\cdot (\vq_1+\vq_2)) \, ,
\end{equation}
omitting the frequencies, which play no rôle for evaluating the $q$ derivatives.
One obtains that 
\begin{align}
  \frac{\p^4}{\p q_1^\mu \p q_1^\nu \p q_2^\rho \p q_2^\sigma} F(\vp,\vq_1,\vq_2) \Big|_{\vq_1=\vq_2 = 0} &= \delta_{\mu\nu}\delta_{\rho\sigma} f_1(p^2) + (\delta_{\mu\rho}\delta_{\nu\sigma} +\delta_{\mu\sigma}\delta_{\nu\rho}) f_2(p^2) \nonumber\\
 & + p_\mu p_\nu p_\rho p_\sigma f_3(p^2) + (\delta_{\mu\nu} p_\rho p_\sigma + \delta_{\rho\sigma}p_\mu p_\nu) f_4(p^2) \nonumber\\
 &+  (\delta_{\mu\rho} p_\nu p_\sigma + \delta_{\mu\sigma}p_\nu p_\rho + \delta_{\nu\rho} p_\mu p_\sigma + \delta_{\nu\sigma}p_\mu p_\rho)  f_5(p^2) 
\end{align}
with $f_1= 4{\cal F}^{(0,2,0,0,0)}-8 {\cal F}^{(0,0,1,0,0)}$, $f_2= {\cal F}^{(0,0,0,2,0)}$, $f_3= {\cal F}^{(0,0,0,0,4)}$, $f_4= 2{\cal F}^{(0,1,0,0,2)}$ and $f_5= {\cal F}^{(0,0,0,1,2)}$. One hence deduces that
\begin{align}
 \frac{\p^4}{\p q_1^\mu \p q_1^\mu \p q_2^\nu \p q_2^\nu} F(\vp,\vq_1,\vq_2) \Big|_{\vq_1=\vq_2 = 0}&= 4 f_1 + 4 f_2+ p^4 f_3 + p^2(8f_4+4 f_5)\nonumber\\
2\frac{\p^4}{\p q_1^\mu \p q_1^\nu \p q_2^\mu \p q_2^\nu} F(\vp,\vq_1,\vq_2) \Big|_{\vq_1=\vq_2 = 0}&= 4 f_1 + 12 f_2+ 2 p^4 f_3 + p^2(8f_4+12 f_5)
\end{align}
The crossed and uncrossed terms seem not to be  proportional in general. They would be  so for instance if the function ${\cal F}$ depends only on the moduli of the wave-vectors, but not on their relative angles. For the structure functions, which involve only two space points, it is not clear which configurations of wave-vectors dominate, and whether this relation could be fulfilled. The detailed analysis of structure functions is left for future investigations.

\section{Conclusion and perspectives}

In this paper, we investigated \twoD forced turbulence, using field theoretical techniques. 
 We unveiled two  extended symmetries of the \NS field theory that were not identified yet. One,  related to time-gauged shift 
 of the response field, exists in both \threeD and \twoD, while the other one, related to time-gauged rotations,  is only realized
  in \twoD. These symmetries bring new exact relations 
 between the correlation functions of the theory through Ward identities, which can be useful in general.
 
We then exploited these Ward identities in  the framework of the \NPRG, within the large-wave number expansion scheme developed in \cite{Canet16,Tarpin18}, to compute some properties of the correlation functions of \twoD isotropic and homogeneous turbulence.
 The leading order term of this expansion can be closed exactly, and allowed us to obtain the  time dependence of
the $2$-point correlation function in the stream formulation at both small and large time delays. This prediction could be tested in numerical simulations or in experiments. The generalization for $n$-point function is left for future work. 
  This exact leading order contribution  explicitly breaks  standard scale invariance.

 At equal times,  the leading order  term vanishes, and one is left with log-corrected power-laws. If one assumes that there is no intermittency, 
then one recovers Kraichnan's logarithms, by unrolling a similar argument within the \NPRG formalism.
 To assess the presence or not of intermittency in equal-time quantities,  we calculated the \NLO term in the large-wave number expansion. We found that almost all the terms are controlled by the symmetries, and that these terms vanish at equal times, and hence cannot generate intermittency. 
  Nevertheless, there remains one contribution, which is not constrained by the symmetries. 
 This contribution could lead to intermittency correction. However,  this correction can be reasonably expected to be much weaker than in \threeD, since in \threeD many other contributions remain. Moreover, the unconstrained contribution could turn out to vanish is some   specific wave-vector configurations, as the ones involved in the calculation of  structure functions. 
    Further works are in progress to  approximate this contribution and estimate the related intermittency correction.
 Let us also emphasize that  the techniques developed in the present work 
 could be useful to study  other hydrodynamical systems, such as passively advected quantities 
 \cite{Adzhemyan1998,Adzhemyan2005,Kupiainen:2006em,Pagani:2015hna,Antonov2015,Hnatic2016,Hnatic2018},
 which is  underway \cite{Canet2019}.

\begin{acknowledgments}
The authors thank B. Delamotte  for fruitful discussions and his contribution in the early stage of this work. N. W. thanks Pedeciba (Programa de desarrollo de las Ciencias B\'asicas, Uruguay)
 and acknowledges funding through  grant from la Comisi\'on Sectorial de Investigaci\'on Cient\'ifica de la Universidad de la Rep\'ublica, Project I+D 2016 (cod 412).
\end{acknowledgments}

\appendix

\section{Extended symmetries and Ward identities}

In this appendix, we explain the general derivation of Ward identities within the \NPRG framework, and then give details on the derivation of the ones associated to the time-gauged Galilean and rotation symmetries in the stream formulation.

\subsection{Ward identities in the \NPRG framework}
\label{app:WardGen}

Let us  illustrate the derivation of a Ward identity on a generic field theory for a  field $\phi$ which can possess multiple components.
Within the \NPRG framework, in the presence of the infrared regulator $\Delta \mathcal{S}_\kappa$, the associated partition function
 is 
\begin{equation}
 \mathcal{Z}_\kappa[ \mathrm{j}] = \int D[\phi] e^{- \mathcal{S}[\phi] - \Delta \mathcal{S}_\kappa[\phi] +  \mathrm{j}   \cdot \phi }
\end{equation} 
Let us consider a change of variable $\phi \to \phi'$ in $\mathcal{Z}_\kappa$ which leaves the functional measure invariant.
Denoting $\delta \mathcal{X}[\phi] = \mathcal{X}[\phi'] - \mathcal{X}[\phi]$, where $\mathcal{X}$ is a generic functional, one has
 \begin{align}
 \mathcal{Z}_\kappa[ \mathrm{j}] &= \int D[\phi'] e^{- \mathcal{S}[\phi'] - \Delta \mathcal{S}_\kappa[\phi'] +  \mathrm{j}   \cdot \phi' }\nonumber\\
  &= \int D[\phi] e^{ -\mathcal{S}[\phi + \delta \phi] - \Delta \mathcal{S}_\kappa[\phi + \delta \phi] +  \mathrm{j}   \cdot \phi +   \mathrm{j}  \cdot \delta \phi}\nonumber\\
  &= \int D[\phi] e^{ -\mathcal{S}[\phi] - \Delta \mathcal{S}_\kappa[\phi] + \mathrm{j}   \cdot \phi} e^{ - \delta(\mathcal{S} + \Delta \mathcal{S}_{\kappa})[\phi] + \mathrm{j}   \cdot \delta \phi} \nonumber\\
  &=  \mathcal{Z}_\kappa[ \mathrm{j}]\mean{ e^{-\delta(\mathcal{S} + \Delta \mathcal{S}_{\kappa})[\phi] + \mathrm{j}  \cdot \delta \phi}}_\mathrm{j}\,,
  \end{align}
where $\mean{\cdot}_\mathrm{j}$ is the mean value in presence of the sources. Since $\phi \to \phi'$ is just a change of variables, the partition function is unchanged, which implies the Ward identity
\begin{equation}
 \mean{ e^{- \delta(\mathcal{S} + \Delta \mathcal{S}_{\kappa})[\phi] +  \mathrm{j}  \cdot \delta \phi}}_\mathrm{j} = 1\,.
\label{eq:protoWard}
\end{equation}

We focus on infinitesimal transformations,  
$\delta\phi = \delta_{\epsilon}\phi = O(\epsilon)$, which are  at most linear in the fields 
\begin{equation}
 \phi_{i}(\bx) \to \phi_i'(\bx) = \phi_i(\bx) + \delta_{\epsilon}\phi_i(\bx),\quad \delta_{\epsilon}\phi_i(\bx) =\epsilon\Big[ \int_{\by}A_{ij}(\bx,\by)\phi_j(\by)+B_i(\bx)\Big]\, ,
\label{eq:defsymlin}
\end{equation}
where $A$ is an operator acting on $\phi$. It follows that $\mean{\delta_\epsilon \phi} = \delta_\epsilon \Phi$, with $\Phi \equiv \mean{\phi}_\mathrm{j}$.
At linear order in $\epsilon$, one obtains from \Eq{eq:protoWard}
\begin{equation}
\mean{ \delta_\epsilon \mathcal{S} }_\mathrm{j} + \mean{\delta_\epsilon \Delta \mathcal{S}_\kappa }_\mathrm{j} + \mathrm{j} \cdot \delta_\epsilon \Phi = 0\,,
\label{eq:protoWardlin}
 \end{equation}
where $\delta_\epsilon \mathcal{X}$ is the part of $\delta \mathcal{X}$ linear in $\epsilon$. Defining
$\mathcal{R}_\kappa^{ij}(\bx,\by) \equiv \ddf{ \Delta \mathcal{S}_{\kappa}}{\phi_i(\bx) }{\phi_j(\by)}\,$, 
 one has
\begin{equation}
\delta_\epsilon \Delta \mathcal{S}_\kappa =\int_{\bx,\by} \mathcal{R}_\kappa^{ij}(\bx,\by)\delta_\epsilon  \phi_i(\bx) \phi_j(\by) \,,
\end{equation}
 since $\Delta \mathcal{S}_\kappa$ is quadratic in the fields and 
\begin{equation}
\mathrm{j}_{i}(\bx) = \df{\Gamma_{\kappa}}{\phi_i(\bx)} + \int_{\by} \mathcal{R}_\kappa^{ij}(\bx,\by) \Phi_{j}(\by)\, ,
\end{equation}
which follows  from the definition of
$\Gamma_{\kappa}$ \eq{eq:legendre}. Using these relations  and \eq{eq:defsymlin}, one deduces from \eq{eq:protoWardlin} the Ward identity
\begin{equation}
 \delta_{\epsilon}\Gamma_\kappa[\Phi] = \mean{\delta_{\epsilon}\mathcal{S}}_\mathrm{j} +\epsilon \int_{\bx,\by,\bz} \mathcal{R}_\kappa^{ij}(\bx,\by) A_{ik}(\bx,\bz) G_{jk}(\by,\bz)\, ,
\label{eq:ward}
\end{equation}
where $\delta_{\epsilon}\Gamma_\kappa[\Phi] =\int_\bx \df{\Gamma_{\kappa}}{\Phi_i(\bx)}\delta_{\epsilon}\Phi_i(\bx)$.
 The second term vanishes if the regulator term is invariant under the transformation. Let us notice that, because of the definition of the Legendre transform, the variation of the regulator under the shift part of the change of variable 
\eq{eq:defsymlin} never enters the Ward identity in this formalism.

For exact symmetries of the action and of the regulator, \Eq{eq:ward} simply translates into 
$\delta_{\epsilon}\Gamma_\kappa[\Phi]= 0$, which means that  $\Gamma_\kappa$ also possesses these symmetries. For extended symmetries, where the variations of the action and of the regulator are non-zero but linear in the fields, the mean and 
the variation commute and the Ward identity reads $\delta_{\epsilon}\Gamma_\kappa[\Phi] = \delta_{\epsilon}\mathcal{S}[\Phi]$, which means that
  the variation of the \EAA
 $\Gamma_\kappa$ is equal to the mean of the variation of $\mathcal{S}$. This provides
non-renormalisation theorems which fix a sector of $\Gamma_\kappa$ to its bare value. 
 The transformations considered in this paper are all pure shifts of the fields, except
 the extended Galilean symmetry and the extended rotations. However, as the regulator is invariant under space translations, rotations, and is instantaneous (delta-correlated in time), it is invariant as well under time-gauged translations and time-gauged rotations, and thus it does  not enter the corresponding Ward identities.

\subsection{Ward identity for the time-gauged Galilean symmetry}
\label{app:WardG}

The set of Ward identities related to the time-gauged Galilean symmetry  for generic vertex functions are derived in \cite{Canet16,Tarpin18} in the velocity formulation.
 We present in this appendix  their derivation in the stream formulation.
The functional  Ward identity for time-gauged Galilean transformation, \Eq{eq:listwardstream} $d)$, reads
\begin{equation}
 \int_{\vx} \Big\{ \big(- \epsilon_{\gamma\beta} x_\gamma \p_t + \p_\beta \Psi\big) \df{\Gamma_\kappa}{\Psi(\bx)} + \p_\beta \bar\Psi \df{\Gamma_\kappa}{\bar \Psi(\bx)} \Big\} = 0 \, .
\label{eq:repeat}
\end{equation}
To deduce the Ward identity for a generic vertex function $\Gamma_\kappa^{(m,n)}$, one takes $m$ functional derivatives of \eq{eq:repeat}
with respect to the stream function $\Psi(\bx_i)$ and $n$ with respect to response response stream $\bar \Psi(\bx_j)$, and then set the fields to zero, which
yields, after multiplying by $\epsilon_{\alpha\beta}$,
\begin{align}
 & \int_{\vx} \Big\{ - x_\alpha \p_t \Gamma_\kappa^{(m+1,n)}(\bx, \{\bx_\ell\}_{1 \leq \ell \leq m+n}) \nonumber\\
&- \epsilon_{\alpha\beta} \sum_{k=1}^{m+n} \delta(t-t_k)\delta^d(\vx-\vx_k)\p_\beta \Gamma_\kappa^{(m,n)}(\{\bx_\ell\}_{1 \leq \ell \leq k-1},\,\bx,\{\bx_\ell\}_{k+1 \leq \ell \leq m+n}) \Big\} =0 \, .
\end{align}
This identity reads in the Fourier space:
\begin{align}
 \frac{\p}{\p q^\alpha} &\tilde \Gamma_\kappa^{(m+1,n)}(\varpi,\vq,\{\bp_\ell\}_{1 \leq \ell \leq m+n})\Big|_{\vq=0} \nonumber\\&= -i \epsilon_{\alpha\beta}\sum_{k=1}^{m+n}\frac{p_k^{\beta}}{\varpi}\tilde \Gamma_\kappa^{(m,n)}(\{\bp_\ell\}_{1 \leq \ell \leq k-1},\,\omega_k+\varpi,\vp_k,\{\bp_\ell\}_{k+1 \leq \ell \leq m+n} )\nonumber\\
&\equiv i \epsilon_{\alpha\beta} \tilde{\mathcal{D}}_\beta(\varpi)\tilde \Gamma_\kappa^{(m,n)}(\{\bp_\ell\}_{1 \leq \ell \leq m+n}) \,.
\label{eq:wardGalstreamtilde}
\end{align}
The corresponding expression for $\bar \Gamma_\kappa^{(m+1,n)}$, which is defined as in \eq{eq:tilde-bar} by extracting the delta functions associated to the conservation of wave-vector and frequency, \ie the invariance under global translations, can be deduced from \eq{eq:wardGalstreamtilde} and reads
\begin{equation}
 \frac{\p}{\p q^\alpha} \bar \Gamma_\kappa^{(m+1,n)}(\varpi,\vq,\{\bp_\ell\}_{1 \leq \ell \leq m+n-1})\Big|_{\vq=0} = i \epsilon_{\alpha\beta}  \mathcal{D}_\beta(\varpi)\bar \Gamma_\kappa^{(m,n)}(\{\bp_\ell\}_{1 \leq \ell \leq m+n-1})\,,
\end{equation}
where the operator $\mathcal{D}_\beta(\varpi)$ is now defined by
\begin{align}
 \mathcal{D}_\alpha(\varpi) &F( \{\bp_\ell\}_{1 \leq \ell \leq n}) \nonumber\\
 &\equiv - \sum_{k=1}^{n} p_k^\alpha
    \Bigg[  \frac{F(\{\bp_\ell\}_{1 \leq \ell \leq k-1}, \omega_k + \varpi,\vp_k, \{\bp_\ell\}_{k+1 \leq \ell \leq n})- F(\{\bp_\ell\})}{\varpi}\Bigg]\,.
\end{align}
The expression shows  explicitly the regularity of the limit $\varpi\to 0$.

\subsection{Ward identity for the time-gauged  rotation symmetries}
\label{app:WardS}

The functional Ward identity associated with the time-gauged rotation, \Eq{eq:listwardstream} $e)$, is given by
\begin{equation}
\int_{\vx} \Big\{ \big(\frac{x^2}{2}\p_t + \epsilon_{\alpha\beta}x_\beta\p_\alpha\Psi\big) \df{\Gamma_\kappa}{\Psi(\bx)} + \epsilon_{\alpha\beta} x_\beta\p_\alpha \bar \Psi \df{\Gamma_\kappa}{\bar \Psi(\bx)} \Big\} = 2 \int_{\vx} \p_t^2 \bar \Psi \, .
\label{eq:repeat2}
\end{equation}
Let us first derive an identity for $\Gamma^{(1,1)}$. Taking
 one derivative with respect to $\bar \Psi(\bx')$,  setting the fields to zero, and noting  that  $\p_\alpha \Gamma_\kappa^{(0,1)}(\bx) = 0$ by translational invariance, one obtains in the Fourier space
\begin{equation}
\frac{\p^2}{\p p^2} \tilde \Gamma_\kappa^{(1,1)}(\bp,\bp')\Big|_{\vp=0} = - 4 i \omega \delta^d(\vp\,')\delta(\omega+\omega')\, .
\end{equation}
This result can be interpreted as the non-renormalisation of the kinetic term in the bare action. 
 
We now derive the identity for a generic vertex function. Taking $m$ functional derivatives of \eq{eq:repeat2} with respect to $\Psi$ and $n$ with respect to $\bar\psi$, one obtains
\begin{align}
& \int_{\vx} \Big\{ \frac{x^2}{2}\p_t \Gamma_\kappa^{(m+1,n)}(\bx, \{\bx_\ell\}_{1 \leq \ell \leq m+n}) \nonumber\\
& - \sum_{k=1}^{m+n} \delta(t-t_k)\delta^d(\vx-\vx_k) \epsilon_{\alpha\beta}x_\beta\p_\alpha \Gamma_\kappa^{(m,n)}(\{\bx_\ell\}_{1 \leq \ell \leq k-1},\,\bx,\{\bx_\ell\}_{k+1 \leq \ell \leq m+n}) \Big\} =0 \, .
\end{align}
One then deduces the general identity in the Fourier space:
\begin{align}
 \frac{\p^2}{\p q^2} &\tilde\Gamma_\kappa^{(m+1,n)}(\bq, \{\bp_\ell\}_{1 \leq \ell \leq m+n})\Big|_{\vq=0}\nonumber\\
 &= \frac{2 i\epsilon_{\alpha\beta}}{\varpi}  \sum_{k=1}^{m+n} p_k^\alpha \frac{\p}{\p p_k^\beta} \tilde\Gamma_\kappa^{(m,n)}( \{\bp_\ell\}_{1 \leq \ell \leq k-1},\,\omega_k + \varpi, \bp_k, \{\bp_\ell\}_{k+1 \leq \ell \leq m+n})\nonumber\\
 &\equiv \tilde{\mathcal{R}}(\varpi)\tilde\Gamma_\kappa^{(m,n)}( \{\bp_\ell\}_{1 \leq \ell \leq m+n})\, .
\end{align}

This identity can be expressed  in terms of the $\bar \Gamma_\kappa^{(m,n)}$,  \ie extracting the delta of conservation, as 
\begin{equation}
 \frac{\p^2}{\p q^2} \bar \Gamma_\kappa^{(m+1,n)}(\varpi,\vq,\{\bp_\ell\}_{1 \leq \ell \leq m+n-1})\Big|_{\vq=0} =\tilde{\mathcal{R}}(\varpi)\bar \Gamma_\kappa^{(m,n)}(\{\bp_\ell\}_{1 \leq \ell \leq m+n-1} )\, ,
\label{eq:temp}
\end{equation}
where the identities $a)$ and $d)$ of \Eq{eq:listwardstreambar} have been used. 
In contrast with the case of Galilean symmetry, this expression does not explicitly show that the limit $\varpi \to 0$ is well-defined. 
 Indeed, extended Galilean
symmetry corresponds to time-gauged space translation and the zero frequency limit is equivalent to time-independent space translation,
 which are implicitly used when passing from the $\tilde \Gamma_\kappa^{(m,n)}$ to the $\bar \Gamma_\kappa^{(m,n)}$. 
 For time-gauged rotations, the
zero frequency limit, \ie  time-independent rotations, are needed. The corresponding Ward identity reads
\begin{equation}
 \epsilon_{\alpha\beta}\sum_{k=1}^{m+n-1} p_k^\alpha \frac{\p}{\p p_k^\beta} \bar \Gamma_\kappa^{(m,n)}( \{\bp_\ell\}_{1 \leq \ell \leq m+n-1}) = 0\,.
\end{equation}
Substracting it from \eq{eq:temp}, one finally obtains an expression in terms of finite differences as for the time-gauged Galilean identity:
\begin{equation}
 \frac{\p^2}{\p q^2} \bar \Gamma_\kappa^{(m+1,n)}(\varpi,\vq,\{\bp_\ell\}_{1 \leq \ell \leq m+n-1})\Big|_{\vq=0} = \mathcal{R}(\varpi) \bar \Gamma_\kappa^{(m,n)}(\{\bp_\ell\}_{1 \leq \ell \leq m+n-1})\, 
\end{equation}
with
\begin{align}
 \mathcal{R}(\varpi) &F(\{\bp_\ell\}_{1 \leq \ell \leq n}) \nonumber\\
 &\equiv 2 i\epsilon_{\alpha\beta}\sum_{k=1}^{n} p_k^\alpha\frac{\p}{\p p_k^\beta} \Bigg[\frac{F(\{\bp_\ell\}_{1 \leq \ell \leq k-1},\,\omega_k+\varpi,\vp_k,\{\bp_\ell\}_{k+1 \leq \ell \leq n} ) - F(\{\bp_\ell\}_{1 \leq \ell \leq n} )}{\varpi}\Bigg]\, ,
\end{align}
where the regularity of the limit  $\varpi\to 0$ is now manifest.

Let us note that similar  subtleties arise when passing  from the $\tilde \Gamma_\kappa^{(m,n)}$ to the $\bar \Gamma_\kappa^{(m,n)}$ for
 the symmetries $b)$ and $c)$ of \Eq{eq:listwardstream}, but the derivation of the corresponding identities is straightforward since  the \rhs is always zero. We finally recapitulate the list of Ward identities for the vertex function $\tilde \Gamma^{(m,n)}$, which are the ones used 
  in  \aref{sec:appLWN}
\begin{align}
 & a) \quad \tilde \Gamma_\kappa^{(m,n)} (\dots, \varpi, \vq, \dots)\Big|_{\vq=0} = 0 \nonumber\\
 & b) \quad \frac{\p}{\p q^i} \tilde \Gamma_\kappa^{(m,n+1)}( \{\bp_\ell\}_{1 \leq \ell \leq m},  \varpi, \vq, \{\bp_\ell\}_{1 \leq \ell \leq n} ) \Big|_{\vq=0} = 0\nonumber\\
 & c) \quad \frac{\p^2}{\p q^2} \tilde \Gamma_\kappa^{(m,n+1)}(\{\bp_\ell\}_{1 \leq \ell \leq m},  \varpi, \vq, \{\bp_\ell\}_{1 \leq \ell \leq n} ) \Big|_{\vq=0} = 0\nonumber\\
 &\quad \quad \quad \quad \text{except } \frac{\p^2}{\p q^2} \tilde \Gamma_\kappa^{(1,1)}(\varpi', {\vq}\,', \varpi, \vq) \Big|_{\vq=0} = 4 i \varpi \delta^d(\vq\,')\delta(\varpi + \varpi')\nonumber\\
 & d) \quad \frac{\p}{\p q^i} \tilde \Gamma_\kappa^{(m+1,n)}(\varpi,\vq,\{\bp_\ell\}_{1 \leq \ell \leq m+n})\Big|_{\vq=0} = i \epsilon_{\alpha\beta}  \tilde{\mathcal{D}}_\beta(\varpi) \tilde \Gamma_\kappa^{(m,n)}(\{\bp_\ell\}_{1 \leq \ell \leq m+n})\nonumber\\
 & e) \quad \frac{\p^2}{\p q^2} \tilde \Gamma_\kappa^{(m+1,n)}({\varpi,\vq},\{\bp_\ell\}_{1 \leq \ell \leq m+n}) \Big|_{\vq=0}  =\tilde{\mathcal{R}}(\varpi) \tilde \Gamma_\kappa^{(m,n)}(\{\bp_\ell\}_{1 \leq \ell \leq m+n})\nonumber\\
 &\quad \quad \quad \quad \text{except } \frac{\p^2}{\p q^2} \tilde \Gamma_\kappa^{(1,1)}(\varpi, \vq, \varpi', \vq\,') \Big|_{\vq=0} = - 4 i \varpi \delta^d(\vq\,')\delta(\varpi + \varpi')\, ,
\label{eq:listwardstreamtilde}
\end{align}
with the two following definitions for the operator $\tilde{\mathcal{D}}_\alpha(\varpi)$ and  $\tilde{\mathcal{R}}(\varpi)$:
\begin{align}
 &\tilde{\mathcal{D}}_\alpha(\varpi) F( \{\bp_\ell\}_{1 \leq \ell \leq n}) \equiv - \sum_{k=1}^{n}\frac{p_k^\alpha}{\varpi}F(\{\bp_\ell\}_{1 \leq \ell \leq k-1},\,\omega_k + \varpi,\vp_k,\{\bp_\ell\}_{k+1 \leq \ell \leq n} )\nonumber\\
&\tilde{\mathcal{R}}(\varpi) F( \{\bp_\ell\}_{1 \leq \ell \leq n}) \equiv \frac{2 i\epsilon_{\alpha\beta}}{\varpi}  \sum_{k=1}^{n} p_k^\alpha \frac{\p}{\p p_k^\beta} F( \{\bp_\ell\}_{1 \leq \ell \leq k-1},\,\omega_k + \varpi, \bp_k, \{\bp_\ell\}_{k+1 \leq \ell \leq n})\, .
\label{eq:defDR}
\end{align}

\section{Large wave-number expansion in the stream formulation}
\label{sec:appLWN}

Let us  derive the expression \eq{eq:flowGnstream} of the flow equation for a generic generalized $n$-point connected correlation function $\tilde G_\psi^{(n)}$ in the regime of large wave-numbers.
 This derivation is the same as the one given  in the velocity formulation in~\cite{Tarpin18},
 but we repeat it in the stream formulation for completeness. The flow equation for a generic $G_\psi^{(n)}$ is obtained by taking $n$ functional derivatives of \eq{eq:dkw} with respect to the sources $\mathrm{j}_{i_k}$, $1 \leq k \leq n$,
 which yields
\begin{align}
 \p_\kappa G^{(n)}_{\psi,i_1\dots i_{n}}[\{\bx_\ell\}_{1 \leq \ell \leq n};\mathrm{j}] &= - \frac{1}{2}\,  \int_{\by_1,\by_2}\! \p_\kappa [\mathcal{R}_\kappa]_{ij}({\by_1-\by_2}) \,\Bigg\{ G^{(n+2)}_{\psi,ij i_1\dots i_{n}}[\by_1,\by_2,\{\bx_\ell\};\mathrm{j}]
\nonumber\\
 & + \sum_{(\{i_\ell\}_1, \{i_\ell\}_2) \atop \#1 + \#2=n} G^{(\#1+1)}_{\psi,i \{i_\ell\}_1}[{\by_1,\{\bx_{\ell}\}_1};\mathrm{j}] G^{(\#2+1)}_{\psi,j \{i_\ell\}_2}  [{\by_2,\{\bx_{\ell}\}_2};\mathrm{j}]\Bigg\} \, ,
\label{eq:A1}
\end{align}
with the indices $i_\ell \in \{1,2\}$ standing for the source  $J$ or $\bar J$,  as in the main text,
 and $(\{i_\ell\}_1, \{i_\ell\}_2)$ indicating all the possible bipartitions of the $n$ indices $\{i_\ell\}_{1 \leq \ell \leq n}$, and $(\{\bx_{\ell}\}_1,\{\bx_{\ell}\}_2)$ the corresponding bipartition in coordinates. 
Finally, $\#1$ and $\#2$ are the cardinals of $\{i_\ell\}_1$ (resp. $\{i_\ell\}_2$).
Focusing  on the first line of \eq{eq:A1}, one can write
\begin{align}
 &\int_{\by_1,\by_2}\! \p_\kappa [\mathcal{ R}_\kappa]_{ij} ({\by_1-\by_2}) \, G^{(n+2)}_{\psi,ij i_1\dots i_{n}}[\by_1,\by_2,\{\bx_\ell\};\mathrm{j}] = \int_{\by_1,\by_2}\! \p_\kappa [\mathcal{ R}_\kappa]_{ij}({\by_1-\by_2}) \nonumber\\
 & \times \Bigg[\int_{\bz_1,\bz_2} G^{(2)}_{\psi,k i}[\bz_1,\by_1;\mathrm{j}]  G^{(2)}_{\psi,\ell j}[\bz_2,\by_2;\mathrm{j}]
  \frac{\delta^2}{\delta \varphi_k(\bz_1)\delta\varphi_\ell(\bz_2)} \nonumber\\
  &+ \int_{\bz} G^{(3)}_{\psi,\ell ij}[\bz,\by_1,\by_2;\mathrm{j}]  \frac{\delta}{ \delta\varphi_\ell(\bz)} \Bigg]
 G^{(n)}_{\psi, i_1\dots i_{n}}[\{\bx_\ell\};\mathrm{j}] \, .
\label{eq:A2}
\end{align}
The derivatives of $G^{(n)}$ with respect to $\varphi$ must be understood as acting on  $G^{(n)}$ expressed as a diagram constructed from $\Gamma_{\kappa}$ vertices. More precisely, $G^{(n)}$ is the sum of all tree diagrams with vertices the $\Gamma^{(k)},k\leq n$ and with edges the propagator $G^{(2)}$, the latter satisfying
\begin{equation}
 G^{(2)}_{\psi, k\ell}[\bx,\by;\mathrm{j}] = \frac{\delta \varphi_k(\bx)}{\delta \mathrm{j}_\ell(\by)} = \left(\frac{\delta \mathrm{j}}{\delta \varphi}\right)^{-1}_{k\ell}(\bx,\by) = [\Gamma^{(2)} + {\cal R_{\kappa}}]^{-1}_{k\ell}[\bx,\by;\varphi]\, .
 \label{eq:g2gam2}
\end{equation}
Furthermore, introducing the differential operator
\begin{equation}
 \tilde \p_\kappa \equiv \p_\kappa R_\kappa \frac{\delta}{\delta R_\kappa} + \p_\kappa N_\kappa \frac{\delta}{\delta N_\kappa} \, ,
\end{equation}
and using the expression \eq{eq:g2gam2}, one has
 \begin{equation}
 \tilde \p_\kappa G^{(2)}_{\psi,k \ell}[\bz_1,\bz_2;\mathrm{j}] = -\int_{\by_1,\by_2} 
  \p_\kappa [\mathcal{R}_\kappa]_{ij}({\by_1-\by_2})  G^{(2)}_{\psi,k i}[\bz_1,\by_1;\mathrm{j}]  G^{(2)}_{\psi,\ell j}[\bz_2,\by_2;\mathrm{j}] \, ,
\label{eq:flowdtildeC}
\end{equation}
which appears in the first term of the \rhs of \eq{eq:A2}.
The second term  in the \rhs of \eq{eq:A2} vanishes when the sources are set to zero, since it is proportional to the flow of the average stream function or the average response stream function.
Indeed, the functions $G^{(1)}_{\psi,i}(\bx)$ are the expectation values of the stream function and response stream function. The expression of their flow can be deduced by 
taking one derivative of \eq{eq:dkw} with respect to a source and setting the sources to zero, which yields
\begin{equation}
 \p_\kappa G^{(1)}_{\psi,\ell}(\bz) = - \frac 1 2 \int_{\by_1,\by_2}  \p_\kappa [{\cal R}_\kappa]_{ij}({\by_1-\by_2})
 G^{(3)}_{\psi,\ell ij}(\bz,\by_1,\by_2)% - G^{(2)}_{\ell i}(\bz,\by_1)G^{(1)}_{j}(\by_2) - G^{(2)}_{\ell j}(\bz,\by_2)G^{(1)}_{i}(\by_1) \Big\}
\, ,
\end{equation}
omitting additional contribution proportional to  $G^{(1)}$. As the average fields $G^{(1)}_{\psi,\ell}(\bz)$ are constant in space coordinates when the sources are set to zero, one can work in the comoving frame
where they are identically zero, and so is their flow $ \p_\kappa G^{(1)}_{\psi,\ell}(\bz)$. By identification,
 one concludes that the second term in the \rhs of \Eq{eq:A2} vanishes when evaluated at zero fields.
Gathering  the previous expressions and setting the fields to zero, the flow equation for $G^{(n)}$ can be rewritten as
\begin{align}
 \p_\kappa G^{(n)}_{\psi,i_1\dots i_{n}}(\{\bx_\ell\}_{1 \leq \ell \leq n}) &= \frac{1}{2}\, \int_{\by_1,\by_2}\! \Bigg\{ 
\tilde \p_\kappa G^{(2)}_{\psi,k l}(\by_1,\by_2) \Big[\frac{\delta^2}{\delta \varphi_k(\by_1)\delta\varphi_\ell(\by_2)} 
G^{(n)}_{\psi,i_1\dots i_{n}}[\{\bx_\ell\}_{1 \leq \ell \leq n};\mathrm{j}]\Big]_{\varphi=0}\nonumber\\
 & - \sum_{(\{i_\ell\}_1, \{i_\ell\}_2) \atop \#1 + \#2=n} G^{(\#1+1)}_{\psi,i \{i_\ell\}_1}({\by_1,\{\bx_{\ell}\}_1})\,\p_\kappa [\mathcal{R}_\kappa]_{ij}({\by_1-\by_2}) \, G^{(\#2+1)}_{\psi,j \{i_\ell\}_2}({\by_2,\{\bx_{\ell}\}_2}) \Bigg\} \, .
\end{align}
This yields in the Fourier space
\begin{align}
 \p_\kappa \tilde G^{(n)}_{\psi,i_1\dots i_{n}}(\{\bp_\ell\}_{1 \leq \ell \leq n}) &=\frac{1}{2}\,  \int_{\bq_1,\bq_2}\! \Bigg\{
\tilde \p_\kappa \tilde G^{(2)}_{\psi,k l}(-\bq_1,-\bq_2) \Big[\frac{\delta^2}{\delta \varphi_k(\bq_1)\delta\varphi_\ell(\bq_2)}
  \tilde G^{(n)}_{\psi,i_1\dots i_{n}}[\{\bp_\ell\};\mathrm{j}]\Big]_{\varphi=0}\nonumber\\
 &  - \sum_{(\{i_\ell\}_1, \{i_\ell\}_2) \atop \#1 + \#2=n} \tilde G^{(\#1+1)}_{\psi,i \{i_\ell\}_1}({\bq_1,\{\bp_{\ell}\}_1}) \p_\kappa [\mathcal{ R}_\kappa]_{ij}(-\bq_1,-\bq_2) \tilde G^{(\#2+1)}_{\psi,j \{i_\ell\}_2}({\bq_2,\{\bp_{\ell}\}_2}) \Bigg\} \, ,
\label{eq:A3}
\end{align}
where in the first line the Fourier transform is meant after the functional derivatives
\begin{equation}
 \Big[\frac{\delta^2}{\delta \varphi_k(\bq_1)\delta\varphi_\ell(\bq_2)}
  \tilde G^{(n)}_{\psi,i_1\dots i_{n}}[\{\bp_{\ell}\};\mathrm{j}]\Big]_{\varphi=0} 
  \equiv \text{FT}\Bigg(\frac{\delta^2}{\delta \varphi_k(\bz_1)\delta\varphi_\ell(\bz_2)} G^{(n)}_{\psi,i_1\dots i_{n}}[\{\bx_{\ell}\};\mathrm{j}]\Big|_{\varphi=0}\Bigg)(\bq_1,\bq_2,\{\bp_{\ell}\}) \,
 \label{eq:symbolic}
\end{equation}
with $\text{FT}(\dots)$ denoting the Fourier transform.

We focus on the flow equation \eq{eq:A3}, and now consider the limit of large wave-numbers, which we define as all external wave-numbers  being large compared to the \RG scale $|\vp_\ell| \gg \kappa$ for $1 \leq \ell \leq n$, as well as all possible  partial sums being large  $\big|\sum_{\ell \in I}\vp_{\ell}| \gg \kappa$, for $I$ a subset of $\{1,\dots n\}$, which means that we exclude  exceptional configurations where a partial sum vanishes. The following proof relies on the presence of the derivative of the regulator term $\p_\kappa [\mathcal{ R}_\kappa]$ in the flow equation \eq{eq:A3}. The key properties of this term are that, on the one hand, it rapidly tends to zero for wave-numbers greater that the \RG scale, and on the other hand, it ensures the analyticity of all vertex functions at any finite $\kappa$. 
Let us examine the second terms of the \rhs of \eq{eq:A3} in this limit. Using invariance under space-time translation, it can be rewritten as
\begin{align}
 & \int_{\bq_1,\bq_2} \sum_{(\{i_\ell\}_1, \{i_\ell\}_2) \atop \#1 + \#2=n} G^{(\#1+1)}_{\psi,i \{i_\ell\}_1}({\bq_1,\{\bp_{\ell}\}_1}) \p_\kappa [\mathcal{ R}_\kappa]_{ij}(-\bq_1,-\bq_2) G^{(\#2+1)}_{\psi,j \{i_\ell\}_2}({\bq_2,,\{\bp_{\ell}\}_2}) \nonumber\\
&= (2\pi)^{3}\delta(\sum_{k=1}^n \omega_k) \delta^2(\sum_{k=1}^n \vp_k)  \int_{\bq} \sum_{(\{i_\ell\}_1, \{i_\ell\}_2) \atop \#1 + \#2=n} \bar G^{(\#1+1)}_{\psi,\{i_\ell\}_1 i}({\{\bp_{\ell}\}_1}) \p_\kappa [\mathcal{ R}_\kappa]_{ij}(\sum \{\vp_{k}\}_1) 
\bar G^{(\#2+1)}_{\psi,j \{i_\ell\}_2}({\{\bp_{\ell}\}_2})\, ,
\end{align}
where $\sum \{\vp_{k}\}_1$ is the sum of all the wave-numbers in $\{\vp_{k}\}_1$. Thus this term
 is proportional to the derivative of the regulator evaluated at a sum of external wave-numbers which is large $\sum \{\vp_{k}\}_1 \gg \kappa$. Hence it is suppressed at least exponentially in the limit of large wave-numbers and can be neglected safely.
Finally, only the first term of \eq{eq:A3} survives in this limit and one obtains \Eq{eq:flowGnstream} of the main text.

\subsection{Flow equation at leading order at unequal times}
\label{app:flowD}

Let us now calculate the leading order term (at unequal times) of the flow equation of a generic correlation function $G^{(n)}$ in the large wave-number expansion, and show that the result obtained in the velocity formulation \cite{Tarpin18} is recovered. This calculation is again analogous to the one in the velocity formulation, but we
 propose here a slightly more condensed derivation.
 As we work exclusively in the stream formulation in this appendix, we drop the $\psi$ index on the correlation function $G^{(n)}_{\psi,\{i_\ell\}}$. 
 The $\psi$ index is used instead to explicitly indicate a $i_\ell=1$ index, \ie a $\psi$ leg. 
We first perform the explicit calculation for the two-point function, and then establish the general expression  for a $n$-point function.

\subsubsection{Two-point function}

 As explained in the main text, the first term of the large wave-number expansion is zero and the leading order term involves two $q$ derivatives and reads for the two-point function
\begin{align}
 \p_\kappa & \tilde  G^{(2)}_{\psi \psi}({\bp_1},{\bp_{2}})\Big|_\mathit{leading}  =  \frac{1}{2}\,  \int_{\bq_1,\bq_2}\!
 \tilde \p_\kappa \tilde  G^{(2)}_{ij}(-\bq_1,-\bq_2) \nonumber\\
 &\times \frac{q_a^\alpha q_b^\beta}{2} \frac{\p^2}{\p q_a^\alpha \p q_b^\beta} \int_{\bk_1, \bk_2} \tilde G_{\psi m}^{(2)}(\bp_1,-\bk_1) \tilde G_{\psi n}^{(2)}(\bp_2,-\bk_2)
 \Big[ -\tilde \Gamma_{ij mn}^{(4)}(\bq_1, \bq_2, \bk_1, \bk_2) \nonumber\\
 & + \int_{\bk_3,\bk_4} \tilde \Gamma_{i m s}^{(3)}(\bq_1, \bk_1, \bk_3) \tilde G_{st}^{(2)}(-\bk_3,-\bk_4)\tilde \Gamma_{j n t}^{(3)}(\bq_2, \bk_2, \bk_4) + (i, \bq_1) \leftrightarrow (j, \bq_2) \Big]_{\vq_1=\vq_2 = 0} \nonumber\\
\label{eq:leadflowstreamG2}
\end{align}
where the double arrow means the permutation of the preceeding term. 
We focus on the stream-stream correlation, but the following derivation holds for generic indices.
The derivatives acting on $ \tilde \Gamma^{(4)}$ can be expressed as
\begin{align}
 \frac{q_a^\alpha q_b^\beta}{2} &\frac{\p^2}{\p q_a^\alpha \p q_b^\beta} \tilde \Gamma_{ij mn}^{(4)}(\bq_1, \bq_2, \bk_1, \bk_2)\Big|_{\vq_1=\vq_2=0} \nonumber\\
 &= \frac{q_1^\alpha q_1^\beta}{2} \frac{\p^2}{\p q_1^\alpha \p q_1^\beta}\tilde \Gamma_{ij mn}^{(4)}(\bq_1, \omega_2, \vec 0, \bk_1, \bk_2)\Big|_{\vq_1=0}+ q_1^\alpha q_2^\beta \frac{\p^2}{\p q_1^\alpha \p q_2^\beta}\tilde \Gamma_{ij mn}^{(4)}(\bq_1, \bq_2, \bk_1, \bk_2)\Big|_{\vq_1=\vq_2=0} \nonumber\\
 &\quad+ \frac{q_2^\alpha q_2^\beta}{2} \frac{\p^2}{\p q_2^\alpha \p q_2^\beta}\tilde \Gamma_{ij mn}^{(4)}(\omega_1, \vec 0, \bq_2, \bk_1, \bk_2)\Big|_{\vq_2=0}\nonumber\\
 &= q_1^\alpha q_2^\beta \frac{\p^2}{\p q_1^\alpha \p q_2^\beta}\tilde \Gamma_{ij mn}^{(4)}(\bq_1, \bq_2, \bk_1, \bk_2)\Big|_{\vq_1=\vq_2=0} \nonumber\\
 &= \delta_{\psi k} \delta_{\psi \ell}\, i^2 q_1^\alpha q_2^\beta\epsilon_{\alpha \mu} \epsilon_{\beta \nu}\tilde{\mathcal{D}}_\mu(\omega_1) \tilde{\mathcal{D}}_\nu(\omega_2) \tilde \Gamma_{mn}^{(2)}(\bk_1, \bk_2)\,,
\end{align}
where  in the second equality, \Eq{eq:listwardstreamtilde} $a)$ --which states that the vertex functions are proportional to the product of their 
wave-numbers-- is used, and in the third equality, \Eq{eq:listwardstreamtilde}  $b)$ and $d)$ are used.
 Thus,  among the different 
combinations of $\vq_1$ and $\vq_2$, only one survives due to the space-independent gauge invariance. 
 
The derivatives acting on the \PR term read
\begin{align}
 \frac{q_a^\alpha q_b^\beta}{2} &\frac{\p^2}{\p q_a^\alpha \p q_b^\beta}\Big[ \tilde \Gamma_{k m s}^{(3)}(\bq_1, \bk_1, \bk_3) \tilde G_{st}^{(2)}(-\bk_3,-\bk_4)\tilde \Gamma_{\ell n t}^{(3)}(\bq_2, \bk_2, \bk_4)\Big]_{\vq_1=\vq_2=0} \nonumber\\
 &= \frac{q_1^\alpha q_1^\beta}{2} \frac{\p^2}{\p q_1^\alpha \p q_1^\beta}\tilde \Gamma_{k m s}^{(3)}(\bq_1, \bk_1, \bk_3)\Big|_{\vq_1=0} \tilde G_{st}^{(2)}(-\bk_3,-\bk_4) \tilde \Gamma_{\ell n t}^{(3)}(\omega_2, \vec 0, \bk_2, \bk_4)\nonumber\\
&\quad+ q_1^\alpha q_2^\beta \frac{\p}{\p q_1^\alpha} \tilde \Gamma_{k m s}^{(3)}(\bq_1, \bk_1, \bk_3) \Big|_{\vq_1=0} \tilde G_{st}^{(2)}(-\bk_3,-\bk_4)  \frac{\p}{ \p q_2^\beta} \tilde \Gamma_{\ell n t}^{(3)}(\bq_2, \bk_2, \bk_4)\Big|_{\vq_2=0} \nonumber\\
 &\quad+ \frac{q_2^\alpha q_2^\beta}{2} \tilde \Gamma_{k m s}^{(3)}(\omega_1, \vec 0, \bk_1, \bk_3) \tilde G_{st}^{(2)}(-\bk_3,-\bk_4) \frac{\p^2}{\p q_2^\alpha \p q_2^\beta} \tilde \Gamma_{\ell n t}^{(3)}(\bq_2, \bk_2, \bk_4)\Big|_{\vq_2=0}\nonumber\\
 &= q_1^\alpha q_2^\beta \frac{\p}{\p q_1^\alpha} \tilde \Gamma_{k m s}^{(3)}(\bq_1, \bk_1, \bk_3) \Big|_{\vq_1=0} \tilde G_{st}^{(2)}(-\bk_3,-\bk_4)  \frac{\p}{ \p q_2^\beta} \tilde \Gamma_{\ell n t}^{(3)}(\bq_2, \bk_2, \bk_4)\Big|_{\vq_2=0} \nonumber\\
 &= \delta_{\psi k} \delta_{\psi \ell}\, i^2 q_1^\alpha q_2^\beta\epsilon_{\alpha \mu} \epsilon_{\beta \nu} 
\tilde{\mathcal{D}}_\mu(\omega_1)\tilde \Gamma_{m s}^{(2)}(\bk_1, \bk_3)  \tilde G_{st}^{(2)}(-\bk_3,-\bk_4) \tilde{\mathcal{D}}_\nu(\omega_2)  \tilde \Gamma_{n t}^{(2)}(\bk_2, \bk_4)\,.
\end{align}
Inserting both expressions in \eq{eq:leadflowstreamG2} leads to 
\begin{align}
 \p_\kappa & \tilde  G^{(2)}_{\psi \psi}({\bp_1},{\bp_{2}})\Big|_\mathit{leading}  =  \frac{1}{2}\,  \int_{\bq_1,\bq_2}\!
 \tilde \p_\kappa \tilde  G^{(2)}_{\psi\psi}(-\bq_1,-\bq_2) \, i^2 q_1^\alpha q_2^\beta\epsilon_{\alpha \mu} \epsilon_{\beta \nu}\nonumber\\
 &\int_{\bk_1, \bk_2} \tilde G_{\psi m}^{(2)}(\bp_1,-\bk_1)\tilde G_{\psi n}^{(2)}(\bp_2,-\bk_2)
 \Big[ -\tilde{\mathcal{D}}_\mu(\varpi_1)\tilde{\mathcal{D}}_\nu(\varpi_2) \tilde \Gamma_{mn}^{(2)}(\bk_1, \bk_2) \nonumber\\
 & + \int_{\bk_3,\bk_4}\tilde{\mathcal{D}}_\mu(\varpi_1)\tilde \Gamma_{m s}^{(2)}(\bk_1, \bk_3)  \tilde G_{st}^{(2)}(-\bk_3,-\bk_4)  \tilde{\mathcal{D}}_\nu(\varpi_2)  \tilde \Gamma_{n t}^{(2)}(\bk_2, \bk_4) +  (\mu, \omega_1) \leftrightarrow (\nu, \omega_2) \Big]\,.
\end{align}
To further simplify  the second term, on can insert the following relation
\begin{equation}
\tilde G_{st}^{(2)}(-\bk_3,-\bk_4) = \int_{\bk_5,\bk_6} \tilde G_{su}^{(2)}(-\bk_3,-\bk_5) \tilde \Gamma_{uv}^{(2)}(\bk_5,\bk_6) \tilde G_{vt}^{(2)}(-\bk_6,-\bk_4) \,,
\end{equation}
such that each $\tilde{\mathcal{D}}(\varpi)\tilde \Gamma^{(2)}$ is enclosed between two $\tilde G^{(2)}$. This combination can  then be rewritten, making explicit the operator $\tilde{\mathcal{D}}$, as
\begin{align}
   \int_{\bk_1,\bk_3} &\tilde G_{\psi m}^{(2)}(\bp_1,-\bk_1)  \tilde{\mathcal{D}}_\mu(\varpi_1)\tilde \Gamma_{m s}^{(2)}(\bk_1, \bk_3) \tilde G_{su}^{(2)}(-\bk_3,-\bk_5)   \nonumber\\
 &= - \int_{\bk_1,\bk_3} \tilde G_{\psi m}^{(2)}(\bp_1,-\bk_1) \Big[\frac{k_1^{\mu}}{\varpi_1}\tilde\Gamma^{(2)}_{ ms } (\nu_1+\varpi_1,\vk_1,\bk_3) + \frac{k_3^{\mu}}{\varpi_1}\tilde\Gamma^{(2)}_{ ms } (\bk_1,\nu_3+\varpi_1,\vk_3) \Big]\tilde G_{su}^{(2)}(-\bk_3,-\bk_5)  \nonumber\\
 &= - \int_{\bk_1}  \frac{k_1^{\mu}}{\varpi} \tilde G_{\psi m}^{(2)}(\bp_1,-\bk_1) \delta_{m u}\delta(\varpi_1+\nu_1-\nu_5)\delta^2(\vk_1-\vk_5) \nonumber\\
&\quad+  \int_{\bk_3} \frac{k_3^{\mu}}{\varpi}  \delta_{\psi s} \delta(\omega_1+\varpi+\nu_3)\delta^2(\vp_1+\vk_3) \tilde G_{su}^{(2)}(-\bk_3,-\bk_5)   \nonumber\\
 &=-\frac{k_5^{\mu}}{\varpi_1}\tilde G^{(2)}_{\psi u}(\bp_1,-\nu_5+\varpi_1,-\vk_5) + \frac{p_1^{\mu}}{\varpi_1}\tilde G^{(2)}_{\psi u}(\omega_1+\varpi_1, \vp_1,-\vk_5) \nonumber\\
 &= - \tilde{\mathcal{ D}}_\mu(\varpi_1)  \tilde G^{(2)}_{\psi u}(\bp_1,-\bk_5)\, ,
 \end{align} 
where the second equality is simplified using that $\tilde G^{(2)}$ and $\tilde \Gamma^{(2)}$ are  the inverse  of each other
\begin{equation}
\int_{\bp_2}\tilde G_{ij}(\bp_1,-\bp_2) \tilde \Gamma_{jk}(\bp_2, -\bp_3) = \delta_{ik}\delta(\nu_1-\nu_3) \delta^d(\vp_1-\vp_3)\,.
\end{equation}

 In fact, this calculation proves the general property 
\begin{align}
\frac{\p}{\p q_\alpha}\df{}{\varphi_k(\bq)}& \tilde G_{ij}^{(2)}[\bp_1,\bp_2;\mathrm{j}]_{\varphi=0}\Big|_{\vq=0}\nonumber\\
&=  - \int_{\bk_1,\bk_2} \tilde G_{i m}^{(2)}(\bp_1,-\bk_1) \frac{\p}{\p q_\alpha} \tilde \Gamma_{k m n}^{3)}(\bq, \bk_1, \bk_2)\Big|_{\vq=0} \tilde G_{n j}^{(2)}(-\bk_2, \bp_2) \nonumber\\
&=  - \int_{\bk_1,\bk_2} \tilde G_{i m}^{(2)}(\bp_1,-\bk_1)   i \epsilon_{\alpha \mu}\tilde{\mathcal{D}}_\mu(\varpi)\tilde \Gamma_{m n}^{(2)}(\bk_1, \bk_2) \tilde G_{n j}^{(2)}(-\bk_2, \bp_2) \nonumber\\
&= i \epsilon_{\alpha\mu} \tilde{\mathcal{ D}}_\mu(\varpi) \tilde G_{ij}^{(2)}(\bp_1,\bp_2)\, ,
\label{eq:DG2}
\end{align}
which will be useful in the general case. For the flow of the two point function, this allows one to write -- relabelling the integration and summation variables for convenience--
\begin{align}
 \p_\kappa & \tilde  G^{(2)}_{\psi \psi}({\bp_1},{\bp_{2}})\Big|_\mathit{leading}  =  \frac{1}{2}\,  \int_{\bq_1,\bq_2}\!
 \tilde \p_\kappa \tilde  G^{(2)}_{\psi\psi}(-\bq_1,-\bq_2) \, i^2 q_1^\alpha q_2^\beta\epsilon_{\alpha \mu} \epsilon_{\beta \nu}\nonumber\\
 &\int_{\bk_1, \bk_2} 
 \Big[ - \tilde G_{\psi m}^{(2)}(\bp_1,-\bk_1)\tilde G_{\psi n}^{(2)}(\bp_2,-\bk_2) \tilde{\mathcal{D}}_\mu(\varpi_1)\tilde{\mathcal{D}}_\nu(\varpi_2) \tilde \Gamma_{mn}^{(2)}(\bk_1, \bk_2) \nonumber\\
 & + \tilde{\mathcal{D}}_\mu(\varpi_1) \tilde G^{(2)}_{\psi m}(\bp_1,-\bk_1)  \Gamma_{mn}^{(2)}(\bk_1,\bk_2)  \tilde{\mathcal{D}}_\nu(\varpi_2)  \tilde G^{(2)}_{\psi n}(\bp_2,-\bk_2) +  (\mu, \omega_1) \leftrightarrow (\nu, \omega_2) \Big]\,.
\end{align}
This expression can be further simplified by making explicit the $\tilde{\mathcal{D}}$ operator, since some terms cancel out between the two terms in brackets. Indeed, the first one reads
\begin{align}
& \int_{\bk_1, \bk_2} \tilde G_{\psi m}^{(2)}(\bp_1,-\bk_1) \tilde G_{\psi n}^{(2)}(\bp_2,-\bk_2) \tilde{\mathcal{D}}_\mu(\varpi_1)\tilde{\mathcal{D}}_\nu(\varpi_2) \tilde \Gamma_{mn}^{(2)}(\bk_1, \bk_2) \nonumber\\
&= \int_{\bk_1, \bk_2} \tilde G_{\psi m}^{(2)}(\bp_1,-\bk_1) \tilde G_{\psi n}^{(2)}(\bp_2,-\bk_2)  \Big[ \frac{k_1^\mu k_1^\nu}{\varpi_1\varpi_2} \tilde \Gamma_{mn}^{(2)}(\nu_1+\varpi_1+\varpi_2,\vk_1, \bk_2) \nonumber\\
&+ \frac{k_1^\mu k_2^\nu}{\varpi_1\varpi_2} \tilde \Gamma_{mn}^{(2)}(\nu_1+\varpi_1,\vk_1, \nu_2+\varpi_2,\vk_2) + (\mu,\varpi_1) \leftrightarrow (\nu,\varpi_2)+ \frac{k_2^\mu k_2^\nu}{\varpi_1\varpi_2} \tilde \Gamma_{mn}^{(2)}(\bk_1,\nu_2+\varpi_1+\varpi_2,\vk_2) \big]\nonumber\\
&= \int_{\bk_1} \frac{k_1^\mu k_1^\nu}{\varpi_1\varpi_2} \tilde G_{\psi m}^{(2)}(\bp_1,-\bk_1) \delta_{\psi m} \delta(\omega_2+\nu_1+\varpi_1+\varpi_2)\delta^2(\vp_2+\vk_1) \nonumber\\
&+  \int_{\bk_2} \frac{k_2^\mu k_2^\nu}{\varpi_1\varpi_2} \tilde G_{\psi n}^{(2)}(\bp_2,-\bk_2) \delta_{\psi n} \delta(\omega_1+\nu_2+\varpi_1+\varpi_2)\delta^2(\vp_1+\vk_2) \nonumber\\
&+ \int_{\bk_1, \bk_2} \tilde G_{\psi m}^{(2)}(\bp_1,-\bk_1) \tilde G_{\psi n}^{(2)}(\bp_2,-\bk_2)\frac{k_1^\mu k_2^\nu}{\varpi_1\varpi_2} \tilde \Gamma_{mn}^{(2)}(\nu_1+\varpi_1,\vk_1, \nu_2+\varpi_2,\vk_2) + (\mu,\varpi_1) \leftrightarrow (\nu,\varpi_2)\nonumber\\
&=\frac{p_2^\mu p_2^\nu}{\varpi_1\varpi_2} \tilde G_{\psi \psi}^{(2)}(\bp_1,\omega_2+\nu_1+\varpi_1+\varpi_2,\vk_2) + \frac{p_1^\mu p_1^\nu}{\varpi_1\varpi_2} \tilde G_{\psi \psi}^{(2)}(\omega_1+\nu_1+\varpi_1+\varpi_2,\vk_1,\bp_2)\nonumber\\
&+ \int_{\bk_1, \bk_2} \tilde G_{\psi m}^{(2)}(\bp_1,-\bk_1) \tilde G_{\psi n}^{(2)}(\bp_2,-\bk_2)\frac{k_1^\mu k_2^\nu}{\varpi_1\varpi_2} \tilde \Gamma_{mn}^{(2)}(\nu_1+\varpi_1,\vk_1, \nu_2+\varpi_2,\vk_2) + (\mu,\varpi_1) \leftrightarrow (\nu,\varpi_2)\, ,
\end{align}
 and the second term is
\begin{align}
&\int_{\bk_1, \bk_2} \tilde{\mathcal{D}}_\mu(\varpi_1) \tilde G^{(2)}_{\psi m}(\bp_1,-\bk_1)  \Gamma_{mn}^{(2)}(\bk_1,\bk_2)  \tilde{\mathcal{D}}_\nu(\varpi_2)  \tilde G^{(2)}_{\psi n}(\bp_2,-\bk_2)\nonumber\\
&= \int_{\bk_1, \bk_2} \Big[ \frac{p_1^\mu}{\varpi_1}  \tilde G^{(2)}_{\psi m}(\omega_1+\varpi_1,\vp_1,-\bk_1) - \frac{k_1^\mu}{\varpi_1}  \tilde G^{(2)}_{\psi m}(\bp_1,-\nu_1+\varpi_1,-\vk_1)\Big]  \nonumber\\
&\times \Gamma_{mn}^{(2)}(\bk_1,\bk_2) \Big[ \frac{p_2^\nu}{\varpi_2}  \tilde G^{(2)}_{\psi n}(\omega_2+\varpi_2,\vp_2,-\bk_2) - \frac{k_2^\nu}{\varpi_2}  \tilde G^{(2)}_{\psi n}(\bp_2,-\nu_2+\varpi_2,-\vk_2)\Big]\nonumber\\
&= \frac{p_1^\mu p_2^\nu}{\varpi_1 \varpi_2} \tilde G^{(2)}_{\psi \psi}(\omega_1+\varpi_1,\vp_1,\omega_2+\varpi_2,\vp_2) \nonumber\\
&+\frac{p_2^\mu p_2^\nu}{\varpi_1\varpi_2} \tilde G_{\psi \psi}^{(2)}(\bp_1,\omega_2+\nu_1+\varpi_1+\varpi_2,\vk_2) + \frac{p_1^\mu p_1^\nu}{\varpi_1\varpi_2} \tilde G_{\psi \psi}^{(2)}(\omega_1+\nu_1+\varpi_1+\varpi_2,\vk_1,\bp_2)\nonumber\\
&+ \int_{\bk_1, \bk_2} \tilde G_{\psi m}^{(2)}(\bp_1,-\nu_1+\varpi_1,-\vk_1) \tilde G_{\psi n}^{(2)}(\bp_2,-\nu_2+\varpi_2,-\vk_2)\frac{k_1^\mu k_2^\nu}{\varpi_1\varpi_2} \tilde \Gamma_{mn}^{(2)}(\bk_1, \bk_2) \,.
\end{align}
The last lines cancel out by shifting the frequencies $\nu_1$ and $\nu_2$ and one is left with
\begin{align}
 \p_\kappa & \tilde  G^{(2)}_{\psi \psi}({\bp_1},{\bp_{2}})\Big|_\mathit{leading}  =  \frac{1}{2}\,  \int_{\bq_1,\bq_2}\!
 \tilde \p_\kappa \tilde  G^{(2)}_{\psi\psi}(-\bq_1,-\bq_2) \, i^2 q_1^\alpha q_2^\beta\epsilon_{\alpha \mu} \epsilon_{\beta \nu}\nonumber\\
&\times \Big[ \frac{p_2^\mu p_2^\nu}{\varpi_1\varpi_2} \tilde G_{\psi \psi}^{(2)}(\bp_1,\omega_2+\nu_1+\varpi_1+\varpi_2,\vk_2) + \frac{p_1^\mu p_1^\nu}{\varpi_1\varpi_2} \tilde G_{\psi \psi}^{(2)}(\omega_1+\nu_1+\varpi_1+\varpi_2,\vk_1,\bp_2)\nonumber\\
&+\frac{p_2^\mu p_2^\nu}{\varpi_1\varpi_2} \tilde G_{\psi \psi}^{(2)}(\bp_1,\omega_2+\nu_1+\varpi_1+\varpi_2,\vk_2) + \frac{p_1^\mu p_1^\nu}{\varpi_1\varpi_2} \tilde G_{\psi \psi}^{(2)}(\omega_1+\nu_1+\varpi_1+\varpi_2,\vk_1,\bp_2) \Big]\nonumber\\
&=\frac{1}{2}\,  \int_{\bq_1,\bq_2}\!
 \tilde \p_\kappa \tilde  G^{(2)}_{v_\mu v_\nu}(-\bq_1,-\bq_2)\mathcal{D}_\mu(\omega_1) \mathcal{D}_\nu(\omega_2)
 \tilde G^{(2)}_{\psi\psi}({\bp_1}, {\bp_{2}})\,,
\end{align}
where the definition of the stream function in Fourier space: ${\tilde v_\mu(\bq) = i \epsilon_{\mu\alpha}q_\alpha}$ has been used in the last equality. This concludes the proof, and shows that  the leading order result obtained in the velocity formulation is recovered in the stream function one for the two-point function.

\subsubsection{$n$-point function}
\label{app:flowDGen}

Let us now consider a generic $n$-point correlation function.  The leading order term of its flow equation reads
\begin{align}
 \p_\kappa \tilde G^{(n)}_{i_1\dots i_{n}}({\bp_1}, \dots, {\bp_{n}})\Big|_\mathit{leading}  &=  \frac{1}{2}\,  \int_{\bq_1,\bq_2}\!
 \tilde \p_\kappa \tilde  G^{(2)}_{ij}(-\bq_1,-\bq_2) \nonumber\\
 &\times \frac{q_a^\alpha q_b^\beta}{2} \frac{\p^2}{\p q_a^\alpha \p q_b^\beta} \left[ \frac{\delta^2}{\delta \varphi_i(\bq_1)\delta\varphi_j(\bq_2)} 
 \tilde G^{(n)}_{i_1\dots i_{n}}[{\bp_1}, \dots, {\bp_{n}};\mathrm{j}] \right]_{\varphi=0}\Bigg|_{\vq_1=\vq_2 = 0} \,.
\label{eq:flowNuneq}
\end{align}
The generalized correlation functions $ \tilde G^{(n)}_{i_1\dots i_{n}}(\{\bp_\ell\})$ can be expressed as the sum over all trees whose edges are the propagators $\tilde G^{(2)}$, whose vertices are the
vertex functions $\tilde \Gamma^{(k)}$ and with external legs carrying momenta (\ie wave-vectors and frequencies) and indices matching the indices of the correlation function: $\{(i_\ell,\bp_\ell)\}$. Symbolically,
\begin{align}
 \tilde G^{(n)}_{i_1 \dots i_ n} [\{\bp_\ell\}_{1 \leq \ell \leq n};\mathrm{j}] &= \sum_{\mathrm{trees}} \alpha_{\mathcal{T}} \tilde{\mathcal{ T}}^{(n)}_{i_1\cdots i_n}[\{\bp_\ell\}]\nonumber\\
 \tilde{\mathcal{ T}}^{(n)}_{i_1\cdots i_n}[\{\bp_\ell\}] &= \int_{\bk_i}\prod_{i=1}^m \mathcal{ E}_i^{\mathcal{T}}[\{\bp_\ell\}_i,\{\bk_\ell\}_i]\,,
\end{align}
where $\alpha_{\mathcal{T}}$ is a combinatorial factor, the $ \mathcal{ E}_i^{\mathcal{T}}$ are the vertex functions and propagators entering the composition of the tree $\tilde{\mathcal{ T}}$ and the integration is done
over all the internal momenta of the diagram. The $\{\bp_\ell\}_i$ which are not empty form a partition of the external momenta $\{\bp_\ell\}_{1 \leq \ell \leq n}$,
and the internal momenta $\{\bk_\ell\}_i$ are chosen such that when a propagator is attached to a vertex function, the sum of the momenta of the propagator and of the
vertex function at the link is zero. Finally, the internal indices of the theory -- here $ i_\ell \in \{ \psi, \bar \psi\}$ -- have been omitted on $ \mathcal{ E}_i^{\mathcal{T}}$ to alleviate notations but follow
 straightforwardly from the partition of momenta. The term in square bracket in the flow equation \eq{eq:flowNuneq} is a sum of tree diagrams where the two functional derivatives have been distributed
\begin{align}
 & \frac{\delta^2}{\delta \varphi_k(\bq_1) \delta \varphi_\ell(\bq_2)} \tilde{\mathcal{ T}}^{(n)}_{i_1\cdots i_n}[\{\bp_\ell\}]_{\varphi=0}\Big|_{\vq_1=\vq_2=0} \nonumber\\
 & = \Bigg\{\int_{\bk_\mathrm{ intern}}\sum_{i,j \atop i \neq j} \Big(\prod_{m\neq i, j} \mathcal{ E}_m^{\mathcal{T}}(\{\bp_\ell\}_m,\{\bk_\ell\}_m)\Big) \nonumber\\
 &\times \frac{\delta}{\delta \varphi_k(\bq_1)} \mathcal{ E}_i^{\mathcal{T}}[\{\bp_\ell\}_i,\{\bk_\ell\}_i]\Big|_{\varphi=0} \frac{\delta}{\delta \varphi_\ell(\bq_2)} \mathcal{ E}_j^{\mathcal{T}}[\{\bp_\ell\}_j,\{\bk_\ell\}_j]\Big|_{\varphi=0} \nonumber\\
 &+\int_{\bk_\mathrm{ intern}} \sum_{i} \Big(\prod_{j\neq i} \mathcal{ E}_j^{\mathcal{T}}(\{\bp_\ell\}_j,\{\bk_\ell\}_j)\Big) \frac{\delta^2}{\delta \varphi_k(\bq_1) \delta \varphi_\ell(\bq_2)} \mathcal{ E}_i^{\mathcal{T}}[\{\bp_\ell\}_i,\{\bk_\ell\}_i]\Big|_{\varphi=0}  \Bigg\}\Big|_{\vq_1=\vq_2=0}\, .
\end{align}
When acting with a field functional derivative, either the derivative hits on a vertex function, giving
\begin{equation}
\frac{\delta}{\delta \varphi_i(\bq_a)} \tilde \Gamma^{(k)}_{i_1 \dots i_k} [\{\bp_\ell\}_{1 \leq \ell \leq k}; \varphi]_{\varphi=0} =\tilde \Gamma^{(k+1)}_{i i_1 \dots i_k} (\bq_a,\{\bp_\ell\}_{1 \leq \ell \leq k})\,
\end{equation}
or the derivative hits on a propagator, giving
\begin{equation}
\frac{\delta}{\delta \varphi_i(\bq_a)} G^{(2)}_{mn}[\bp_1,\bp_2;\mathrm{j}]_{\varphi=0} = - \int_{\bk_1,\bk_2} G^{(2)}_{mu}(\bp_1, - \bk_1) \tilde \Gamma^{(3)}_{i u v}(\bq_a, \bk_1, \bk_2) G^{(2)}_{vn}(-\bk_2,\bp_2)\,,
\end{equation}
with $a \in \{1,2\}$. Distributing the two $\vq$-derivatives in the flow equation \eq{eq:flowNuneq}, the terms with $a=b$, which correspond to   two $\vq$-derivatives acting on the same leg of the diagram, are zero, 
because a vertex function with a wave-number set to zero vanishes. Thus one 
 is left with one derivative hitting on each leg:
\begin{align}
 \p_\kappa \tilde G^{(n)}_{i_1\dots i_{n}}(\{\bp_\ell\}_{1 \leq \ell \leq n})\Big|_\mathit{leading}  &=  \frac{1}{2}\,  \int_{\bq_1,\bq_2}\!
 \tilde \p_\kappa \tilde  G^{(2)}_{ij}(-\bq_1,-\bq_2) \nonumber\\
 &\times q_1^\alpha q_2^\beta \frac{\p^2}{\p q_1^\alpha \p q_2^\beta} \left[ \frac{\delta^2}{\delta \varphi_i(\bq_1)\delta\varphi_j(\bq_2)} 
 \tilde G^{(n)}_{i_1\dots i_{n}}[\{\bp_\ell\};\mathrm{j}] \right]_{\varphi=0}\Bigg|_{\vq_1=\vq_2 = 0} \,.
\end{align}

Next, we need to prove the following property
\begin{align}
 \frac{\p^2}{\p q_1^\alpha \p q_2^\beta} & \Big[ \frac{\delta^2}{\delta \varphi_i(\bq_1)\delta\varphi_j(\bq_2)} 
 \tilde G^{(n)}_{i_1\dots i_{n}}[\{\bp_\ell\}_{1 \leq \ell \leq n};\mathrm{j}] \Big]_{\varphi=0}\Bigg|_{\vq_1=\vq_2 = 0} \nonumber\\
&= \delta_{i\psi}\delta_{j\psi} i^2 \epsilon_{\alpha\mu} \epsilon_{\beta\nu} \tilde{\mathcal{D}}_\mu(\varpi_1) \tilde{\mathcal{D}}_\nu(\varpi_2) \tilde G^{(n)}_{i_1\dots i_{n}}(\{\bp_\ell\})\, .
\label{eq:ClosureUneq}
\end{align}
For this, let us examine the case of only one functional derivative and subsequent wave-number derivative applied to a generic tree $\tilde{\mathcal{T}}^{(n)}$, which reads
\begin{align}
\frac{\p}{\p q_a^\alpha} \frac{\delta}{\delta \varphi_i(\bq_a)} &\tilde{\mathcal{ T}}^{(n)}_{i_1\cdots i_n}[\{\bp_\ell\}]_{\varphi=0}\Big|_{\vq_a=0} \nonumber\\
&= \int_{\bk_\mathrm{ intern}}\sum_{i=1}^{m} \Bigg(\prod_{j=1 \atop j\neq i}^m \mathcal{ E}_j^{\mathcal{T}}(\{\bp_\ell\}_j,\{\bk_\ell\}_j)\Bigg) \frac{\p}{\p q_a^\alpha} \frac{\delta}{\delta \varphi_i(\bq_a)} \mathcal{ E}_i^{\mathcal{T}}[\{\bp_\ell\}_i,\{\bk_\ell\}_i]\,,
\end{align}
giving either
\begin{equation}
\frac{\p}{\p q_a^\alpha} \frac{\delta}{\delta \varphi_i(\bq_a)} \tilde \Gamma^{(k)}_{i_1 \dots i_k} [\{\bp_\ell\}_{1 \leq \ell \leq k}; \varphi]_{\varphi=0} \Big|_{\vq_a=0} = \delta_{i\psi} i \epsilon_{\alpha\mu}\tilde{\mathcal{D}}_\mu(\varpi_a) \tilde \Gamma^{(k)}_{i_1 \dots i_k} (\{\bp_\ell\})\, ,
\end{equation}
if $\mathcal{ E}_i^{\mathcal{T}}$ is a vertex function, or
\begin{equation}
\frac{\p}{\p q_a^\alpha} \frac{\delta}{\delta \varphi_i(\bq_a)} G^{(2)}_{mn}[\bp_1,\bp_2;\mathrm{j}]_{\varphi=0}\Big|_{\vq_a=0} =  \delta_{i\psi} i \epsilon_{\alpha\mu} \tilde{\mathcal{D}}_\mu(\varpi_a) \tilde G^{(2)}_{mn}(\bp_1,\bp_2)\,,
\end{equation}
if it is a propagator, using the property \eq{eq:DG2}. This shows that
\begin{align}
\frac{\p}{\p q_a^\alpha} \frac{\delta}{\delta \varphi_i(\bq_a)} &\tilde{\mathcal{ T}}^{(n)}_{i_1\cdots i_n}[\{\bp_\ell\}]_{\varphi=0}\Big|_{\vq_a=0} \nonumber\\
&=  \delta_{i\psi} i \epsilon_{\alpha\mu} \int_{\bk_\mathrm{ intern}}\sum_{i=1}^{m} \Bigg(\prod_{j=1 \atop j\neq i}^m \mathcal{ E}_j^{\mathcal{T}}(\{\bp_\ell\}_j,\{\bk_\ell\}_j)\Bigg) \tilde{\mathcal{D}}_\mu(\varpi_a) \mathcal{ E}_i^{\mathcal{T}}(\{\bp_\ell\}_i,\{\bk_\ell\}_i)\,,
\end{align}
which means that the operator $\tilde{\mathcal{D}}_\mu(\varpi_a)$ is distributed on the elements of the tree according to the Leibniz rule. 
 Furthermore, because of conservation of momenta, the \rhs is equivalent to  the operator $\tilde{\mathcal{D}}_\mu$ acting on the tree $\tilde{\mathcal{ T}}^{(n)}$, that is only on the external legs $\{\bp_\ell\}$, \ie
\begin{align}
 \tilde{\mathcal{D}}_\mu(\varpi_a) \int_{\bk} \mathcal{ E}_1^{\mathcal{T}}(\{\bk_\ell\}_1, -\bk) \mathcal{ E}_2^{\mathcal{T}}(\bk, \{\bk_\ell\}_2) &= \int_{\bk} \Big[ \tilde{\mathcal{D}}_\mu(\varpi_a) \mathcal{ E}_1^{\mathcal{T}}(\{\bk_\ell\}_1, -\bk) \mathcal{ E}_2^{\mathcal{T}}(\bk, \{\bk_\ell\}_2) \nonumber\\
 &+ \mathcal{ E}_1^{\mathcal{T}}(\{\bk_\ell\}_1, -\bk) \tilde{\mathcal{D}}_\mu(\varpi_a) \mathcal{ E}_2^{\mathcal{T}}(\bk, \{\bk_\ell\}_2)\Big]\,.
\label{eq:leibnizD}
 \end{align}
This relation can be simply established by expanding  the \rhs,
\begin{align}
 \int_{\bk} &\Big[ \tilde{\mathcal{D}}_\mu(\varpi_a) \mathcal{ E}_i^{\mathcal{T}}(\{\bk_\ell\}_{\ell \in I}, -\bk) \mathcal{ E}_j^{\mathcal{T}}(\bk, \{\bk_\ell\}_{\ell \in J}) + \mathcal{ E}_i^{\mathcal{T}}(\{\bk_\ell\}_{\ell \in I}, -\bk)\tilde{\mathcal{D}}_\mu(\varpi_a) \mathcal{ E}_j^{\mathcal{T}}(\bk, \{\bk_\ell\}_{\ell \in J})\Big]\nonumber\\
 &= - \int_{\bk} \Big[ \sum_{i \in I} \frac{k_i^\mu}{\varpi_a}  \mathcal{ E}_i^{\mathcal{T}}(\varpi_i + \omega_a, \vk_i, \{\bk_\ell\}_{\ell \in I\textbackslash i}, -\bk) \mathcal{ E}_j^{\mathcal{T}}(\bk, \{\bk_\ell\}_{\ell \in J}) \nonumber\\
 &+ \sum_{j \in J}\frac{ k_j^\mu}{\varpi_a} \mathcal{ E}_i^{\mathcal{T}}(\{\bk_\ell\}_{\ell \in I}, - \bk) \mathcal{ E}_j^{\mathcal{T}}(\bk, \varpi_j + \omega_a, \vk_j, \{\bk_\ell\}_{\ell \in J\textbackslash j})\nonumber\\
 &+ \frac{k^\mu}{\varpi_a} \mathcal{ E}_i^{\mathcal{T}}(\{\bk_\ell\}_{\ell \in I}, -\varpi + \omega_a, -\vk) \mathcal{ E}_j^{\mathcal{T}}(\bk, \{\bk_\ell\}_{\ell \in J}) \nonumber\\
 &- \mathcal{ E}_i^{\mathcal{T}}(\{\bk_\ell\}_{\ell \in I},-\bk) \frac{k^\mu}{\varpi_a} \mathcal{ E}_j^{\mathcal{T}}(\varpi + \omega_a, \vk, \{\bk_\ell\}_{\ell \in J}) \Big]\,, 
\end{align}
and shifting the appropriate frequency to show that  the two last terms cancel out, which proves \eq{eq:leibnizD}.
One hence obtains
\begin{align}
\frac{\p}{\p q_a^\alpha} \frac{\delta}{\delta \varphi_i(\bq_a)} \tilde{\mathcal{ T}}^{(n)}_{i_1\cdots i_n}[\{\bp_\ell\}]_{\varphi=0}\Big|_{\vq_a=0} &
=\delta_{i\psi} i \epsilon_{\alpha\mu} \tilde{\mathcal{D}}_\mu(\varpi_a) \int_{\bk_\mathrm{ intern}}\prod_{i=1 }^m \mathcal{ E}_i^{\mathcal{T}}(\{\bp_\ell\}_i,\{\bk_\ell\}_i)\nonumber\\
&=\delta_{i\psi} i \epsilon_{\alpha\mu} \tilde{\mathcal{D}}_\mu(\varpi_a) \tilde{\mathcal{ T}}^{(n)}_{i_1\cdots i_n}(\{\bp_\ell\})\,.
\end{align}

To prove \eq{eq:ClosureUneq}, one still needs to check that the same property holds for two functional derivatives and their subsequent wave-number derivatives. Let us examine this contribution
\begin{align}
 &\frac{\p^2}{\p q_1^\alpha \p q_2^\beta } \frac{\delta^2}{\delta \varphi_i(\bq_1) \delta \varphi_j(\bq_2)} \tilde{\mathcal{ T}}^{(n)}_{i_1\cdots i_n}[\{\bp_\ell\}]_{\varphi=0}\Big|_{\vq_1=\vq_2=0} \nonumber\\
 & =\frac{\p^2}{\p q_1^\alpha \p q_2^\beta } \Bigg\{\int_{\bk_\mathrm{ intern}}\sum_{k,k' \atop k \neq k'} \Big(\prod_{m\neq k, k'} \mathcal{ E}_m^{\mathcal{T}}(\{\bp_\ell\}_m,\{\bk_\ell\}_m)\Big) \nonumber\\
 &\times \frac{\delta}{\delta \varphi_i(\bq_1)} \mathcal{ E}_k^{\mathcal{T}}[\{\bp_\ell\}_k,\{\bk_\ell\}_k]\Big|_{\varphi=0} \frac{\delta}{\delta \varphi_j(\bq_2)} \mathcal{ E}_{k'}^{\mathcal{T}}[\{\bp_\ell\}_{k'},\{\bk_\ell\}_{k'}]\Big|_{\varphi=0} \nonumber\\
 &+\int_{\bk_\mathrm{ intern}} \sum_{k} \Big(\prod_{k'\neq k} \mathcal{ E}_{k'}^{\mathcal{T}}(\{\bp_\ell\}_{k'},\{\bk_\ell\}_{k'})\Big) \frac{\delta^2}{\delta \varphi_i(\bq_1) \delta \varphi_j(\bq_2)} \mathcal{ E}_k^{\mathcal{T}}[\{\bp_\ell\}_k,\{\bk_\ell\}_k]\Big|_{\varphi=0}  \Bigg\}\Big|_{\vq_1=\vq_2=0}\, .
\end{align}
First applying one of the wave-number derivative -- say $\vq_2$ --, the $\varphi_j(\bq_2)$ functional derivatives can be replaced by $\delta_{j\psi} i \epsilon_{\beta\nu} \tilde{\mathcal{D}}_\nu(\varpi_2)$. This is possible
  for the second term in curly brackets as well because the object to which the derivative is applied, $\frac{\delta}{\delta \varphi_i(\bq_1)} \mathcal{ E}_k^{\mathcal{T}}[\{\bp_\ell\}_k,\{\bk_\ell\}_k]\Big|_{\varphi=0}$ is  a tree with the $\vq_1$ leg amputated, for which
 the derivative property of $\tilde{\mathcal{D}}_\mu(\varpi_a)$ \eq{eq:leibnizD} holds. Thus, one obtains 
\begin{align}
 &\frac{\p^2}{\p q_1^\alpha \p q_2^\beta } \frac{\delta^2}{\delta \varphi_i(\bq_1) \delta \varphi_j(\bq_2)} \tilde{\mathcal{ T}}^{(n)}_{i_1\cdots i_n}[\{\bp_\ell\}]_{\varphi=0}\Big|_{\vq_1=\vq_2=0} \nonumber\\
 &=\delta_{j\psi} i \epsilon_{\beta\nu} \frac{\p}{\p q_1^\alpha}\Bigg\{\int_{\bk_\mathrm{ intern}}\sum_{k,k' \atop k \neq k'} \Big(\prod_{m\neq k, k'} \mathcal{ E}_m^{\mathcal{T}}(\{\bp_\ell\}_m,\{\bk_\ell\}_m)\Big) \nonumber\\
 &\times \frac{\delta}{\delta \varphi_i(\bq_1)} \mathcal{ E}_k^{\mathcal{T}}[\{\bp_\ell\}_k,\{\bk_\ell\}_k]\Big|_{\varphi=0} \tilde{\mathcal{D}}_\nu(\varpi_2)\mathcal{ E}_{k'}^{\mathcal{T}}(\{\bp_\ell\}_{k'},\{\bk_\ell\}_{k'}) \nonumber\\
 &+ \int_{\bk_\mathrm{ intern}}\sum_{k} \Big(\prod_{k'\neq k} \mathcal{ E}_{k'}^{\mathcal{T}}(\{\bp_\ell\}_{k'},\{\bk_\ell\}_{k'})\Big)\tilde{\mathcal{D}}_\nu(\varpi_2)\Big[ \frac{\delta}{\delta \varphi_i(\bq_1)} \mathcal{ E}_k^{\mathcal{T}}[\{\bp_\ell\}_k,\{\bk_\ell\}_k]\Big|_{\varphi=0}\Big]  \Bigg\}\Bigg|_{\vq_1=0}\nonumber\\
 &=\delta_{j\psi} i \epsilon_{\beta\nu} \frac{\p}{\p q_1^\alpha}\Bigg\{ \tilde{\mathcal{D}}_\nu(\varpi_2)\Big[\int_{\bk_\mathrm{ intern}} \sum_{k} \Big(\prod_{k'\neq k} \mathcal{ E}_{k'}^{\mathcal{T}}(\{\bp_\ell\}_{k'},\{\bk_\ell\}_{k'})\Big)\frac{\delta}{\delta \varphi_i(\bq_1)} \mathcal{ E}_k^{\mathcal{T}}[\{\bp_\ell\}_k,\{\bk_\ell\}_k]\Big|_{\varphi=0}\Big]  \Bigg\}\Bigg|_{\vq_1=0} \, .
\end{align}
Making explicit the operator $\tilde{\mathcal{D}}_\nu(\varpi_2)$ yields
\begin{align}
 &\frac{\p^2}{\p q_1^\alpha \p q_2^\beta } \frac{\delta^2}{\delta \varphi_i(\bq_1) \delta \varphi_j(\bq_2)} \tilde{\mathcal{ T}}^{(n)}_{i_1\cdots i_n}[\{\bp_\ell\}]_{\varphi=0}\Big|_{\vq_1=\vq_2=0} \nonumber\\
 &=- \delta_{j\psi} i \epsilon_{\beta\mu} \frac{\p}{\p q_1^\alpha} \Bigg\{\frac{q_1^\nu}{\varpi_2} \Big[\int_{\bk_\mathrm{ intern}} \sum_{k} \Big(\prod_{k'\neq k} \mathcal{ E}_{k'}^{\mathcal{T}}(\{\bp_\ell\}_{k'},\{\bk_\ell\}_{k'})\Big)\frac{\delta}{\delta \varphi_i(\varpi_1+\varpi_2,\vq_1)} \mathcal{ E}_k^{\mathcal{T}}[\{\bp_\ell\}_k,\{\bk_\ell\}_k]\Big|_{\varphi=0}\Big] \nonumber\\
 &+ \sum_{k=1}^{n}\frac{p_k^\nu}{\varpi_2}  \Big[ \int_{\bk_\mathrm{ intern}}\sum_{m} \Big(\prod_{m'\neq m} \mathcal{ E}_{m'}^{\mathcal{T}}(\{\bp_\ell\}_{m'}\textbackslash k_+,\{\bk_\ell\}_{m'})\Big)\frac{\delta}{\delta \varphi_i(\bq_1)} \mathcal{ E}_m^{\mathcal{T}}[\{\bp_\ell\}_m\textbackslash k_+,\{\bk_\ell\}_m]\Big|_{\varphi=0}\Big]  \Bigg\}\Bigg|_{\vq_1=0}\,,
\end{align} 
with the shorthand notation $\{\bp_\ell\}_m\textbackslash k_+ = \{ \nu_k + \omega_2, \vp_k, \bp_{\ell \neq k}\}_m$ if $\bp_k \in \{\bp_\ell\}_m$ and else $\{\bp_\ell\}_m\textbackslash k_+=\{\bp_\ell\}_m$. Distributing the $\vq_1$ derivative  gives
\begin{align}
 &\frac{\p^2}{\p q_1^\alpha \p q_2^\beta } \frac{\delta^2}{\delta \varphi_k(\bq_1) \delta \varphi_\ell(\bq_2)} \tilde{\mathcal{ T}}^{(n)}_{i_1\cdots i_n}[\{\bp_\ell\}]_{\varphi=0}\Big|_{\vq_1=\vq_2=0} \nonumber\\
 &=- \delta_{j\psi} i \epsilon_{\beta\nu} \Bigg\{ \frac{\delta_{\alpha\nu}}{\varpi_2} \Big[ \int_{\bk_\mathrm{ intern}}\sum_{k} \Big(\prod_{k'\neq k} \mathcal{ E}_{k'}^{\mathcal{T}}(\{\bp_\ell\}_{k'},\{\bk_\ell\}_{k'})\Big)\frac{\delta}{\delta \varphi_i(\varpi_1+\varpi_2,\vec 0)} \mathcal{ E}_k^{\mathcal{T}}[\{\bp_\ell\}_k,\{\bk_\ell\}_k]\Big|_{\varphi=0}\Big]_{\vq_1=0} \nonumber\\
 &+ \delta_{i\psi} i \epsilon_{\alpha\mu} \sum_{k=1}^{n}\frac{p_k^\nu}{\varpi_2}  \Big[ \int_{\bk_\mathrm{ intern}}\sum_{m} \Big(\prod_{m'\neq m} \mathcal{ E}_{m'}^{\mathcal{T}}(\{\bp_\ell\}_{m'}\textbackslash k_+,\{\bk_\ell\}_{m'})\Big)\tilde{\mathcal{D}}_\mu(\varpi_1) \mathcal{ E}_m^{\mathcal{T}}(\{\bp_\ell\}_m\textbackslash k_+,\{\bk_\ell\}_m)\Big]  \Bigg\}\nonumber\\
 &=-  \delta_{i\psi} \delta_{j\psi} i^2  \epsilon_{\alpha\mu} \epsilon_{\beta\nu} \sum_{k=1}^{n}\frac{p_k^\nu}{\varpi_2} \Big[ \tilde{\mathcal{D}}_\mu(\varpi_1) \int_{\bk_\mathrm{ intern}} \prod_{m} \mathcal{ E}_m^{\mathcal{T}}(\{\bp_\ell\}_m\textbackslash k_+,\{\bk_\ell\}_m) \Big]\nonumber\\
 &= \delta_{i\psi}\delta_{j\psi} i^2 \epsilon_{\alpha\mu} \epsilon_{\beta\nu} \tilde{\mathcal{D}}_\nu(\varpi_2)  \tilde{\mathcal{D}}_\mu(\varpi_1)\tilde{\mathcal{ T}}^{(n)}_{i_1\cdots i_n}(\{\bp_\ell\})\,,
\end{align}
which  proves the property \eq{eq:ClosureUneq}. 
The flow equation \eq{eq:flowNuneq} for the two-point function at leading order is finally given by 
\begin{align}
 \p_\kappa \tilde G^{(n)}_{\psi,i_1\dots i_{n}}(\{\bp_\ell\}_{1 \leq \ell \leq n})\Big|_\mathit{leading}  &=  \frac{1}{2}\,  \int_{\bq_1,\bq_2}\!
 \tilde \p_\kappa \tilde  G^{(2)}_{\psi \psi}(-\bq_1,-\bq_2) i^2 q_1^\alpha q_2^\beta \epsilon_{\alpha \mu} \epsilon_{\beta \nu} \tilde{\mathcal{D}}_\mu(\omega_1) \tilde{\mathcal{D}}_\nu(\omega_2)
 \tilde G^{(n)}_{\psi,i_1\dots i_{n}}(\{\bp_\ell\})\nonumber\\
 &= \frac{1}{2}\,  \int_{\bq_1,\bq_2}\!
 \tilde \p_\kappa \tilde  G^{(2)}_{v_\mu v_\nu}(-\bq_1,-\bq_2)\tilde{\mathcal{D}}_\mu(\omega_1) \tilde{\mathcal{D}}_\nu(\omega_2)
 \tilde G^{(n)}_{\psi,i_1\dots i_{n}}(\{\bp_\ell\})\,, 
\end{align}
which is closed, \ie it does not involve higher-order correlation functions. This closure has been achieved without any  approximation apart 
 from the large wave-number limit. 

\subsection{Flow equation at \NLO at equal times}
\label{app:flowR}

As presented in the main text, the leading order term in the flow equation vanishes when all times are equal (\ie when integrated over all the external frequencies). In this appendix, we hence calculate the \NLO  term,
focusing on equal-times correlations, and show that all the terms which are controlled by the extended symmetries vanish.

The NLO term of the flow equation for the generalized correlation function reads
\begin{align}
 \p_\kappa \int_{\{\omega_\ell\}} &\tilde G^{(n)}_{i_1\dots i_{n}}(\{\bp_\ell\}_{1 \leq \ell \leq n})\Big|_\mathit{NLO}  =  \frac{1}{2}\,  \int_{\bq_1,\bq_2}\!
 \tilde \p_\kappa \tilde  G^{(2)}_{ij}(-\bq_1,-\bq_2) \nonumber\\
 &\times \int_{\{\omega_\ell\}} \frac{q_a^\mu q_b^\nu q_c^\rho q_d^\sigma}{4!} \frac{\p^4}{\p q_a^\mu \p q_b^\nu \p q_c^\rho \p q_d^\sigma} 
 \left[ \frac{\delta^2}{\delta \varphi_i(\bq_1)\delta\varphi_j(\bq_2)} 
 \tilde G^{(n)}_{i_1\dots i_{n}}[\{\bp_\ell\};\mathrm{j}] \right]_{\varphi=0}\Bigg|_{\vq_1=\vq_2 = 0} \,,
\end{align}
where as before $a,\,b,\,c,\,d$ take value in $\{1,2\}$. As in the previous calculation \aref{app:flowD}, the four wave-number
derivatives can be classified according to the respective number of $\vq_1$ and $\vq_2$ derivatives. 
Using the same argument  as in  \sref{eq:nexttoleading}, if the 
four derivatives are $\vq_1$ (resp. four $\vq_2$), this contribution is zero because the vertex function with the wave-number $\vq_2$ (resp. $\vq_1$) vanishes when this wave-number is set to zero.  Thus we consider below separately
 the two remaining cases: three $\vq_1$ and one $\vq_2$ (and equivalently one $\vq_1$ and three $\vq_2$), and finally two $\vq_1$ and two $\vq_2$. 

\subsubsection{Contributions 1-3 and 3-1}

Although no Ward identity exists for the third wave-number derivative of a vertex function, the time-gauged Galilean Ward identity can still be  used on the leg with one $q$ derivative and the proof of \aref{app:flowD} for one $\psi$ derivative can be carried through to show that one  obtains an operator  $\tilde{\mathcal{D}}$ acting on the external legs of the whole diagram
\begin{align}
 \p_\kappa \int_{\{\omega_\ell\}} &\tilde G^{(n)}_{i_1\dots i_{n}}(\{\bp_\ell\}_{1 \leq \ell \leq n})\Big|_{\mathit{NLO},1-3}  =  \frac{1}{2}\,  \int_{\bq_1,\bq_2}\!
 \tilde \p_\kappa \tilde  G^{(2)}_{\psi j}(-\bq_1,-\bq_2) \nonumber\\
 &\times  \int_{\{\omega_\ell\}} \frac{q_1^\mu q_2^\nu q_2^\rho q_2^\sigma}{3!} \frac{\p^3}{\p q_2^\nu \p q_2^\rho \p q_2^\sigma} 
 \Bigg\{i \epsilon_{\mu\alpha}\tilde{\mathcal{D}}_\alpha(\varpi_1) \Big[\frac{\delta}{\delta\varphi_j(\bq_2)} 
 \tilde G^{(n)}_{i_1\dots i_{n}}[\{\bp_\ell\};\mathrm{j}] \Big]_{\varphi=0}\Bigg\}_{\vq_2 = 0} \,.
 \label{eq:streamNLO13}
\end{align}
 Distributing the $\vq_2$ derivatives, one obtains two types of terms: either all $\vq_2$ derivatives act on the term in square bracket or one of
them acts on the operator $\tilde{\mathcal{D}}$
\begin{align} 
 & \int_{\{\omega_\ell\}}\frac{q_1^\mu q_2^\nu q_2^\rho q_2^\sigma}{3!} \frac{\p^3}{\p q_2^\nu \p q_2^\rho \p q_2^\sigma} 
 \Bigg\{i \epsilon_{\mu \alpha}\tilde{\mathcal{D}}_\alpha(\varpi_1) \Big[\frac{\delta}{\delta\varphi_j(\bq_2)} 
 \tilde G^{(n)}_{i_1\dots i_{n}}[\{\bp_\ell\};\mathrm{j}] \Big]_{\varphi=0}\Bigg\}_{\vq_2 = 0} \nonumber\\
 &= \int_{\{\omega_\ell\}} \frac{q_1^\mu q_2^\nu q_2^\rho q_2^\sigma}{3!} i \epsilon_{\mu\alpha} \tilde{\mathcal{D}}_\alpha(\omega_1)\Big|_{\vq_2 = 0}\frac{\p^3}{\p q_2^\nu \p q_2^\rho \p q_2^\sigma} 
\Big[\frac{\delta}{\delta\varphi_j(\bq_2)} 
 \tilde G^{(n)}_{i_1\dots i_{n}}[{\bp_1}, \dots, {\bp_{n}};\mathrm{j}] \Big]_{\varphi=0}\Big|_{\vq_2 = 0}\nonumber\\
 &\quad +  \int_{\{\omega_\ell\}} \frac{q_1^\mu q_2^\nu q_2^\rho q_2^\sigma}{2} i \epsilon_{\mu \alpha} \frac{\p}{\p q_2^\nu} \tilde{\mathcal{D}}_\alpha(\omega_1)\Big|_{\vq_2 = 0}\frac{\p^2}{\p q_2^\rho \p q_2^\sigma} \Big[\frac{\delta}{\delta\varphi_j(\bq_2)} 
 \tilde G^{(n)}_{i_1\dots i_{n}}[\{\bp_\ell\};\mathrm{j}] \Big]_{\varphi=0}\Big|_{\vq_2 = 0}\nonumber\\
 &= \int_{\{\omega_\ell\}}\frac{q_1^\mu q_2^\nu q_2^\rho q_2^\sigma}{2} i \epsilon_{\mu \alpha} \delta_{\nu \alpha}  \frac{1}{\omega_1} \frac{\p^2}{\p q_2^\rho \p q_2^\sigma} \Big[\frac{\delta}{\delta\varphi_j(\varpi_2+\varpi_1, \vq_2)} 
 \tilde G^{(n)}_{i_1\dots i_{n}}[\{\bp_\ell\};\mathrm{j}] \Big]_{\varphi=0}\Big|_{\vq_2 = 0}\, .
\end{align}
In the first term of the first equality,  $\tilde{\mathcal{D}}\big|_{\vq_2 = 0}$ shifts only the frequencies associated to 
$\vp_1$ and $\vp_2$, thus this term is zero due to the conservation of wave-number of the object in square bracket and the integration in frequency. The $\vq_2$ derivative on $\tilde{\mathcal{D}}$ selects the frequency shift on the $\bq_2$ leg, which does not vanishes.
However, this term is proportional to $\epsilon_{\mu\nu} q_1^\mu q_2^\nu$. Hence, within \Eq{eq:streamNLO13}, the conservation of wave-number of $G^{(2)}_{i j}(-\bq_1,-\bq_2)$ gives $\vq_1 + \vq_2 = 0$, and  thus $\epsilon_{\mu\nu} q_1^\mu q_2^\nu = - \epsilon_{\mu\nu} q_2^\mu q_1^\nu = 0$, and this term gives no contribution either to the flow equation.
Hence, all the contributions with three $q_1$ derivatives and one $q_2$, or   three $q_2$ derivatives and one $q_1$ vanish
\begin{equation}
\p_\kappa \int_{\{\omega_\ell\}} \tilde G^{(n)}_{i_1\dots i_{n}}(\{\bp_\ell\}_{1 \leq \ell \leq n})\Big|_{\mathit{NLO},1-3}  = 0\, .
\end{equation}

\subsubsection{Contributions 2-2}

The only remaining contribution in the NLO term of the flow equation involves an equal number of $q_1$ and $q_2$ derivatives
\begin{align}
& \p_\kappa \int_{\{\omega_\ell\}}  \tilde G^{(n)}_{i_1\dots i_n}(\{\bp_\ell\}_{1 \leq \ell \leq n})\Big|_{\mathit{NLO}} = \p_\kappa \int_{\{\omega_\ell\}}  \tilde G^{(n)}_{i_1\dots i_n}(\{\bp_\ell\}_{1 \leq \ell \leq n})\Big|_{\mathit{NLO},2-2}  \nonumber\\
 &=  \frac{1}{2}\,  \int_{\bq_1,\bq_2}\!
 \tilde \p_\kappa \tilde  G^{(2)}_{ij}(-\bq_1,-\bq_2)\nonumber\\
&\quad\times \int_{\{\omega_\ell\}} \frac{q_1^\mu q_1^\nu q_2^\rho q_2^\sigma}{4} \frac{\p^4}{\p q_1^\mu \p q_1^\nu \p q_2^\rho \p q_2^\sigma} \Big[ \frac{\delta^2}{\delta \varphi_i(\bq_1)\delta\varphi_j(\bq_2)} 
 \tilde G^{(n)}_{i_1\dots i_{n}}[{\{\bp_\ell\}};\mathrm{j}] \Big]_{\varphi=0} \Bigg|_{\vq_1=\vq_2 = 0}\,.
\label{eq:streamflow22}
\end{align}
Using the invariance under space translation and rotation in the $\vq_1,\vq_2$ integrals, one can write
\begin{align}
 \,   \int_{\vq_1,\vq_2}\!
 \tilde \p_\kappa \tilde  G^{(2)}_{ij}(-\bq_1,-\bq_2) \frac{q_1^\mu q_1^\nu q_2^\rho q_2^\sigma}{4}
&=  \frac{1}{4}\,  \int_{\vq}\!
 \tilde \p_\kappa \bar{\tilde  G}^{(2)}_{ij}(-\omega_1,-\omega_2, q^2)  q^\mu q^\nu q^\rho q^\sigma\nonumber\\
&= (\delta_{\mu\nu}\delta_{\rho\sigma} + \delta_{\mu\rho}\delta_{\nu\sigma} + \delta_{\mu\sigma}\delta_{\nu\rho}) \tilde K_{ij}(\omega_1,\omega_2)\,,
\end{align}
with
\begin{equation}
\tilde K_{ij}(\omega_1,\omega_2) \equiv \frac{1}{32}\,  \int_{\vq}\!
 \tilde \p_\kappa \bar{\tilde  G}^{(2)}_{ij}(-\omega_1,-\omega_2, q^2)  (q^2)^2\,,
\end{equation}
where the notation $\bar{\tilde  G}$  indicates that the delta of conservation has been extracted for the wave-numbers only, and not for the frequencies.
The expression  \eq{eq:streamflow22} hence comprises two contributions
\begin{align}
 \p_\kappa &\int_{\{\omega_\ell\}}   \tilde G^{(n)}_{i_1\dots i_n}(\{\bp_\ell\}_{1 \leq \ell \leq n})\Big|_{\mathit{NLO}} =  \frac{1}{2}\,  \int_{\varpi_1,\varpi_2}\!
\tilde K_{ij}(\varpi_1,\varpi_2) \nonumber\\
&\times \int_{\{\omega_\ell\}} \Big(\frac{\p^4}{\p q_1^\mu \p q_1^\mu \p q_2^\nu \p q_2^\nu} + 2 \frac{\p^4}{\p q_1^\mu \p q_2^\mu \p q_1^\nu \p q_2^\nu} \Big)  \Big[ \frac{\delta^2}{\delta \varphi_i(\bq_1)\delta\varphi_j(\bq_2)} 
 \tilde G^{(n)}_{i_1\dots i_{n}}[\{\bp_\ell\};\mathrm{j}] \Big]_{\varphi=0} \Bigg|_{\vq_1=\vq_2 = 0}\, ,
\end{align} 
 referred to as the uncrossed and the crossed ones, according to whether the $\vq_1$ derivative is contracted with the other $\vq_1$ derivative or with the $\vq_2$ derivative.  

We show in the next sections,
first on the example of the two-point function, and then for a generic $n$-point function,
that the uncrossed contribution can be closed exactly using the Ward identities associated with the new symmetries
\begin{align}
 \p_\kappa &  \tilde G^{(n)}_{i_1 \dots i_n}(\{\bp_\ell\}_{1 \leq \ell \leq n})\Big|_{\mathit{uncrossed}}\nonumber\\
&= \frac{1}{2}\, \int_{\varpi_1,\varpi_2}\! \tilde K_{ij}(\varpi_1,\varpi_2) \frac{\p^4}{\p q_1^\mu \p q_1^\mu \p q_2^\nu \p q_2^\nu}  \Big[ \frac{\delta^2}{\delta \varphi_i(\bq_1)\delta\varphi_j(\bq_2)} 
 \tilde G^{(n)}_{i_1\dots i_{n}}[\{\bp_\ell\};\mathrm{j}] \Big]_{\varphi=0} \Bigg|_{\vq_1=\vq_2 = 0}\nonumber\\
&=\frac{1}{2}\,   \int_{\varpi_1,\varpi_2}\! \tilde K_{\psi \psi}(\varpi_1,\varpi_2)  \tilde{\mathcal{R}}(\varpi_1)  \tilde{\mathcal{R}}(\varpi_2) \tilde G^{(n)}_{i_1\dots i_{n}}(\{\bp_\ell\})  \,,
\label{eq:flowNLOuncrossed}
\end{align}
where $\tilde{\mathcal{R}}$ is defined in \Eq{eq:defDR}. It follows that this contribution also vanishes when integrated over the external frequencies. However, the crossed contribution is not controlled by these Ward identities, and we have not been able to further constrain
 this last remaining term. 

\subsubsection{Uncrossed derivatives contribution for the flow of the two-point function}

In this section, we show that the uncrossed contribution to the flow of $G^{(2)}_{\psi \psi}({\bp_1},{\bp_{2}})$ can be closed exactly. 
Distributing the $\vq$-derivatives, this contribution reads
\begin{align}
 \p_\kappa &\int_{\omega_1,\omega_2}  \tilde G^{(2)}_{\psi \psi}({\bp_1},{\bp_{2}})\Big|_{\mathit{uncrossed}}
=\frac{1}{2}\,  \int_{\varpi_1,\varpi_2}\! \tilde K_{ij}(\varpi_1,\varpi_2) \int_{\omega_1,\omega_2} \frac{\p^4}{\p q_1^\mu \p q_1^\mu \p q_2^\nu \p q_2^\nu}  \nonumber\\
 & \int_{\bk_1, \bk_2} \tilde G_{\psi m}^{(2)}(\bp_1,-\bk_1) \tilde G_{\psi n}^{(2)}(\bp_2,-\bk_2)
 \Big[ -\tilde \Gamma_{ij mn}^{(4)}(\bq_1, \bq_2, \bk_1, \bk_2)\nonumber\\
 & + \int_{\bk_3,\bk_4} \tilde \Gamma_{i m s}^{(3)}(\bq_1, \bk_1, \bk_3) \tilde G_{st}^{(2)}(-\bk_3,-\bk_4) \tilde \Gamma_{j n t}^{(3)}(\bq_2, \bk_2, \bk_4) + (i, \bq_1) \leftrightarrow (j, \bq_2) \Big]_{\vq_1=\vq_2 = 0}\nonumber\\
&= \frac{1}{2}\,  \int_{\varpi_1,\varpi_2}\! \tilde K_{ij}(\varpi_1,\varpi_2) \int_{\omega_1,\omega_2}  \nonumber\\
 & \int_{\bk_1, \bk_2} \tilde G_{\psi m}^{(2)}(\bp_1,-\bk_1) \tilde G_{\psi n}^{(2)}(\bp_2,-\bk_2)
 \Big[ - \frac{\p^4}{\p q_1^\mu \p q_1^\mu \p q_2^\nu \p q_2^\nu} \tilde \Gamma_{ij mn}^{(4)}(\bq_1, \bq_2, \bk_1, \bk_2)\nonumber\\
 & + \int_{\bk_3,\bk_4}  \frac{\p^2}{\p q_1^\mu \p q_1^\mu}\tilde \Gamma_{i m s}^{(3)}(\bq_1, \bk_1, \bk_3) \tilde G_{st}^{(2)}(-\bk_3,-\bk_4)  \frac{\p^2}{\p q_2^\nu \p q_2^\nu}\tilde \Gamma_{j n t}^{(3)}(\bq_2, \bk_2, \bk_4) + (i, \bq_1) \leftrightarrow (j, \bq_2) \Big]_{\vq_1=\vq_2 = 0}\nonumber\\
&= \frac{1}{2}\,  \int_{\varpi_1,\varpi_2}\! \tilde K_{\psi\psi}(\varpi_1,\varpi_2) \int_{\omega_1,\omega_2}  \nonumber\\
 & \int_{\bk_1, \bk_2} \tilde G_{\psi m}^{(2)}(\bp_1,-\bk_1) \tilde G_{\psi n}^{(2)}(\bp_2,-\bk_2)
 \Big[ - \tilde{\mathcal{R}}(\varpi_1)\tilde{\mathcal{R}}(\varpi_2) \tilde \Gamma_{mn}^{(2)}(\bk_1, \bk_2)\nonumber\\
 & + \int_{\bk_3,\bk_4}  \tilde{\mathcal{R}}(\varpi_1) \tilde \Gamma_{m s}^{(2)}(\bk_1, \bk_3) \tilde G_{st}^{(2)}(-\bk_3,-\bk_4)  \tilde{\mathcal{R}}(\varpi_2) \tilde \Gamma_{ n t}^{(2)}( \bk_2, \bk_4) + (\varpi_1) \leftrightarrow (\varpi_2) \Big] \, .\nonumber\\
\label{eq:streamflow22R}
\end{align}
Following the same lines as for the operator $\tilde{\mathcal{D}}$, one can show  that the terms in square bracket can be expressed as the operator $\tilde{\mathcal{R}}$ acting on the external legs of the original diagram. 
 Using that $\tilde G^{(2)}$ and $\tilde \Gamma^{(2)}$ are inverse of each other,  the second term of \eq{eq:streamflow22R} can be rewritten as two combinations of $\tilde G^{(2)} \tilde{\mathcal{R}}(\varpi_a) \tilde \Gamma^{(2)} \tilde G^{(2)}$ attached  by a  $\tilde \Gamma^{(2)}$. 
 For each of them, one has
\begin{align}
   \int_{\bk_1,\bk_3} &\tilde G_{\psi m}^{(2)}(\bp_1,-\bk_1)  \tilde{\mathcal{R}}(\varpi_1)\tilde \Gamma_{m s}^{(2)}(\bk_1, \bk_3) \tilde G_{su}^{(2)}(-\bk_3,-\bk_5)   \nonumber\\
 &= \frac{2 i \epsilon_{\alpha\beta}}{\varpi_1}\int_{\bk_1,\bk_3} \tilde G_{\psi m}^{(2)}(\bp_1,-\bk_1) \Big[k_1^{\alpha} \frac{\p}{\p k_1^\beta} \tilde\Gamma^{(2)}_{ ms } (\nu_1+\varpi_1,\vk_1,\bk_3) \nonumber\\
&\quad+ k_3^{\alpha} \frac{\p}{\p k_3^\beta} \tilde\Gamma^{(2)}_{ ms } (\bk_1,\nu_3+\varpi_1,\vk_3) \Big]\tilde G_{su}^{(2)}(-\bk_3,-\bk_5)  \nonumber\\
 &= \frac{2 i \epsilon_{\alpha\beta}}{\varpi_1}\Big[ \int_{\bk_1}  \tilde G_{\psi m}^{(2)}(\bp_1,-\bk_1) k_1^{\alpha} \frac{\p}{\p k_1^\beta}\delta_{m u}\delta(\varpi_1+\nu_1-\nu_5)\delta^2(\vk_1-\vk_5) \nonumber\\
&\quad+  \int_{\bk_3}  \tilde G_{su}^{(2)}(-\bk_3,-\bk_5)  k_3^{\alpha} \frac{\p}{\p k_3^\beta}  \delta_{\psi s} \delta(\omega_1+\varpi+\nu_3)\delta^2(\vp_1+\vk_3) \Big]  \nonumber\\
 &=- k_5^{\alpha} \frac{\p}{\p k_5^\beta} \tilde G^{(2)}_{\psi u}(\bp_1,-\nu_5+\varpi_1,-\vk_5) - p_1 ^{\alpha} \frac{\p}{\p p_1^\beta} \tilde G^{(2)}_{\psi u}(\omega_1+\varpi_1, \vp_1,-\vk_5) \nonumber\\
 &= - \tilde{\mathcal{ R}}(\varpi_1)  \tilde G^{(2)}_{\psi u}(\bp_1,-\bk_5)\, ,
 \end{align} 
where in the second equality, $\tilde G^{(2)}$ and $\tilde \Gamma^{(2)}$ have been contracted,  and in the third, an integration by part
 is performed. We have thereby established the equivalent  property \eq{eq:DG2} of $\tilde{\mathcal{D}}$ for $\tilde{\mathcal{R}}$:
\begin{align}
\frac{\p^2}{\p q_\mu \p q_\mu}\df{}{\varphi_k(\bq)}& \tilde G_{ij}^{(2)}[\bp_1,\bp_2;\mathrm{j}]_{\varphi=0}\Big|_{\vq=0} =\tilde{\mathcal{ R}}(\varpi) \tilde G_{ij}^{(2)}(\bp_1,\bp_2)\, .
\label{eq:RG2}
\end{align}
Inserting this result in the last line of \eq{eq:streamflow22R} yields
\begin{align}
 \p_\kappa &\int_{\omega_1, \omega_2}  \tilde G^{(2)}_{\psi \psi}({\bp_1},{\bp_{2}})\Big|_{\mathit{uncrossed}}
= 
\frac{1}{2}\,  \int_{\varpi_1,\varpi_2}\!
 \tilde K_{\psi \psi}(\varpi_1,\varpi_2) \int_{\omega_1, \omega_2} \int_{\bk_1, \bk_2}  \nonumber\\
 & \Big[ - G_{\psi u}^{(2)}(\bp_1,-\bk_1)G_{\psi v}^{(2)}(\bp_2,-\bk_2) \tilde{\mathcal{R}}(\varpi_1) \tilde{\mathcal{R}}(\varpi_2) \tilde \Gamma_{uv}^{(2)}( \bk_1, \bk_2) \nonumber\\
&+ \tilde{\mathcal{R}}(\varpi_1) G_{\psi u}^{(2)}(\bp_1,-\bk_1) \Gamma_{uv}^{(2)}(\bk_1, \bk_2) \tilde{\mathcal{R}}(\varpi_2)  G_{\psi v}^{(2)}(\bp_2, -\bk_2) + (\varpi_1) \leftrightarrow (\varpi_2) \Big] \, .\nonumber\\
\label{eq:flotstreamG2uncrossed}
\end{align}
 In fact, this structure is the same as the one appearing with the operator $\tilde{\mathcal{D}}$ for the leading order at unequal times. It generalizes as well
 for any correlation functions, as shown in the next section. 
Let us examine separately the two terms in square brackets. The first term reads
\begin{align}
&\int_{\bk_1, \bk_2}  G_{\psi u}^{(2)}(\bp_1,-\bk_1)G_{\psi v}^{(2)}(\bp_2,-\bk_2) \tilde{\mathcal{R}}(\varpi_1) \tilde{\mathcal{R}}(\varpi_2) \tilde \Gamma_{uv}^{(2)}( \bk_1, \bk_2) \nonumber\\
&= \frac{2 i \epsilon_{\alpha\beta}}{\varpi_2} \int_{\bk_1, \bk_2}  G_{\psi u}^{(2)}(\bp_1,-\bk_1)G_{\psi v}^{(2)}(\bp_2,-\bk_2) \nonumber\\
&\quad\times\tilde{\mathcal{R}}(\varpi_1) \Big[ k_1^\alpha \frac{\p}{\p k_1^\beta}  \tilde \Gamma_{uv}^{(2)}( \nu_1+\varpi_2, \vk_1, \bk_2) + k_2^\alpha \frac{\p}{\p k_2^\beta}  \tilde \Gamma_{uv}^{(2)}(\bk_1,\nu_2+\varpi_2,  \vk_2)\Big]\nonumber\\
&= -\frac{4 \epsilon_{\alpha\beta} \epsilon_{\rho\sigma}}{\varpi_1 \varpi_2} \int_{\bk_1, \bk_2}  G_{\psi u}^{(2)}(\bp_1,-\bk_1)G_{\psi v}^{(2)}(\bp_2,-\bk_2)\nonumber\\
&\quad \times\Bigg\{k_1^\rho \frac{\p}{\p k_1^\sigma} \Big[ k_1^\alpha \frac{\p}{\p k_1^\beta}  \tilde \Gamma_{uv}^{(2)}( \nu_1+\varpi_1+\varpi_2, \vk_1, \bk_2) + k_2^\alpha \frac{\p}{\p k_2^\beta}  \tilde \Gamma_{uv}^{(2)}(\nu_1+\varpi_1,\vk_1,\nu_2+\varpi_2,  \vk_2)\Big]\nonumber\\
& + k_2^\rho \frac{\p}{\p k_2^\sigma}\Big[ k_1^\alpha \frac{\p}{\p k_1^\beta}  \tilde \Gamma_{uv}^{(2)}( \nu_1+\varpi_2, \vk_1, \nu_2+\varpi_1, \vk_2) + k_2^\alpha \frac{\p}{\p k_2^\beta}  \tilde \Gamma_{uv}^{(2)}(\bk_1,\nu_2+\varpi_1+\varpi_2,  \vk_2)\Big] \Bigg\}\nonumber\\
&= -\frac{4 \epsilon_{\alpha\beta} \epsilon_{\rho\sigma}}{\varpi_1 \varpi_2} \Bigg\{ \int_{\vk_1 } G_{\psi \psi}^{(2)}(\bp_1,\omega_2+\varpi_1 + \varpi_2, -\vk_1) k_1^\rho \frac{\p}{\p k_1^\sigma} \Big[ k_1^\alpha \frac{\p}{\p k_1^\beta}  \delta^2(\vp_2+ \vk_1) \Big] \nonumber\\
&\quad+ \int_{\vk_2 } G_{\psi \psi}^{(2)}(\omega_1+\varpi_1 + \varpi_2,-\vk_2, \bp_2) k_2^\rho \frac{\p}{\p k_2^\sigma} \Big[ k_2^\alpha \frac{\p}{\p k_2^\beta}  \delta^2(\vp_1+ \vk_2) \Big] \nonumber\\
 &+ \int_{\bk_1, \bk_2}  G_{\psi u}^{(2)}(\bp_1,-\bk_1)G_{\psi v}^{(2)}(\bp_2,-\bk_2) k_1^\alpha k_2^\rho \frac{\p^2}{\p k_1^\beta \p k_2^\sigma}  \tilde \Gamma_{uv}^{(2)}( \nu_1+\varpi_2, \vk_1, \nu_2+\varpi_1, \vk_2) + (\varpi_1) \leftrightarrow (\varpi_2) \Bigg\}\nonumber\\
&= -\frac{4 \epsilon_{\alpha\beta} \epsilon_{\rho\sigma}}{\varpi_1 \varpi_2} \Bigg\{  p_2^\alpha \frac{\p}{\p p_2^\beta} \Big[ p_2^\rho \frac{\p}{\p p_2^\sigma} G_{\psi \psi}^{(2)}(\bp_1,\omega_2+\varpi_1 + \varpi_2, \vp_2) \Big]  +  p_1^\alpha \frac{\p}{\p p_1^\beta} \Big[ p_1^\rho \frac{\p}{\p p_1^\sigma} G_{\psi \psi}^{(2)}(\omega_1+\varpi_1 + \varpi_2,\vp_1, \bp_2) \Big] \nonumber\\
 &+ \int_{\bk_1, \bk_2}  G_{\psi u}^{(2)}(\bp_1,-\bk_1)G_{\psi v}^{(2)}(\bp_2,-\bk_2) k_1^\alpha k_2^\rho \frac{\p^2}{\p k_1^\beta \p k_2^\sigma}  \tilde \Gamma_{uv}^{(2)}( \nu_1+\varpi_2, \vk_1, \nu_2+\varpi_1, \vk_2) + (\varpi_1) \leftrightarrow (\varpi_2) \Bigg\}\,,
\label{eq:flotstreamG2uncrossed1PI}
\end{align}
using integration by part twice for the first two terms in the last equality. For the second term, one has
 \begin{align}
&\int_{\bk_1, \bk_2} \tilde{\mathcal{R}}(\varpi_1) G_{\psi u}^{(2)}(\bp_1,-\bk_1) \Gamma_{uv}^{(2)}(\bk_1,\bk_2) \tilde{\mathcal{R}}(\varpi_2)  G_{\psi v}^{(2)}(\bp_2,-\bk_2)\nonumber\\
&=-\frac{4 \epsilon_{\alpha\beta}\epsilon_{\rho\sigma}}{\varpi_1\varpi_2}  \int_{\bk_1,\bk_2}\Big[  k_1^\alpha \frac{\p}{\p k_1^\beta} G_{\psi u}^{(2)}(\bp_1,-\nu_1+\varpi_1,-\vk_1)+ p_1^\alpha \frac{\p}{\p p_1^\beta} G_{\psi u}^{(2)}(\vp_1,\omega_1+\varpi_1,-\bk_1)\Big]\tilde \Gamma_{uv}^{(2)}(\bk_1,\bk_2)\nonumber\\
 & \times\Big[  k_2^\rho \frac{\p}{\p k_2^\sigma} G_{\psi v}^{(2)}(\bp_2,-\nu_2+\varpi_2,-\vk_2) + p_2^\rho \frac{\p}{\p p_2^\sigma} G_{\psi v}^{(2)}(\omega_2+\varpi_2,\vp_2,-\bk_2) \Big] \nonumber\\
&=-\frac{4 \epsilon_{\alpha\beta}\epsilon_{\rho\sigma}}{\varpi_1\varpi_2} \Bigg\{   \int_{\bk_1,\bk_1} k_1^\alpha \frac{\p}{\p k_1^\beta} G_{\psi u}^{(2)}(\bp_1,-\varpi_1+\varpi_1,-\vk_1)  \tilde \Gamma_{uv}^{(2)}(\bk_1,\bk_2)  k_2^\rho \frac{\p}{\p k_2^\sigma} G_{\psi v}^{(2)}(\bp_2,-\varpi_2+\varpi_2,-\vk_2) \nonumber\\
&+  \int_{\vk_1}  p_2^\rho \frac{\p}{\p p_2^\sigma} \delta^\sigma(\vp_2+\vk_1)  k_1^\alpha \frac{\p}{\p k_1^\beta} G_{\psi \psi}^{(2)}(\bp_1,\omega_2+\varpi_1+\varpi_2,-\vk_1)\nonumber\\
 &+  \int_{\vk_2}  p_1^\alpha \frac{\p}{\p p_1^\beta} \delta^\sigma(\vp_1+\vk_2)\Big[  k_2^\rho \frac{\p}{\p k_2^\sigma} G_{\psi \psi}^{(2)}(\omega_1+\varpi_1+\varpi_2,-\vk_2,\bp_2) + p_2^\rho \frac{\p}{\p p_2^\sigma} G_{\psi \psi}^{(2)}(\omega_2+\varpi_2,\vp_2,\omega_1+\varpi_1,-\vk_2) \Big] \Bigg\}\,.
\end{align}
Integrating by part and shifting the frequencies in the first term in the curly bracket and exchanging the $\vk_1$ (resp. $\vk_2$) integral with the $\vp_2$ (resp. $\vp_1$) derivative in the two last terms, one obtains
 \begin{align}
&\int_{\bk_1, \bk_2} \tilde{\mathcal{R}}(\varpi_1) G_{\psi u}^{(2)}(\bp_1,-\bk_1) \Gamma_{uv}^{(2)}(\bk_1,\bk_2) \tilde{\mathcal{R}}(\varpi_2)  G_{\psi v}^{(2)}(\bp_2,-\bk_2)\nonumber\\
&=-\frac{4 \epsilon_{\alpha\beta}\epsilon_{\rho\sigma}}{\varpi_1\varpi_2} \Bigg\{   \int_{\bk_1,\bk_1} G_{\psi u}^{(2)}(\bp_1,-\bk_1) k_1^\alpha  k_2^\rho  \frac{\p^2}{\p k_1^\beta \p k_2^\sigma} \tilde \Gamma_{uv}^{(2)}(\varpi_1+\varpi_1,\vk_1,\varpi_2+\varpi_2,\vk_2)  G_{\psi v}^{(2)}(\bp_2,-\bk_2) \nonumber\\
&+  p_2^\rho  \frac{\p}{\p  p_2^\sigma} \Big[ p_2^\alpha \frac{\p}{\p p_2^\beta} G_{\psi \psi}^{(2)}(\bp_1,\omega_2+\varpi_1+\varpi_2,\vp_2)\Big]\nonumber\\
 &+  p_1^\alpha  \frac{\p}{\p p_1^\beta} \Big[ p_1^\rho \frac{\p}{\p p_1^\sigma} G_{\psi \psi}^{(2)}(\omega_1+\varpi_1+\varpi_2,\vp_1,\bp_2)\Big]+  p_1^\alpha p_2^\rho \frac{\p^2}{\p p_1^\beta \p p_2^\sigma} G_{\psi \psi}^{(2)}(\omega_1+\varpi_2,\vp_1,\omega_2+\varpi_1,\vp_2)  \Bigg\}\,.
\label{eq:flotstreamG2uncrossed1PR}
\end{align}
 Inserting the expressions \eq{eq:flotstreamG2uncrossed1PI} and \eq{eq:flotstreamG2uncrossed1PR} into \eq{eq:flotstreamG2uncrossed} finally leads to the expected result
\begin{align}
 \p_\kappa & \tilde G^{(2)}_{\psi \psi}({\bp_1},{\bp_{2}})\Big|_{\mathit{uncrossed}}
= 
\frac{1}{2}\,  \int_{\varpi_1,\varpi_2}\!
\tilde K_{\psi \psi}(\varpi_1,\varpi_2) \nonumber\\
 & -\frac{4 \epsilon_{\alpha\beta}\epsilon_{\rho\sigma}}{\varpi_1\varpi_2} \Big[  p_1^\alpha  \frac{\p}{\p p_1^\beta} \Big[ p_1^\rho \frac{\p}{\p p_1^\sigma} G_{\psi \psi}^{(2)}(\omega_1+\varpi_1+\varpi_2,\vp_1,\bp_2)\Big] +  p_2^\rho  \frac{\p}{\p  p_2^\sigma} \Big[ p_2^\alpha \frac{\p}{\p p_2^\beta} G_{\psi \psi}^{(2)}(\bp_1,\omega_2+\varpi_1+\varpi_2,\vp_2)\Big] \nonumber\\
&+ p_1^\alpha p_2^\rho \frac{\p^2}{\p p_1^\beta \p p_2^\sigma} G_{\psi \psi}^{(2)}(\omega_1+\varpi_2,\vp_1,\omega_2+\varpi_1,\vp_2)  + (\varpi_1) \leftrightarrow (\varpi_2) \Big]\nonumber\\
&= \frac{1}{2}\,  \int_{\varpi_1,\varpi_2}\!
K_{\psi \psi}(\varpi_1,\varpi_2) \tilde{\mathcal{R}}(\varpi_1)  \tilde{\mathcal{R}}(\varpi_2) G_{\psi \psi}^{(2)}(\bp_1,\bp_2)\,.
\end{align}

\subsubsection{Uncrossed derivatives contribution for the flow of a generic $n$-point function}

Let us show that the result of the previous section for $\tilde G^{(2)}_{\psi \psi}({\bp_1},{\bp_{2}})$ can be generalized for any generic correlation function, \ie that
\begin{align}
\frac{\p^4}{\p q_1^\mu \p q_1^\mu \p q_2^\nu \p q_2^\nu}  &\Big[ \frac{\delta^2}{\delta \varphi_i(\bq_1)\delta\varphi_j(\bq_2)} 
 \tilde G^{(n)}_{i_1\dots i_{n}}[\{\bp_\ell\}_{1 \leq \ell \leq n};\mathrm{j}] \Big]_{\varphi=0} \Bigg|_{\vq_1=\vq_2 = 0}\nonumber\\
&=\delta_{i\psi}\delta_{j\psi}\tilde{\mathcal{R}}(\varpi_1)  \tilde{\mathcal{R}}(\varpi_2) \tilde G^{(n)}_{i_1\dots i_{n}}(\{\bp_\ell\})  \,.
\label{eq:ClosureUncrossed}
\end{align}
Following the same procedure as for the leading order at unequal times, let us first examine the action of only one functional derivative and subsequent two wave-number derivatives applied to a tree $\tilde{\mathcal{T}}^{(n)}$ composing $\tilde G^{(n)}_{i_1\dots i_{n}}(\{\bp_\ell\})$.
Using the property \eq{eq:RG2} demonstrated in the previous section, one  readily obtains
\begin{align}
\frac{\p^2}{\p q_a^\mu \p q_a^\mu} \frac{\delta}{\delta \varphi_i(\bq_a)} &\tilde{\mathcal{ T}}^{(n)}_{i_1\cdots i_n}[\{\bp_\ell\}]_{\varphi=0}\Big|_{\vq_a=0} \nonumber\\
&= \delta_{i\psi}\int_{\bk_\mathrm{ intern}}\sum_{k=1}^{m} \Bigg(\prod_{k'=1 \atop k'\neq k}^m \mathcal{ E}_{k'}^{\mathcal{T}}(\{\bp_\ell\}_{k'},\{\bk_\ell\}_{k'})\Bigg) \tilde{\mathcal{R}}(\varpi_a) \mathcal{ E}_k^{\mathcal{T}}(\{\bp_\ell\}_k,\{\bk_\ell\}_k)\,.
\end{align}
Thus, one only needs to show that the operator $\tilde{\mathcal{R}}(\varpi_a)$ verifies as well the Leibniz rule.
\begin{align}
 \tilde{\mathcal{R}}(\varpi_a) \int_{\bk} \mathcal{ E}_1^{\mathcal{T}}(\{\bk_\ell\}_1, -\bk) \mathcal{ E}_2^{\mathcal{T}}(\bk, \{\bk_\ell\}_2) &= \int_{\bk} \Big[ \tilde{\mathcal{R}}(\varpi_a) \mathcal{ E}_1^{\mathcal{T}}(\{\bk_\ell\}_1, -\bk) \mathcal{ E}_2^{\mathcal{T}}(\bk, \{\bk_\ell\}_2) \nonumber\\
 &+ \mathcal{ E}_1^{\mathcal{T}}(\{\bk_\ell\}_1, -\bk) \tilde{\mathcal{R}}(\varpi_a)  \mathcal{ E}_2^{\mathcal{T}}(\bk, \{\bk_\ell\}_2)\Big]\,.
\label{eq:leibnizR}
 \end{align}
This can be checked by inspection of the \rhs, which reads
\begin{align}
 \int_{\bk} &\Big[ \tilde{\mathcal{R}}(\varpi_a) \mathcal{ E}_i^{\mathcal{T}}(\{\bk_\ell\}_{\ell \in I}, -\bk) \mathcal{ E}_j^{\mathcal{T}}(\bk, \{\bk_\ell\}_{\ell \in J}) + \mathcal{ E}_i^{\mathcal{T}}(\{\bk_\ell\}_{\ell \in I}, -\bk) \tilde{\mathcal{R}}(\varpi_a)  \mathcal{ E}_j^{\mathcal{T}}(\bk, \{\bk_\ell\}_{\ell \in J})\Big]\nonumber\\
 &= \frac{2 i \epsilon_{ab}}{\varpi_a}\int_{\bk} \Big[ \sum_{i \in I} k_i^a \frac{\p}{\p k_i^b}  \mathcal{ E}_i^{\mathcal{T}}(\nu_i + \varpi_a, \vk_i, \{\bk_\ell\}_{\ell \in I\textbackslash i}, -\bk) \mathcal{ E}_j^{\mathcal{T}}(\bk, \{\bk_\ell\}_{\ell \in J}) \nonumber\\
 &+ \sum_{j \in J} k_j^a \frac{\p}{\p k_j^b} \mathcal{ E}_i^{\mathcal{T}}(\{\bk_\ell\}_{\ell \in I}, - \bk) \mathcal{ E}_j^{\mathcal{T}}(\bk, \nu_j + \varpi_a, \vk_j, \{\bk_\ell\}_{\ell \in J\textbackslash j})\nonumber\\
 &+ k^a \frac{\p}{\p k^b} \mathcal{ E}_i^{\mathcal{T}}(\{\bk_\ell\}_{\ell \in I}, -\nu + \varpi_a, -\vk) \mathcal{ E}_j^{\mathcal{T}}(\bk, \{\bk_\ell\}_{\ell \in J}) \nonumber\\
 &+ \mathcal{ E}_i^{\mathcal{T}}(\{\bk_\ell\}_{\ell \in I},-\bk) k^a \frac{\p}{\p k^b} \mathcal{ E}_j^{\mathcal{T}}(\nu + \varpi_a, \vk, \{\bk_\ell\}_{\ell \in J}) \Big]\,.
\end{align}
Integrating by part in $\vk$ and shifting the associated frequency, the two last terms cancel out, proving \eq{eq:leibnizR}. One concludes that
\begin{align}
\frac{\p^2}{\p q_a^\mu \p q_a^\mu} \frac{\delta}{\delta \varphi_i(\bq_a)} \tilde{\mathcal{ T}}^{(n)}_{i_1\cdots i_n}[\{\bp_\ell\}]_{\varphi=0}\Big|_{\vq_a=0} = \tilde{\mathcal{R}}(\omega_a) \tilde{\mathcal{ T}}^{(n)}_{i_1\cdots i_n}(\{\bp_\ell\})\,.
\end{align}

The remaining task to  prove \eq{eq:ClosureUncrossed} is to show that the same property holds for two functional derivatives and their subsequent wave-number derivatives. As for the leading order at unequal time,  distributing the two $\vq_2$ derivatives and setting $\vq_2$ to zero, one then applies the property \eq{eq:leibnizR}  to show that the resulting $\tilde{\mathcal{R}}(\varpi_2)$ can be pulled out
 of the remaining diagram
\begin{align}
 &\frac{\p^4}{\p q_1^\mu \p q_1^\mu \p q_2^\nu \p q_2^\nu} \frac{\delta^2}{\delta \varphi_i(\bq_1) \delta \varphi_j(\bq_2)} \tilde{\mathcal{ T}}^{(n)}_{i_1\cdots i_n}[\{\bp_\ell\}]_{\varphi=0}\Big|_{\vq_1=\vq_2=0} \nonumber\\
 &=\delta_{j\psi}\frac{\p^2}{\p q_1^\mu \p q_1^\mu }\Bigg\{ \tilde{\mathcal{R}}(\varpi_2)\Big[\int_{\bk_\mathrm{ intern}} \sum_{k} \Big(\prod_{k'\neq k} \mathcal{ E}_{k'}^{\mathcal{T}}(\{\bp_\ell\}_{k'},\{\bk_\ell\}_{k'})\Big)\frac{\delta}{\delta \varphi_i(\bq_1)} \mathcal{ E}_k^{\mathcal{T}}[\{\bp_\ell\}_k,\{\bk_\ell\}_k]\Big|_{\varphi=0}\Big]  \Bigg\}\Bigg|_{\vq_1=0}\, .
\end{align}
Expressing $ \tilde{\mathcal{R}}(\varpi_2)$ explicitly and distributing the $\vq_1$ derivative gives
\begin{align}
 &\frac{\p^4}{\p q_1^\mu \p q_1^\mu \p q_2^\nu \p q_2^\nu} \frac{\delta^2}{\delta \varphi_i(\bq_1) \delta \varphi_j(\bq_2)} \tilde{\mathcal{ T}}^{(n)}_{i_1\cdots i_n}[\{\bp_\ell\}]_{\varphi=0}\Big|_{\vq_1=\vq_2=0} \nonumber\\
 &= \delta_{j\psi} \frac{2 i \epsilon_{\rho\sigma}}{\varpi_2}\frac{\p^2}{\p q_1^\mu \p q_1^\mu } \Bigg\{ q_1^\rho \frac{\p}{\p q_1^\sigma} \Big[\int_{\bk_\mathrm{ intern}} \sum_{k} \Big(\prod_{k'\neq k} \mathcal{ E}_{k'}^{\mathcal{T}}(\{\bp_\ell\}_{k'},\{\bk_\ell\}_{k'})\Big)\frac{\delta}{\delta \varphi_i(\varpi_1+\varpi_2,\vq_1)} \mathcal{ E}_k^{\mathcal{T}}[\{\bp_\ell\}_k,\{\bk_\ell\}_k]\Big|_{\varphi=0}\Big] \nonumber\\
 &+ \sum_{k=1}^{n} p_k^\rho \frac{\p}{\p p_k^\sigma} \Big[ \int_{\bk_\mathrm{ intern}}\sum_{m} \Big(\prod_{m'\neq m} \mathcal{ E}_{m'}^{\mathcal{T}}(\{\bp_\ell\}_{m'}\textbackslash k_+,\{\bk_\ell\}_{m'})\Big)\frac{\delta}{\delta \varphi_i(\bq_1)} \mathcal{ E}_m^{\mathcal{T}}[\{\bp_\ell\}_m\textbackslash k_+,\{\bk_\ell\}_m]\Big|_{\varphi=0}\Big]  \Bigg\}\Bigg|_{\vq_1=0}\nonumber\\
 &=\delta_{j\psi} \frac{2 i \epsilon_{\rho\sigma}}{\varpi_2}\Bigg\{ 2 \delta_{\mu\rho} \frac{\p^2}{\p q_1^\mu \p q_1^\sigma }  \Big[ \int_{\bk_\mathrm{ intern}}\sum_{k} \Big(\prod_{k'\neq k} \mathcal{ E}_{k'}^{\mathcal{T}}(\{\bp_\ell\}_{k'},\{\bk_\ell\}_{k'})\Big)\frac{\delta}{\delta \varphi_i(\varpi_1+\varpi_2,\vq_1)} \mathcal{ E}_k^{\mathcal{T}}[\{\bp_\ell\}_k,\{\bk_\ell\}_k]\Big|_{\varphi=0}\Big]_{\vq_1=0} \nonumber\\
 &+ \delta_{i\psi} \sum_{k=1}^{n} p_k^\rho \frac{\p}{\p p_k^\sigma}   \Big[ \int_{\bk_\mathrm{ intern}}\sum_{m} \Big(\prod_{m'\neq m} \mathcal{ E}_{m'}^{\mathcal{T}}(\{\bp_\ell\}_{m'}\textbackslash k_+,\{\bk_\ell\}_{m'})\Big)\tilde{\mathcal{R}}(\varpi_1) \mathcal{ E}_m^{\mathcal{T}}(\{\bp_\ell\}_m\textbackslash k_+,\{\bk_\ell\}_m)\Big]  \Bigg\}\nonumber\\
 &=  \delta_{i\psi} \delta_{j\psi}\frac{2 i \epsilon_{\rho\sigma}}{\varpi_2} \sum_{k=1}^{n}  p_k^\rho \frac{\p}{\p p_k^\sigma} \Big[ \tilde{\mathcal{R}}(\varpi_1) \int_{\bk_\mathrm{ intern}} \prod_{m} \mathcal{ E}_m^{\mathcal{T}}(\{\bp_\ell\}_m\textbackslash k_+,\{\bk_\ell\}_m) \Big]\nonumber\\
 &= \delta_{i\psi}\delta_{j\psi} \tilde{\mathcal{R}}(\varpi_2)  \tilde{\mathcal{R}}(\varpi_1)\tilde{\mathcal{ T}}^{(n)}_{i_1\cdots i_n}(\{\bp_\ell\})\,,
\end{align}
where the first term in the second equality vanishes by antisymmetry of $\epsilon_{\rho\sigma}$.  This proves the property \eq{eq:ClosureUncrossed}, and yields \eq{eq:flowNLOuncrossed}.

\bibliographystyle{unsrt}
%\bibliography{../../Biblio-KPZ/biblioKPZ2}

\end{document}